%% file: SubsystemMWC_arXiv_v2.tex
\documentclass[11pt,a4paper]{article}

\input{preamble.tex}
\begin{document}

\begin{center}
    	{\Large\textsc{Higher-form entanglement asymmetry.}} \\[0.2em] 
        {\Large\textsc{Part I. The limits of symmetry breaking}} \\ 
	\bigskip
	Francesco Benini,\kern-3pt\textsuperscript{\textit{a,b}}\kern5pt Eduardo García-Valdecasas,\kern-3pt\textsuperscript{\textit{a,b}}\kern5pt and Stathis Vitouladitis\textsuperscript{\textit{c}} \\
	\bigskip\bigskip
	\footnotesize{
		{\normalsize \textsuperscript{\textit{a}}}SISSA, Via Bonomea 265, Trieste 34136, Italy \\[0.8em] 
		{\normalsize \textsuperscript{\textit{b}}}INFN -- Sezione di Trieste, Via Valerio 2, 34127 Trieste, Italy \\[0.8em]
		{\normalsize \textsuperscript{\textit{c}}}\textrm{Physique Théorique et Mathématique, \\ Université Libre de Bruxelles \& International Solvay Institutes, \\ CP 231, 1050 Brussels, Belgium} \\
	}
	\bigskip
	\href{mailto:fbenini@sissa.it}{\small \sf fbenini@sissa.it}\qquad \href{mailto:egarciav@sissa.it}{\small \sf egarciav@sissa.it}\qquad \href{mailto:stathis.vitouladitis@ulb.be}{\small \sf stathis.vitouladitis@ulb.be}
\end{center}

\begin{abstract}
    \noindent Entanglement asymmetry is a relative entropy that faithfully diagnoses symmetry breaking in quantum states, possibly within a spatial subregion. In this work, we extend such framework to higher-form symmetries and compute entanglement asymmetry in theories with spontaneously-broken continuous zero- and higher-form symmetries. One of our central results is an entropic Coleman--Mermin--Wagner theorem, for 0- and $p$-form symmetries, valid also on subregions, which forbids spontaneous breaking of continuous $p$-form symmetries in spacetime dimensions $d\leq p+2$. Our theorem not only qualifies symmetry breaking, it also quantifies it: spontaneous breaking triggers a nonvanishing entanglement asymmetry that grows monotonically towards the infrared, and counts the number of Goldstone fields. Along the way, we derive standalone results concerning the entanglement entropy and asymmetry of Goldstone bosons and gauge fields. In particular, we find a closed-form expression for the Rényi asymmetries of a compact scalar field on spherical subregions in three and four spacetime dimensions, and for higher-form gauge fields in higher dimensions.
\end{abstract}

\pagebreak
    \vspace{10pt}
    \tableofcontents
\pagebreak

\section{Introduction and summary}

Over the past decade, our understanding of quantum many-body systems and quantum field theory (QFT) has advanced substantially thanks to the generalisation of symmetry far beyond the standard textbook treatment. From a modern point of view, symmetries are given by the set of topological operators of a theory --- each of which implements a symmetry action --- together with a prescription of what happens when these operators meet. Standard symmetries are implemented by codimension-$1$ operators whose fusion obeys a group law. A first generalisation consists in allowing topological operators of higher codimension, generating so-called \(p\)-form symmetries \cite{Gaiotto:2014kfa}. Besides, topological operators may fail to satisfy a traditional group-like fusion law. In particular, they may not have an inverse. This gives rise to noninvertible symmetries \cite{Bhardwaj:2017xup, Chang:2018iay}. The general case involves combinations of the two, whereby the operators and their fusion rules furnish a higher fusion category, see \cite{Cordova:2022ruw, Schafer-Nameki:2023jdn, Shao:2023gho} for reviews. This shift in viewpoint has brought a wide array of physical phenomena under the umbrella of symmetry.

A second significant thread of progress has emerged from the interaction with quantum information theory. In this regard, entanglement has come to function as one of the most incisive probes of QFT. In two-dimensional conformal field theory it famously counts degrees of freedom \cite{Holzhey:1994we, Calabrese:2004eu}. More generally, it provides irreversibility theorems for the renormalisation group (RG) in various dimensions \cite{Casini:2004bw, Casini:2012ei, Casini:2016udt, Casini:2017vbe}, even in the presence of boundaries or defects \cite{Casini:2016fgb, Casini:2018nym, Casini:2022bsu, Casini:2023kyj}. Moreover, it serves as a diagnostic of topological order \cite{Kitaev:2005dm, Levin:2006zz}. Perhaps most notably, entanglement has become indispensable in efforts to understand quantum gravity. See, for instance, \cite{Faulkner:2022mlp} for a review.

It is then natural to investigate entanglement in the presence of symmetries; see, for example, \cite{Goldstein:2017bua, Murciano:2020vgh, Xavier:2018kqba, Choi:2024wfm, Das:2024qdx}. A particularly noteworthy development appeared in \cite{Ares:2022koq}, which introduced \emph{entanglement asymmetry} as a diagnostic to characterise symmetry breaking in states using methods of entanglement entropy. This quantity is closely related to previously defined notions in quantum information theory \cite{Vaccaro:2007drw, Gour:2009hng, Marvian:2014awa}, in particular in the resource theory of frameness, and in algebraic QFT \cite{Casini:2019kex, Casini:2020rgj, Magan:2020ake, Magan:2021myk}, where it is an example of an \textit{entropic order parameter}.

Entanglement asymmetry has already been explored across a broad range of physical regimes. In condensed-matter systems it appeared in studies of explicit or spontaneous symmetry breaking, matrix product states, integrable models, and quantum circuits \cite{Murciano:2023qrv, Ferro:2023sbn, Fossati:2024xtn, Lastres:2024ohf, Fossati:2024ekt, Capizzi:2023yka, Capizzi:2023xaf, Rylands:2023yzx, Murciano:2023qrv, Klobas:2024mlb, Ares:2023kcz, Turkeshi:2024juo, Liu:2024kzv, Chalas:2024wjz}. Its out-of-equilibrium behaviour is particularly interesting, as it revealed a ``quantum Mpemba effect'' (see \cite{Ares:2025onj} and references therein), later observed experimentally \cite{Joshi:2024sup}. In high-energy physics it has been examined in conformal field theories \cite{Chen:2023gql, Kusuki:2024gss, Benini:2024xjv, Fujimura:2025rnm, Benini:2025lav}, in certain two-dimensional gauge theories \cite{Florio:2025xup}, and in toy models of evaporating black holes \cite{Ares:2023ggj, Chen:2024lxe, Russotto:2024pqg, Russotto:2025cpn}.

To fix ideas, we briefly recall the definition of entanglement asymmetry in the simplest setting, involving ordinary (i.e., zero-form and invertible) symmetries. Suppose we have a quantum mechanical theory that enjoys a global symmetry given by a group \(G\). Consider a state, described by a trace-normalised density matrix \(\rho\), which can be pure or mixed, defined on a subsystem or on the entire system. If one wants to test whether or not the symmetry is broken in the state \(\rho\), one can compare it to its symmetrisation, which by construction preserves the symmetry:
\begin{equation}
\label{eq:rhoG}
	\rho_\sfS \coloneqq \frac{1}{\abs{G}} \, \sum_{g\in G} \, U_g\; \rho\; U_g^\dagger ~.
\end{equation}
Here \(U_g\) are the operators that implement the symmetry action on the Hilbert space. Eqn.~\cref{eq:rhoG} then describes a trace-preserving quantum channel that projects to the neutral singlet representations of \(G\). For brevity we will sometimes refer to this map \(\rho\mapsto\rho_\sfS\) as a \emph{symmetriser}. The above formula is written for a discrete group, but the generalisation to continuous symmetries is straightforward and it involves an integral in place of the sum.

A simple and, as it turns out, useful way to compare the two states \(\rho\) and \(\rho_\sfS\) is the relative entropy between them:
\begin{equation}
	\entas_G(\rho) = \relent{\vphantom{\big|}\rho}{\rho_\sfS} = \tr \bigl( \rho \, \qty(\log \rho - \log \rho_\sfS) \bigr) ~.
\end{equation}
This is entanglement asymmetry. By the properties of relative entropy, we immediately have \(\entas_G\geq 0\) on any state and moreover \(\entas_G = 0\) if and only if \(\rho=\rho_\sfS\). Hence it is a faithful measure of symmetry breaking. Moreover, since the map \(\rho\mapsto \rho_\sfS\) is a projector, \(\entas_G\) admits an alternative expression as a difference of von Neumann entropies:
\begin{equation}
\label{eq:entas-def}
	\entas_G(\rho) = \ent\qty(\rho_\sfS)-\ent(\rho) ~.
\end{equation}
This equivalent presentation has certain computational benefits --- of which the main one is enabling Rényi versions of the asymmetry, by means of the replica trick.

While originally studied on the lattice, entanglement asymmetry displays a number of important features when imported into QFT. To begin with, as a relative entropy, it is free of ultraviolet (UV) divergences, at least whenever the symmetry breaking is spontaneous, or explicit but soft. Furthermore, it depends solely on the state and not on the Hamiltonian, and so it captures universal data. Universal and UV-finite observables are notoriously scarce in QFT, which makes this one particularly valuable.

Another remarkable trait is the following. Monotonicity of relative entropy \cite{Lindblad:1975kmh, Uhlmann:1977mon} implies that entanglement asymmetry cannot increase under any quantum channel \(\cE\) that commutes with the symmetriser:
\begin{equation}
	\entas_G(\rho) \geq \entas_G \bigl( \cE(\rho) \bigr) ~.
\end{equation}
Consider now the situation of typical interest in QFT: a state \(\rho_\aent\) defined on a region \(\aent\). For any subregion \(\cB\subseteq \aent\), the partial trace over the complement of \(\cB\) is such a trace-preserving quantum channel.%
\footnote{Here we implicitly assume a factorisation of the Hilbert space. This requires care, but can be achieved. See for instance \cite{Ohmori:2014eia}.}
Denoting the resulting reduced state on \(\cB\) by \(\rho_\cB\), entanglement asymmetry satisfies:
\begin{equation}
	\entas_G(\rho_\aent) \geq \entas_G(\rho_\cB) ~,
\end{equation}
for any \(\cB\subseteq \cA\) and for all states. This is already an interesting structural constraint. Specialising further to a specific class of regions, it becomes even more powerful. For instance, let \(\cA\) be a spherical ball of radius \(R_\cA\) and \(\cB\) a smaller concentric ball of radius \(R_\cB < R_\cA\). It is clear that tracing out the complement of \(\cB\) naturally realises an RG flow. Writing \(\entas_G(R)\) for the entanglement asymmetry of the reduced state (which we leave implicit) on a ball of radius \(R\), it follows that:
\begin{equation}
\label{eq:entasgrowth}
	\dv{R} \, \entas_G(R) \geq 0 ~.
\end{equation}
The above establishes entanglement asymmetry as an RG monotone.%
\footnote{We thank the participants in the workshop ``Entanglement asymmetry across energy scales'', and especially Victor Godet, for discussions on this topic. We note also a related argument in \cite{Casini:2020rgj}. Although we used spherical regions as the most illustrative and natural case, a broader class of shapes, \emph{star domains} (a mild generalisation of convex sets), also admits an RG flow interpretation. On the other hand, it was shown in \cite{Benini:2025lav} in explicit examples that R\'enyi asymmetries are not always monotonic in the subsystem size.}

Entanglement asymmetry detects symmetry breaking for both ground and excited states. Whenever \(\rho\) is the ground state, the question is one of \emph{spontaneous} symmetry breaking (SSB). In the standard Landau paradigm, SSB is diagnosed in effective field theory by a local operator, an \emph{order parameter}, whose expectation value is nonvanishing in the broken phase. One of the standard textbook accounts of SSB goes as follows. Start with the underlying theory at finite volume \(V\) and turn on a symmetry-breaking source \(h\). Calculate the expectation value \(\ev{\cO}_{V,h}\) of an order parameter \(\cO\). If 
\begin{equation}
	\lim_{h\to 0} \, \lim_{V\to\infty} \ev{\cO}_{V,h} \neq 0 ~,
\end{equation}
the theory exhibits SSB. Note that the limits do not commute: at finite \(V\), sending \(h\to 0\) first forces \(\ev{\cO}=0\) by selection rules. Let us now give an alternative way of detecting SSB through the asymmetry. Consider a generic (clustering, excited) symmetry breaking density matrix \(\rho_V\) at finite volume \(V\) and evolve it in Euclidean time \(\beta\), with the system's Hamiltonian:
\begin{equation}
	\rho_{V,\beta} = \frac{\ex{-\beta H}\, \rho_V\, \ex{-\beta H}}{\tr(\rho_V\, \ex{-2\beta H})} ~. 
\end{equation}
The limit \(\beta\to\infty\) projects to the vacuum. At finite volume this is unique, but taking \(V\to\infty\) first can lead to SSB. Hence SSB occurs whenever
\begin{equation}
	\lim_{\beta\to\infty}\lim_{V\to\infty} \entas_G\qty(\rho_{V,\beta}) \neq 0 ~. 
\end{equation}
This is the precise sense in which \(\entas_G\) is a good entropic order parameter (using the terminology of \cite{Casini:2020rgj}).

In recent years, a wide class of phases has emerged where symmetries break spontaneously but no local order parameter exists, known as ``beyond Landau’’ phases \cite{Wegner:1971app, Wen:1989iv, Wen:2007joe, Savary:2016ksw, Nandkishore:2018sel}. There is an extensive body of work that attempts to incorporate such phases in a suitable extension of the Landau paradigm \cite{Nussinov:2006iva, Nussinov:2009zz, Kovner:1992pu, Iqbal:2020msy, Zhao:2020vdn, Iqbal:2021rkn, Liu:2024znj, Bhardwaj:2023fca, Bhardwaj:2024wlr}. Nonetheless, the approach of entanglement asymmetry that we are advocating here shows its first clear advantage: regardless of whether the phase is Landau or not, as long as one can identify a ground state of a quantum system, \(\entas\) provides a faithful measure of SSB.

A particularly important subclass of such phases of matter are those that break higher-form symmetries \cite{Gaiotto:2014kfa}. From the high-energy perspective, they are central because of their deep ties to confinement \cite{Polyakov:1975rs}. Here, local order parameters are replaced by extended operators: line operators or, in general, \(p\)-dimensional objects, whose perimeter-law expectation values signal spontaneous breaking of a 1-form (or more generally \(p\)-form) symmetry. Meanwhile, in condensed-matter settings, phases that break discrete higher-form symmetries play an equally prominent role: they underlie deconfined gapped phases, known as \emph{topological orders}. In that case the link to information-theoretic language is especially direct: their ground states exhibit long-range entanglement \cite{Wen:1989iv}. 

Our first goal in this paper is to adapt entanglement asymmetry to generalised symmetries. Here we focus on higher-form symmetries. For noninvertible symmetries (with particular focus on two-dimensional systems) this task was undertaken in \cite{Benini:2025lav, AliAhmad:2025bnd}.%
\footnote{The framework of \cite{Benini:2025lav} is in principle applicable to finite higher-form symmetries as well. However, the formalism we present here proves more suited to our applications. See also \cite{Schafer-Nameki:2025fiy} for a discussion of mixed states and noninvertible symmetries.}
A benefit of this approach is that it provides an entanglement-based characterisation of (spontaneous) symmetry breaking for both continuous and discrete symmetries. Moreover, it can do so directly in the low-energy effective field theory, dispensing of the need of a lattice realisation. In the present paper we focus on continuous symmetries. In a follow-up paper \cite{benini_garcia-valdecasas_vitouladitis_toappear} we will elaborate on discrete symmetries and anomalies. We also note a related work that recently appeared \cite{Lamas:2025eay}, which discusses entanglement asymmetry for (finite) higher-form symmetries on the lattice.

\subsubsection*{Entropic symmetry-breaking theorems}

A resounding result in quantum field theory is the theorem due to Mermin--Wagner \cite{Mermin:1966fe} and Coleman \cite{Coleman:1973ci} (hereby the \emph{CMW theorem}) that continuous symmetries cannot break spontaneously in spacetime dimensions \(d\leq 2\). A similar statement exists for higher-form symmetries \cite{Gaiotto:2014kfa, Lake:2018dqm}, stating that a continuous \(p\)-form symmetry cannot break spontaneously when \(d\leq p+2\). As an application of our formalism, we examine the SSB patterns of quantum field theories with continuous symmetries and establish an \emph{entropic} CMW theorem both for 0-form and higher-form symmetries. A benefit of our entropic theorems is that they are also \emph{quantitative}: we show how SSB is controlled by the system size. 

Let us state this result more precisely. We consider effective field theories (EFTs) in which a continuous symmetry --- Abelian or non-Abelian, 0-form or \(p\)-form --- is realised nonlinearly, and compute the entanglement asymmetry of a pure clustering ground state \(\rho\). We start by considering the entanglement asymmetry of the full system (as opposed to a subsystem). We find that it behaves as follows (stated  here in the general case of a \(p\)-form symmetry):
\begin{equation}
\label{eq:CMWteaser1}
	\entas(\rho) = 
	\begin{cases} 
		0 ~,& d\leq p+2 ~, \\[0.2em]
	\dfrac{\cN (d-p-2)}{2} \, \log(\mu R) ~, & d>p+2 ~.
	\end{cases}
\end{equation}
Here \(R\) is the linear system size, taken to infinity in accordance with the thermodynamic limit, and \(\mu\) is an energy scale: the UV cutoff of the EFT.%
\footnote{We also consider symmetry breaking of \(p\)-form symmetries in CFTs. In that case, instead of by a cutoff, symmetry breaking is controlled by the ratio between the size of a \((d{-}p{-}1)\)-dimensional manifold and that of a compact \(p\)-dimensional manifold, in accordance with SSB patterns of higher-form symmetries \cite{Lake:2018dqm, Cordova:2019bsd, Maeda:2025ycr}.}
This is a first incarnation of our entropic CMW theorem: in \(d \leq p+2\), the symmetry is unbroken and the ground state is unique. But in \(d > p+2\), there exists a moduli space of symmetry-breaking vacua. Entanglement asymmetry detects this by picking up a logarithmic growth.

The dimension of this moduli space is encoded in the coefficient \(\cN\) of the logarithmic scaling. Specifically, we find that \(\cN\) is the number of broken generators of the symmetry under consideration. For instance, in the case of \(\U(1)\) symmetry breaking we find  
\begin{equation}
	\cN_{\U(1)} = 1 ~,
\end{equation}
while for a general pattern \(G\to H\) the coefficient is:
\begin{equation}
\label{eq:NGH}
	\cN_{G/H} = \dim(G/H) = \frac{\dim G}{\dim H} ~.
\end{equation}
In the 0-form case, our results match with results obtained on the lattice using excited matrix product states \cite{Capizzi:2023xaf}. 

More strikingly, this result provides an indirect proof of a conjecture by Metlitski and Grover \cite{Metlitski:2011pr},%
\footnote{See also \cite{Kallin:2011utz, Kulchytskyy:2015yea}.}
and in doing so, it clarifies some key features of entanglement asymmetry. Let us elaborate on that. In \cite{Metlitski:2011pr} the authors computed the entanglement entropy for Goldstone modes associated with the symmetry breaking pattern \(\O(N)\to\O(N-1)\). Based on their results, they conjectured that the entanglement entropy of Goldstone fields should contain a universal logarithmic term:
\begin{equation}
	\ent_\t{Goldstone} = a \, \qty(\frac{R}{\epsilon})^{d-2} + b \log(\mu R) + \t{const} ~.
\end{equation}
Here \(\epsilon\) is a short-distance cutoff and \(\mu\) is the EFT scale. The area-law coefficient \(a\) is non-universal, but \(b\) was conjectured to be universal: 
\begin{equation}
\label{eq:b-Metlitski-Grover}
	b = \frac{\cN_{G/H} (d-2)}{2} ~,
\end{equation}     
with \(\cN_{G/H}\) as in \cref{eq:NGH}. Here we find that 
\begin{enumerate*}[label=(\alph*)]
	\item the entanglement asymmetry \emph{isolates} the universal logarithmic behaviour, and
	\item the coefficient is precisely \cref{eq:b-Metlitski-Grover}.
\end{enumerate*} 
Moreover, our results extend to higher-form symmetries. In this case the role of the Goldstones is played by higher-form fields: photons and \(p\)-form generalisations \cite{Gaiotto:2014kfa, Hofman:2018lfz, Lake:2018dqm}. There is again a logarithmic divergence in their entanglement entropy, with coefficient
\begin{equation}
	b_{p\t{-form}} = \frac{\cN (d-p-2)}{2} ~, 
\end{equation}
where \(\cN\) is now the number of broken \(\U(1)\) symmetries. Crucially, $\cN$ counts broken generators, not the individual polarisation modes of the associated Goldstone fields.

\subsubsection*{Subregion theorems}

A further advantage of entanglement asymmetry is that it can detect symmetry breaking in a subsystem. This can either consist of a subset of the degrees of freedom, or of the degrees of freedom associated with a spatial subregion. The latter is the case that we focus on in the present work. Having a local description of symmetry breaking could be very useful in situations where some of the degrees of freedom are inaccessible. A prototypical example of such a situation is in quantum gravity, where degrees of freedom are often hidden behind a black hole or a cosmological horizon.

We utilise entanglement asymmetry to explore such situations in quantum field theory and we establish a subregion CMW theorem. For concreteness we focus on \(\U(1)\) \(p\)-form symmetries. To get a flavour of our result, suppose that \(\rho\) is a clustering ground state at infinite volume, and \(\rho_\aent\) is its restriction to a subregion \(\aent\) of size \(R\). The entanglement asymmetry of that state has the following behaviour:
\begin{equation}
\label{eq:CMWteaser2}
	\entas_{\gf{\U(1)}{p}}(\rho_\cA) = 
	\begin{cases} 
		0 ~,& d\leq p+2 ~, \\[.2em]
	f(\mu R) ~, & d>p+2 ~.
	\end{cases}
\end{equation}
Here \(f(\mu R)\) is a monotonic function of \(\mu R\),%
\footnote{In particular, we analytically compute the exact entanglement asymmetry of a compact scalar field on a spherical subregion $\aent$ in $d=3$ and $d=4$ dimensions.}
which interpolates between no symmetry breaking at small scales:
\begin{equation} 
	f(\mu R) \,\sim\, 0 \qquad\qq{when} \mu R \ll  1 ~,
\end{equation}	
and full symmetry breaking at large scales, when the subregion covers the entire Cauchy slice:
\begin{equation}
	f(\mu R) \,\sim\, \frac{d-p-2}{2}\;\log(\mu R) \qquad\qq{when} \mu R \gg  1 ~.
\end{equation}
In other words, for $d > p+2$, the subregion entanglement asymmetry coincides at leading order with the one of the full system when the region is big, and it vanishes when the region is small. Recalling that the inverse size of the subregion, $R^{-1}$, is an RG flow scale is clarifying. In the deep infrared (IR) the entanglement asymmetry coincides  with the one of the full system. In the ultraviolet (UV) the symmetry is restored and the entanglement asymmetry vanishes.

The rest of the paper is organised as follows. In \cref{sec:higher-form-asymmetry} we extend the definition of entanglement asymmetry to higher-form symmetries and review the path integral methods used to compute it. In \cref{sec:entropic-CMW-global} we consider  global states, i.e. no subsystem. We canonically quantise a free $p$-form gauge field in $d$ dimensions and compute entanglement asymmetry in the ground state. The results reproduce the $p$-form CMW theorem. In \cref{subsec:nonabelian} the computation is extended to non-Abelian $0$-form symmetries. In \cref{sec:subregion0-form} we restrict to a spatial subregion in the Abelian $0$-form symmetry case and compute the Rényi asymmetries of the ground state using path integral methods. The results match those of the global state in the appropriate limit and extend the CMW theorem to spatial subregions. In \cref{sec:subregionp-form} we generalise these results to the $p$-form case. We close this work in \cref{sec:discussion} with a discussion of open problems and promising avenues to be explored. Some technical results are relegated to the \cref{App:TopRepMan,app:electrostatics,app:smooth-Bent}.


\section{Entanglement asymmetry for higher-form symmetries}
\label{sec:higher-form-asymmetry}

The aim of this section is to extend the notion of entanglement asymmetry to encompass higher-form symmetries in $d$ spacetime dimensions. This generalisation falls into place once we rewrite the definition \cref{eq:rhoG} in a  more geometric language. For concreteness, let us take \(G\) to be an Abelian continuous compact group. We begin with global states, namely those defined on the entire $(d-1)$-dimensional spatial slice \(\Sigma\). On such a slice, the topological defects that implement a standard 0-form symmetry are labelled by a \((d-1)\)-cycle in the homology group \(\H_{d-1}(\Sigma;G)\) with coefficients in \(G\). Since we are primarily concerned with connected spatial slices (the generalisation to multiple connected components being straightforward), we have the isomorphism \(\H_{d-1}(\Sigma;G)\cong G\). 

With this identification, the symmetrised density matrix can be written as
\begin{equation}
\label{eq:rhoG-hom}
	\rho_\sfS = \int_{\H_{d-1}(\Sigma;G)} \dd{\mu(a)} \, U_a\; \rho\; U_a^\dagger ~.
\end{equation}
The measure
\begin{equation}
	\dd{\mu(a)} = \frac{\dd{a}}{\vol \bigl( \H_{d-1}(\Sigma;G) \bigr)}
\end{equation}
is the pushforward of the Haar measure of \(G\) by the map implementing the isomorphism \(G\xlongrightarrow{\sim}\H_{d-1}(\Sigma;G)\). It is normalised so that the total measure is unity, ensuring that \(\rho_\sfS\) remains trace-normalised. When \(G\) is discrete, its Haar measure is naturally the counting measure and \cref{eq:rhoG-hom} reduces to \cref{eq:rhoG}.%
\footnote{Let us stress that, although the symmetrisation in eqn.~\cref{eq:rhoG-hom} might resemble a higher gauging \cite{Roumpedakis:2022aik} of $G$ on $\Sigma$, in fact it is a distinct operation. Indeed, if we define the condensate $C$ as such a higher gauging, $C = \int_{H_{d-1}(\Sigma;G)} d\mu(a) \, U_a$, then the symmetrisation $\rho_\sfS$ is \emph{not} $C \rho \, C$.}

Now suppose we focus not on global states, but rather on reduced states \(\rho_\aent\) on a subregion \(\aent \subset \Sigma\). In general the Hilbert space does not factorise between subregions. A fix was proposed in \cite{Ohmori:2014eia}: one rather considers a linear map from the total Hilbert space to a factorised Hilbert space. The latter can be obtained by cutting a tubular neighbourhood around \(\pd\aent\), and implementing suitable boundary conditions on this thickened entangling surface. We will discuss this in greater detail in the following sections; for the time being let us simply assume we have achieved such a factorisation. 

With suitable boundary conditions on the regulated entangling surface, the symmetry operators can act on the factorised Hilbert space:  
\begin{equation}
	U_{a}(\Sigma) \,\to\, U_a(\aent) \; U_a(\coaent) ~,
\end{equation}
where \(\coaent=\Sigma\setminus\aent\) denotes the complement of \(\aent\). If the theory allows for symmetric boundary conditions, such a factorisation is immediate.%
\footnote{Even when symmetric simple boundary conditions are not available, for instance because of the presence of anomalies \cite{Jensen:2017eof, Thorngren:2020yht}, factorisation can be achieved by averaging over the orbit of boundary conditions \cite{Benini:2025lav}.}
For notational clarity we do not denote explicitly the boundary conditions, but one needs to keep in mind that such a choice was made. At the end of the day, the setup becomes equivalent to the standard one of \cite{Ares:2022koq}, which assumes that the charge operators decompose as $Q(\Sigma) = Q\qty(\aent)\otimes\id+\id\otimes Q\qty(\coaent)$. In this expression, \(Q\qty(\aent)\) acts solely on the Hilbert space of \(\aent\), while \(Q(\coaent)\) acts on the Hilbert space of the complement. Then, the symmetrised state is obtained by averaging over topological defects that can end on \(\pd \aent\). Such defects are labelled by \emph{relative cycles} \(a\in\H_{d-1}(\aent,\pd \aent;G)\). This fits consistently with the fact that
\begin{equation}
	\H_{d-1}(\aent, \pd \aent; G) \cong G^{b_0(\aent)} ~,
\end{equation} 
where \(b_0(\aent)\) is the number of connected components of \(\aent\). In this language, the symmetrised density matrix is expressed as
\begin{equation}
\label{eq:rhoRG}
	\rho_{\aent,\sfS} = \int_{\H_{d-1}(\aent,\pd \aent;G)} \dd{\mu(a)} \, U_a\; \rho_\aent\; U_a^\dagger ~.
\end{equation}
While for ordinary symmetries this framework may look like overkill, it will prove useful by making the passage to higher-form symmetries completely natural. Let us make the transition smooth by temporarily returning to global states on the entire spatial slice.

For higher-form symmetries, the previous discussion goes mostly unchanged. The main difference is that now topological operators are labelled by group-valued \((d-p-1)\)\nobreakdash-cycles \(a\in\H_{d-p-1}(\Sigma;G)\). Symmetrising the density matrix with respect to a $p$-form symmetry is therefore a straightforward generalisation of \cref{eq:rhoRG}:%
\footnote{The description in terms of homology becomes slightly imprecise in the presence of anomalies: while $H_{d-p-1}(\Sigma; G)$ is an Abelian group, the product of operators $U_a$ may become non-Abelian \cite{Benini:2025lav}. For instance, consider a $\bbZ_2$ 1-form symmetry with self-anomaly in 2+1 dimensions. On the torus $\Sigma = T^2$ the operators $U_x$ and $U_y$, corresponding to the nontrivial $\bbZ_2$ line along the two cycles of the torus, anticommute. Therefore, in the identification of the topological operators on $\Sigma$ with the elements of $H_{d-p-1}(\Sigma; G)$ there might be ambiguous phases. Those phases, however, cancel out in \cref{eq:rho-sym-def} between $U_a$ and $U_a^\dag$.}
\begin{equation}
\label{eq:rho-sym-def}
	\rho_\sfS = \int_{\H_{d-p-1}(\Sigma;G)} \dd{\mu(a)} \, U_a\; \rho\; U_a^\dagger ~.
\end{equation}
Inserting this symmetrised density matrix into the definition \cref{eq:entas-def} yields the entanglement asymmetry, giving a seamless prescription for computing it in the case of higher-form symmetries. One can immediately check that \(\sfS:\rho\mapsto\rho_\sfS\) is a projector, and hence it satisfies the desired properties of entanglement asymmetry:
\begin{align}
	& \entas_{\gf{G}{p}}(\rho) \geq 0~, \qquad\qq{and} \\ 
	& \entas_{\gf{G}{p}}(\rho) = 0\quad \Leftrightarrow\quad \rho = \rho_\sfS ~. 
\end{align}

Rather than calculating directly the entanglement asymmetry, it is often more convenient to calculate its close cousins: the \emph{Rényi asymmetries}. These are defined in terms of the Rényi entropies as
\begin{equation}
\label{eq:Rényis}
    \entas_{\n,\gf{G}{p}}(\rho) \coloneqq \frac{1}{1-\n} \log\frac{\tr(\rho_\sfS^\n)}{\tr(\rho^\n)} ~,
\end{equation}
and are indexed by an integer \(\n \geq 2\). The symmetrised contribution takes the following form:
\begin{equation}
    \tr(\rho_\sfS^\n) = \int_{\H_{d-p-1}(\Sigma;G)} \, \prod_{i=1}^\n\dd \mu_i(a_i) \, \cE_\n(\vec{a}) ~,
\end{equation}
where $\vec{a}$ denotes the $\n$-component vector of elements $a_i \in\H_{d-p-1}(\Sigma;G) $. The $\cE_\n(\vec{a})$ are the \textit{charged moments}:
\begin{equation}
\label{eq:ChargedMoments}
    \cE_\n(\vec{a}) = \tr [ \, \prod_{i=1}^\n U_{a_i}\, \rho\, U_{a_i}^\dagger] ~.
\end{equation}
The connection to entanglement asymmetry comes naturally through the replica trick. Specifically, once the Rényi asymmetries are calculated, by analytically continuing \(\n\) and taking the limit \(\n\to 1\) one recovers the entanglement asymmetry:
\begin{equation}
\label{Eq:replicatrick}
    \entas_{\gf{G}{p}}(\rho) = \lim_{\n\to 1} \; \entas_{\n,\gf{G}{p}}(\rho) ~.
\end{equation}

\subsection{Path integral methods}
\label{sec:pathintegralmethods}

When dealing with reduced density matrices associated with spatial subregions, we will employ a Euclidean path integral representation. This is a standard tool in the computation of entanglement entropies in quantum field theory (see  \cite{Callan:1994py} for a classical work and \cite{hartman2015lectures} for a gentle introduction), which recently has also been used to study entanglement asymmetries in systems with 0-form symmetries \cite{Capizzi:2023yka, Fossati:2024xtn, Benini:2024xjv}. We briefly review the construction here, both to fix our notation and to highlight the ingredients relevant for later sections. This will form the backbone of the calculations presented in \cref{sec:subregion0-form,sec:subregionp-form}.

A Euclidean path integral on an open manifold $X_d$ with boundary $\partial X_d = \Sigma_{d-1}$ prepares a state $\ket{\Psi}$ on $\Sigma_{d-1}$. Likewise, a path integral with two boundaries defines an (unnormalised) density matrix $\rho$. The density matrix $\rho = \ketbra{\Psi}$ of a pure state is built from a path integral with two identical disconnected open components with opposite orientation, each preparing $\ket{\Psi}$ and $\bra{\Psi}$, respectively. In pictures:
\begin{equation}
\label{eq:rho}
	\rho =  \vcenter{\hbox{\def\svgwidth{6em}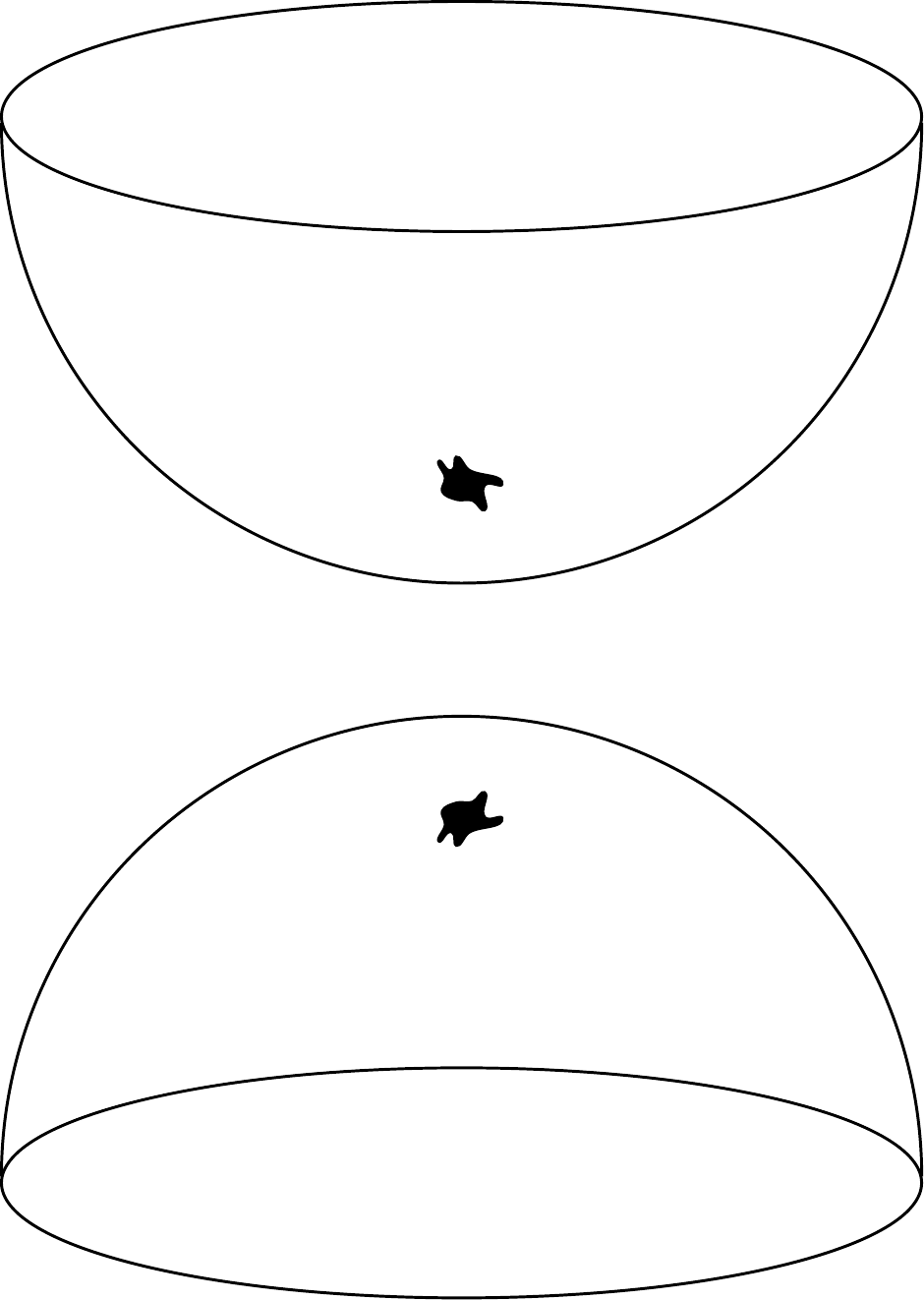}} \;.
\end{equation}
The black blob denotes any nontrivial geometry or operator insertions. Taking the trace of the density matrix amounts to gluing the two boundaries, $\partial X_d = \Sigma \cup \overline{\Sigma}$. A reduced density matrix $\rho_\mathcal{A}$ for a subregion $\mathcal{A} \subset \Sigma$ is computed by taking a partial trace of $\rho$ over the complement $\mathcal{A}^\prime$ of $\mathcal{A}$. In the path-integral language, this is implemented by a partial gluing along $\mathcal{A^\prime} \subset \Sigma$ and $\overline{\mathcal{A^\prime}} \subset \overline{\Sigma}$. As mentioned before, an important subtlety is that in taking the partial trace one needs a factorisable Hilbert space, that is $\mathcal{H}_{\Sigma} = \mathcal{H}_{\mathcal{A}} \otimes \mathcal{H}_{\mathcal{A}^\prime}$. This is not generically the case because of UV divergences and gauge constraints. In practice, one can regulate the entangling surface \(\pd\aent\) by thickening it to \(\cR_\varepsilon \coloneqq \pd\aent\times \closed{0}{\varepsilon}\) and prescribing boundary conditions \(\sfb\)  on \(\pd\cR_\varepsilon\) \cite{Ohmori:2014eia}. This provides a map:
\begin{equation}
\label{eq:brick}
	\iota_\sfb:\cH_\Sigma \,\to\, \cH_{\aent}^{\sfb}\otimes \cH_{\coaent}^{\sfb} ~.
\end{equation} 
The resulting entanglement entropies generically depend on the boundary condition. However, as we will discuss in due time, the results presented in this work are independent of the boundary conditions. We will therefore ignore their implications for entanglement asymmetry in this work, which we leave for the future \cite{benini_garcia-valdecasas_vitouladitis_toappear}.%
\footnote{For 0-form symmetries in 2d CFTs, independence of entanglement asymmetry from the regulating boundary conditions was discussed in \cite{Benini:2025lav}.}
Let us illustrate the path integral methods for a circular slice $\Sigma = S^1$ and a segment subregion $\mathcal{A}$. In pictures, the path integral computing a reduced density matrix is
\begin{equation}
\label{eq:ReducedDensity}
	\rho_\aent \,=\;\;
    \vcenter{\hbox{\footnotesize \def\svgwidth{6em}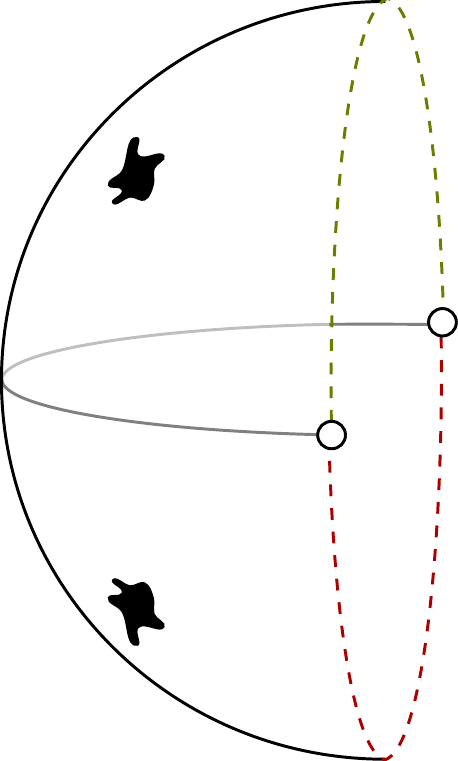}} \quad\;\;,
\end{equation}
where we have denoted the subregion $\mathcal{A}^\prime$, along which the gluing is performed, by a grey line. The trace of the reduced density matrix can be taken in the above picture by gluing $\aent$ with $\overline{\aent}$, denoted in dashed coloured lines, to close the spherical shell. In computing the Rényi asymmetries \cref{eq:Rényis} we will have to deal with traces of products of $\n$ reduced density matrices: $\tr(\rho_\mathcal{A}^\n)$. These are constructed by gluing cyclically $\n$ copies of $\rho_\aent$. Pictorially,
\begin{equation}
\label{eq:Traces}
	\tr(\rho_\mathcal{A}^\n) \,=\;\; 
    \vcenter{\hbox{\footnotesize \def\svgwidth{10em}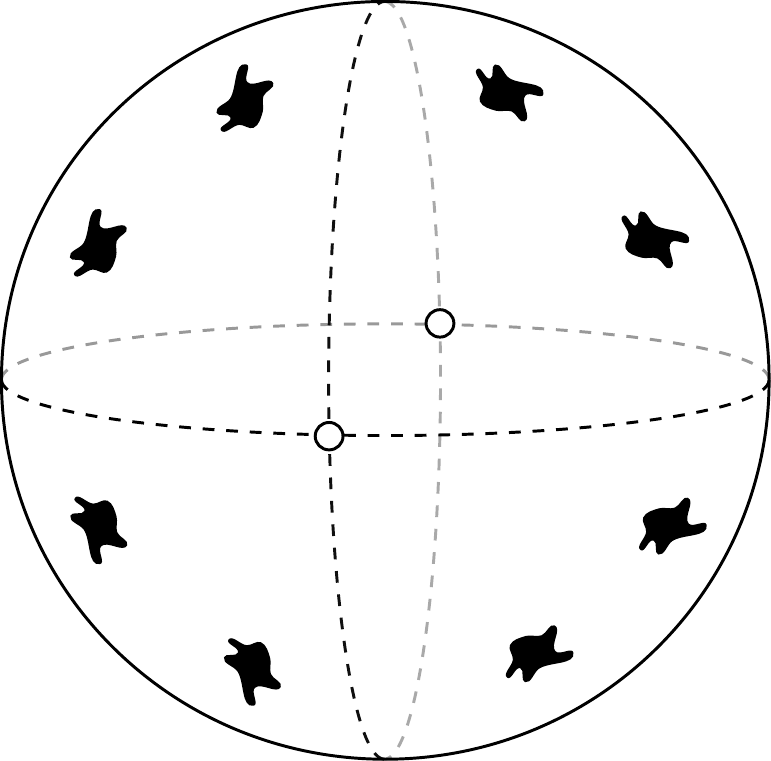}} \qquad\qquad \text{(here for $\n=4$)} ~.
\end{equation}
Here the dashed lines represent the gluings between reduced density matrices. The path integral expression of the charged moments \cref{eq:ChargedMoments} is found by decorating \cref{eq:Traces} with symmetry operators inserted along the appropriate cycles. For the case at hand, the only possibility is to insert $0$-form symmetry defects $U_{a_i}$ and  $U_{a_i}^\dag$ along $\aent_i$ and $\overline{\aent_i}$ with $i=1,...,\n$, respectively. Pictorially,
\begin{equation}
\label{eq:ChargedMoment}
	\tr(\rho_\mathcal{A}^\n) \,=\;\;
    \vcenter{\hbox{\footnotesize \def\svgwidth{16em}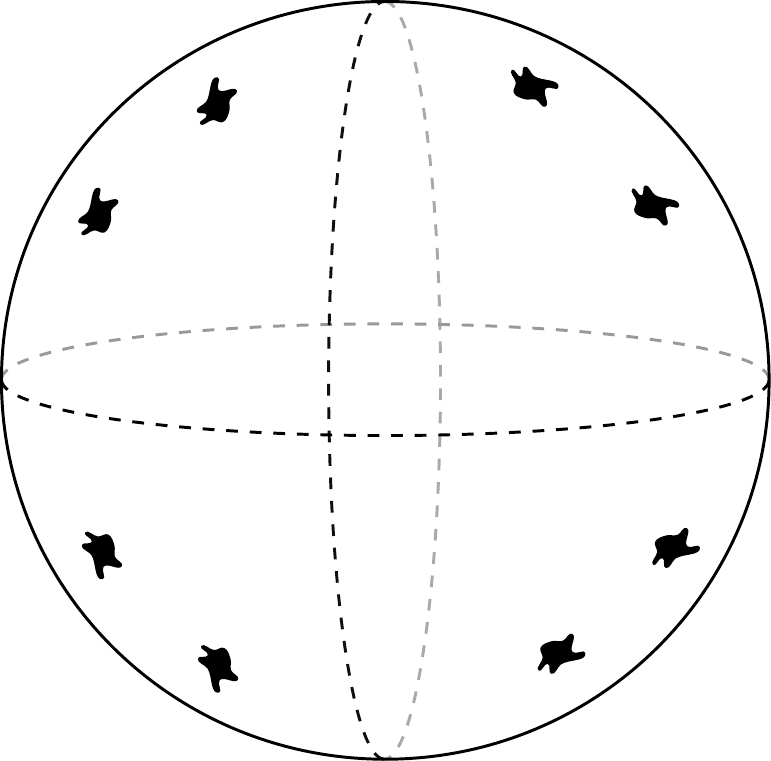}} \quad.
\end{equation}
By taking clear-cut and symmetric boundary conditions, as we will do in \cref{sec:subregion0-form,sec:subregionp-form}, the operators $U_{a_i}$ and $U_{a_i}^\dagger$ may fuse together, detach from the boundary, and move into the bulk until they act on the blobs and eventually shrink to zero size. The result is a path integral without symmetry defects but where the blobs have been acted upon by them.

\section{An entropic Coleman--Mermin--Wagner theorem}
\label{sec:entropic-CMW-global}

It is a classic result \cite{Mermin:1966fe, Hohenberg:1967zz, Coleman:1973ci} that a continuous global symmetry in $d \leq 2$ spacetime dimensions is necessarily preserved in the ground state. Following the literature, we call this result the Coleman--Mermin--Wagner (CMW) theorem. Most standard derivations show that a putative order parameter, $v$, must vanish in two dimensions. In \cite{Mermin:1966fe} this is the magnetisation of the Heisenberg model, while in \cite{Coleman:1973ci} it is the commutator of the conserved charge with a generic scalar operator. With the advent of generalised symmetries, the CMW theorem has been extended to higher-form symmetries: a continuous $p$-form symmetry in $d$ spacetime dimensions must remain unbroken in the ground state if $d \leq p+2$ \cite{Gaiotto:2014kfa, Lake:2018dqm}. In what follows we revisit this classic problem from a different angle, computing the entanglement asymmetry in the ground state of a would-be Goldstone $p$-form gauge field.

We consider a $\U(1)$ $p$-form symmetry, denoted as $\gf{\U(1)}{p}_\t{e}$, of a QFT on a Euclidean (closed) $d$-dimensional spacetime $X_d$, and assume it to be spontaneously broken. The universal IR effective description of such a phase is provided by a free $p$-form Maxwell theory. Its lowest-derivative action, in Euclidean signature, takes the form:
\begin{equation}
\label{Eq:GenMaxwell}
    S = \frac{\lambda\,\mu^\Delta}{2} \int_{X_d} \f{f}{p+1} \wedge \star \, \f{f}{p+1} ~.
\end{equation}
Here $\f{f}{p+1}$ is the $(p+1)$-form field strength. Being closed, it admits a decomposition
\begin{equation}
    \f{f}{p+1} = \f{f}{p+1}^{(h)} + \dd{\f{a}{p}} ~,
\end{equation}
where $\f{f}{p+1}^{(h)} \in 2\pi\, H^{p+1}(X_d; \mathbb{Z})$ is the harmonic part, while $\f{a}{p}$ is a globally-defined $\U(1)$ $p$-form gauge connection. The coupling \(\lambda\) is dimensionless, while \(\mu\) is an arbitrary energy scale. We also introduced the shorthand $\Delta = d-2(p+1)$.%
\footnote{Note that the action has no scale when $\Delta=0$: the theory is indeed conformal. This is for example the case of a compact scalar in \(d=2\) or free electromagnetism in \(d=4\). Whenever \(\Delta\neq 0\), the theory has a scale, and flows to a scale-invariant theory at the two ends of the RG flow. Note, however, that the fixed points are not always conformal \cite{El-Showk:2011xbs, Nakayama:2013is, Lee:2021obi}.}
In this effective theory the $\gf{\U(1)}{p}_\t{e}$ electric symmetry is realised nonlinearly, through shifts of the gauge connection:
\begin{equation}
    \f{a}{p} \mapsto \f{a}{p} + 2\pi \,\f{\Omega}{p}~, \qquad\quad \f{\Omega}{p} \in H^p\bigl( X_d; \U(1) \bigr) ~.
\end{equation}

Besides the electric \(\gf{\U(1)}{p}_\t{e}\) symmetry, the effective action \cref{Eq:GenMaxwell} also enjoys an emergent $\gf{\U(1)}{d-p-2}_\t{m}$ magnetic symmetry. This symmetry is likewise spontaneously broken in the phase of interest. Moreover, the two symmetries are linked by a mixed anomaly. It is precisely this interplay that guarantees the existence of massless excitations, even in regimes where the CMW theorem rules out spontaneous symmetry breaking \cite{Delacretaz:2019brr,Hinterbichler:2024cxn}. For our purposes, however, the magnetic symmetry will play almost no role and we will not mention it further. Hence, in what follows we denote the electric symmetry simply as \(\gf{\U(1)}{p}\).

\subsection{Quantisation of \tps{p}{p}-form gauge theory}
\label{ssec:quantisation}

We now turn to the quantisation of the theory on a general spatial slice $\Sigma$, following \cite{Fliss:2023uiv, Hofman:2024oze}. Denoting the coordinate transversal to \(\Sigma\) by \(t\), the field strength decomposes as%
\footnote{We use the mostly-plus convention in Lorentzian signature, so that $\star \, (\dd{t} \wedge \f{\omega}{p}) = - \star_\Sigma \f{\omega}{p}$ and $\star \, \omega_p = (-1)^p \, \dd{t} \wedge \star_\Sigma \, \f{\omega}{p}$ for a $p$-form $\f{\omega}{p}$ on $\Sigma$.}
\begin{equation}
    \f{f}{p+1} = \dd{t}\w\f{E}{p} + (-1)^{d(p+1)}\star_\s{\Sigma}\f{B}{d-p-2} ~,
\end{equation}
where \(\f{E}{p}\) and \(\f{B}{d-p-2}\) are forms on \(\Sigma\). They are the electric and magnetic fields, respectively, as we first learn them in electrodynamics. In terms of these, the Hamiltonian on \(\Sigma\) can be written as 
\begin{equation}
\label{eq:Hamiltonian-p-form}
    H_\Sigma = \frac{\lambda\, \mu^\Delta}{2}\qty(\norm{\f{E}{p}}^2+\norm{\f{B}{d-p-2}}^2) ~,
\end{equation} 
where the norm denotes $\norm{\omega}^2 = \int_\Sigma \omega \w\star_\s{\Sigma}\, \omega $. 

Already at the semiclassical level, both electric and magnetic fluxes are quantised.%
\footnote{Magnetic fluxes are obviously quantised as follows from elementary differential geometry. For electric fluxes, a transparent way to see that they are quantised is to consider an open path integral preparing the state in question. The path-integral sums over magnetic fluxes $\oint\! f \in 2\pi \mathbb{Z}$. This sum generates a delta function that quantises the electric fluxes.}
For concreteness let us fix bases \(\set{C^i_{p+1}}\) and \(\set{\hat{C}^{\hat{\jmath}}_{d-p-1}}\) of \((p+1)\)- and \((d-p-1)\)-cycles in $\Sigma$, and define the magnetic and electric fluxes, respectively, as 
\begin{align}
    \Phi_\t{m}^i &\coloneqq \int_{C^i_{p+1}} \f{f}{p+1} = (-1)^{d(p+1)}\int_{C^i_{p+1}} \star_\s{\Sigma} \, \f{B}{d-p-2}~, \qq{and} \\[10pt]
    \Phi_\t{e}^{\hat{\jmath}} &\coloneqq \int_{C^{\hat{\jmath}}_{d-p-1}} \star \, \f{f}{p+1} = - \int_{C^{\hat{\jmath}}_{d-p-1}} \star_\s{\Sigma} \, \f{E}{p} ~.
\end{align}
Flux quantisation then implies that:
\begin{equation}
    \Phi_\t{m}^i = 2\pi\,m^i~, \qq{and} \Phi_\t{e}^{\hat{\jmath}} = \frac{1}{\lambda\, \mu^\Delta} \, k^{\hat{\jmath}}~,\qq{with} m^i,k^{\hat{\jmath}} \in \Z ~. 
\end{equation} 
One of the main observations of \cite{Freed:2006ya, Freed:2006yc} is that, whenever \(\Sigma\) has no torsion in its homology --- an assumption we adopt henceforth --- passing to the quantum theory is relatively straightforward. Namely, the Hilbert space over \(\Sigma\) is simultaneously graded by electric and magnetic fluxes. 

The topological sector of a state is thus specified by its fluxes through the chosen cycles. As discussed in \cite{Hofman:2024oze}, excited states within each topological sector can be reached by applying creation operators. The lowest lying states, i.e. those annihilated by all lowering operators, are completely characterised by the fluxes \(\vec{m}\) and \(\vec{k}\). Denoting these states as $\ket{\vec{m},\vec{k}}$, they satisfy:
\begin{align}
    \Phi_\t{m}^i\ket{\vec{m},\vec{k}} &=   2\pi\, m^i \ket{\vec{m},\vec{k}} ~, \\ 
    \Phi_\t{e}^{\hat{\jmath}}\ket{\vec{m},\vec{k}} &=  \frac{k^{\hat{\jmath}}}{\lambda \mu^\Delta}\ket{\vec{m},\vec{k}} ~.
\end{align}
These states are taken to be orthonormal. The energy of the state \(\ket{\vec{m},\vec{k}}\), as measured by the Hamiltonian \cref{eq:Hamiltonian-p-form}, is
\begin{equation}
    E_{\vec{m},\vec{k}} = 2\pi^2 \lambda \mu^\Delta\, m_i G_{ij} m_j + \frac{1}{2\lambda\mu^\Delta} k_{\hat{\imath}} \hat{G}_{\hat{\imath}\hat{\jmath}} k_{\hat{\jmath}} ~.
\end{equation}
In this formula, $G_{ij}$ and $\hat{G}_{\hat{\imath}\hat{\jmath}}$ are the induced metrics on $\H^{p+1}(\Sigma,\mathbb{Z})$ and $\H^{d-p-1}(\Sigma; \mathbb{Z})$, respectively, and there is an implicit summation over repeated indices (see \cite{Fliss:2023uiv} for details). Following the terminology of \cite{Fliss:2023uiv, Hofman:2024oze, Vitouladitis:2025zoy}, we will sometimes refer to the states \(\ket{\vec{m}, \vec{k}}\) as ``primary'' states, as they furnish highest-weight states of a Kac--Moody algebra.

For our purposes it will suffice to consider spatial slices of the form
\begin{equation}
    \Sigma = S_r^p \times S_R^{d-p-1} ~,
\end{equation}
where the subscripts $r$ and $R$ denote the radii of the spheres.%
\footnote{\label{foot:radius}%
For simplicity, here we define the ``radius'' $r$ such that $S_r^p$ has volume $r^p$, and similarly for $S_R^{d-p-1}$.}
For generic $d$ and $p$, all states have $m=0$ and therefore they are fully specified by the electric flux on $S^{d-p-1}$.%
\footnote{\label{foot:special-d}%
In some cases more care is needed.
If \(d = 2p+1\), one can turn on electric fluxes on both spheres, but in the limit of interest the extra states have energies that diverge in units of $1/R$ and thus do not become vacua.
On the other hand, if \(d=2p+2\), nonvanishing magnetic flux on $S^{d-p-1}_R$ is allowed, and the situation is subtler. States carrying magnetic flux (or mixed electric–magnetic flux) have vanishing energy at infinite volume. Moreover, a \(\vartheta\)-term \(\sim \vartheta\, \f{f}{p+1}\wedge \f{f}{p+1}\) may be turned on. Restricting to \( \vartheta=0 \) and clustering states, it turns out that only the electric flux contributes to the
asymmetry, and therefore we omit magnetic or dyonic states in what follows.} We denote these states by $\ket{k}$, indicating that they carry electric flux \(k/(\lambda \mu^\Delta)\). Such states are prepared by a Euclidean path integral on $\mathbb{R}_- \times \Sigma$ with an insertion of a charge \(k\) Wilson line on $S^p$ at some Euclidean time $\tau_0$: $W_k(\tau_0,S^p)$.%
\footnote{In fact, one needs to dress the Wilson lines with squeezing operators, see \cite{Hofman:2024oze}. We will omit this step as it will not affect our results.}
We note that the states $\ket{k}$ are eigenstates of the electric $\gf{\U(1)}{p}$ symmetry. Indeed, the symmetry operator
\begin{equation}
    U_\alpha\qty(S^{d-p-1}) = \exp( \ii \alpha \, \lambda \mu^\Delta \int_{S^{d-p-1}} \star \, \f{f}{p+1})
\end{equation}
acts on the states as:
\begin{equation}
    U_\alpha\qty(S^{d-p-1}) \ket{k} = \ex{\ii\alpha k} \ket{k} ~. 
\end{equation}
As detailed above, these states are also energy eigenstates with energy:
\begin{equation}
\label{eq:energies}
    E_k = \frac{1}{R} \, \frac{k^2}{2\lambda} \, \frac{\qty(\mu r)^p}{(\mu R)^{d-p-2}} ~.
\end{equation}
An interesting question is whether these states are degenerate. Generically they are not, but if we consider the limit in which $\mu R \to \infty$ while keeping $\mu r$ fixed, the energy of all states goes to zero in units of $1/R$ as long as $d > p+2$:
\begin{equation}
    R \, E_k \to 0 \quad\quad \text{if} \quad\quad d > p+2 ~.
\end{equation}
Let us pause with this result. This is a decompactification limit, in which $R \gg r$ while $\mu r $ is kept fixed. In other words, the volume of $S^p_r$, which is where the Wilson lines $W_k(\tau_0,S^p)$ are inserted to prepare the states, is kept fixed. Overall, this is a large volume limit akin to the standard one used for $0$-form symmetries, to which it reduces in the $p=0$ case.%
\footnote{Note that, indeed, a spontaneously broken $p$-form symmetry does not lead to a degeneracy of states on $\mathbb{R}^{d-1}$, but rather on spatial slices of the form $M_{p}\times M_{d-p-1}$ such that $M_{d-p-1}$ is very large  \cite{Cordova:2019bsd}.} We conclude that in the large volume limit the ground state is degenerate if and only if $d > p+2$. This agrees with the higher-form Coleman--Mermin--Wagner theorem, but does not yet constitute a proof, since we have not addressed whether the \(\gf{\U(1)}{p}\) symmetry is spontaneously broken or not.

We end by commenting on the nongeneric cases where more states are available due to the interplay between $p$ and $d$. In the case $d=2p+1$, where electric flux on \(\S^p_r\) is also allowed, the result is unchanged since, in the limit, the extra states have energies that diverge in units of $1/R$. On the contrary, the states with only electric flux on $S^{d-p-1}_R$ have energies that, in units of $1/R$, vanish for $p\geq 2$ (indicating degenerate vacua at infinite volume), are constant for $p=1$ (indicating a scale-invariant effective theory with a single vacuum), and diverge for $p=0$ (indicating a gapped effective theory). On the other hand, in the case $d=2p+2$, the theory is conformal and there is no energy scale $\mu$ (indeed $\Delta = 0$). Moreover, magnetic flux can be threaded through \(\S^{d-p-1}_R\). The states with magnetic flux on $S_R^{d-p-1}$, or dyonic states with both electric and magnetic flux, have energies that scale in the same way with $R$ as those of the purely electric states. We can still take the limit
\begin{equation}
    \frac{R}{r} \,\to\, \infty ~.
\end{equation}
For $p \geq 1$, all energies go to zero in units of $1/R$ and thus all dyonic states \(\ket{k,m}\) are degenerate at infinite volume. This hints at the magnetic and electric symmetries --- which have the same degree in this case --- being both spontaneously broken. We will come back to this point in the next section. For $d=2$ and $p=0$, instead, no dimensionless parameter exists. In this case the energies remain fixed in units of $1/R$ and there is no limit in which the states become degenerate, indicating that the symmetry remains unbroken in a scale-invariant theory with a single vacuum.

\subsection{Asymmetry in \tps{p}{p}-form gauge theory}

We have yet to address the spontaneous breaking (or not) of the \(\gf{\U(1)}{p}\) symmetry, which we do in this section by explicitly computing the entanglement asymmetry.  For this purpose we introduce a new basis of states (see also \cite{Vitouladitis:2025zoy} for a related discussion):
\begin{equation}
\label{eq:ClusteringState}
    \ket{\theta} = \sum\nolimits_{k\in \mathbb{Z}} \ex{\ii k\theta} \ket{k} ~.
\end{equation}
The \(\gf{\U(1)}{p}\) symmetry acts on these states nontrivially:
\begin{equation}
\label{eq:symaction}
    U_\alpha\qty(S^{d-p-1}) \ket{\theta} =  \ket{\theta + \alpha} ~.
\end{equation}
More importantly, the states $\ket{\theta}$ are the ones that satisfy cluster decomposition and are thus the relevant ones in order to study the spontaneous breaking of the \(\gf{\U(1)}{p}\) symmetry.%
\footnote{Consider for simplicity the case $p=0$ for $d>2$. A vertex operator $V_q(x)= \ex{\ii q\phi(x)}$ of charge $q$ acts as $V_q(x) \, \ket{k} \sim \ket{k+q}$ on the states with fixed flux, which implies $\bra{k} V_q(x) \ket{k} = 0$. This is not surprising: the flux sourced by $V_q(x)$ has nowhere to go in $S_R^{d-1}$. Conversely, 
\begin{equation}
    \bra{k} V_{q_1}(0) \, V_{q_2}(x) \ket{k} \,\propto\, \delta_{q_1, -q_2} \, \exp(-\frac{q_1 q_2}{\lambda \abs{\mu\, x}^{d-2}}) ~,
\end{equation}
which is nonvanishing as $\abs{x} \to \infty$ when $q_1 = -q_2$. This violates cluster decomposition. Similar arguments hold for generic $p$. In contrast, one can check that cluster decomposition is preserved on $S^{d-p-1}_R$ in the $\ket{\theta}$ states. In those cases without vacuum degeneracy one always achieves cluster decomposition after projection to the unique vacuum.}
An interesting property of the states $\ket{\theta}$ is that generically they are not energy eigenstates, but become eigenstates with zero energy in the large volume limit explored in the previous section, as long as $d > p+2$. That is the case where we expect the symmetry to be spontaneously broken in the large volume limit and the states $\ket{\theta}$ are good candidates to describe the symmetry breaking vacua. However, these states are only delta-function normalisable. Inspired by computations in quenched CFTs \cite{Calabrese:2006rx, Calabrese:2007rg, Takayanagi:2010wp, Cardy:2014rqa}, topological order and boundary CFT \cite{Das:2015oha, Wong:2017pdm, Lou:2019heg, Fliss:2020cos, Fliss:2023uiv}, we regulate the states using the quadratic Hamiltonian \cref{eq:Hamiltonian-p-form}:
\begin{equation}
\label{eq:regulated}
    \ket{\theta_\epsilon} \coloneqq \ex{-\epsilon R\, H_\Sigma} \ket{\theta} ~. 
\end{equation}
Also these states transform nontrivially under the \(\gf{\U(1)}{p}\) symmetry action, but they are normalisable. What was sacrificed in the process is orthogonality of the states at different angles, but this is something we can keep track of. It follows from $\theta \sim \theta + 2\pi$ and \cref{eq:symaction} that, as long as $d>p+2$ so that these states are degenerate in the large volume limit, the moduli space of vacua is a circle $S^1$.

This brings us to the computation of entanglement asymmetry of the states $\ket{\theta_\epsilon}$. The corresponding trace-normalised pure density matrix is
\begin{equation}
\label{eq:tr-normalised-abelian-rho}
    \rho_\theta = \frac{\ketbra{\theta_\epsilon}}{\Theta\qty(q_\epsilon)} ~.
\end{equation}
We introduced the Jacobi theta constants
\begin{equation}
    \Theta(q) = \sum\nolimits_{n\in \mathbb{Z}} q^{n^2} ~,
\end{equation}
and we defined
\begin{equation}
    q_\epsilon = \exp(-\frac{\epsilon}{\lambda} \frac{\qty(\mu r)^p}{(\mu R)^{d-p-2}})
\end{equation}
as a shorthand. According to \cref{eq:rho-sym-def}, the symmetrised density matrix takes the form
\begin{equation}
\label{eq:rho-theta-S}
    \rho_{\theta,\sfS} = \int_{0}^{2\pi} \frac{\dd{\alpha}}{2\pi} \; U_\alpha\qty(S^{d-p-1}) \, \rho_\theta \, U^\dag_\alpha\qty(S^{d-p-1}) = \int_{0}^{2\pi} \frac{\dd{\alpha}}{2\pi} \; \rho_{\theta + \alpha} ~.
\end{equation}
The integral can be evaluated, yielding
\begin{equation}
\label{eq:rho-theta-S thermal}
    \rho_{\theta, \sfS} = \frac1{\Theta\qty(q_\epsilon)} \, \sum\nolimits_{k\in \bbZ} \; e^{-2\epsilon R \, E_k} \ketbra{k} ~.
\end{equation}
To compute the entanglement asymmetry we employ the replica trick, as in \cref{Eq:replicatrick}. A straightforward check shows that \(\tr(\rho_\theta^\n)=1\) and that the symmetrised trace evaluates to
\begin{equation}
    \tr( \rho_{\theta,\sfS}^\n ) = \frac{\Theta\qty(q_\epsilon^\n)}{\Theta\qty(q_\epsilon)^\n}~. 
\end{equation}
Substituting this expression into \cref{eq:Rényis} we obtain the Rényi asymmetries:
\begin{equation}
    \entas_{\n, \, \gf{\U(1)}{p}} \bigl( \rho_\theta \bigr) = \frac{1}{1-\n}\log\frac{\Theta \qty(q_\epsilon^\n)}{\Theta\qty(q_\epsilon)^\n}~.
\end{equation}
The limit \(\n\to 1\) can be carried out explicitly to yield finally  
the entanglement asymmetry:
\begin{equation}
    \entas_{\gf{\U(1)}{p}} \bigl( \rho_\theta \bigr) =  \log\Theta\qty\big(q_\epsilon)- q_\epsilon \log q_\epsilon\;\frac{\Theta'\qty(q_\epsilon)}{\Theta\qty(q_\epsilon)} ~,
\end{equation}
where \(\Theta'\) denotes the derivative of the Jacobi theta constant with respect to its argument. Let us briefly comment on the case $d= 2p+2$. The clustering states analogue to \cref{eq:ClusteringState} have two sums, one over electric and one over magnetic charges, with different angles $\theta_\t{e}$ and $\theta_\t{m}$. Denote the regularised state $\ket{\theta_\t{e}, \theta_\t{m}}_\epsilon$. In the large volume limit for $p\geq 1$ these states are degenerate and the moduli space is a torus instead of a circle. However, acting with the electric symmetry still generates a circle $U_\alpha \ket{\theta_\t{e}, \theta_\t{m}}_\epsilon = \ket{\theta_\t{e} + \alpha, \theta_\t{m}}_\epsilon$ and the corresponding entanglement asymmetry is the one we just computed.

\subsubsection{Various physical limits}

Despite its simplicity, the model has many available knobs we can adjust to probe interesting physics. The behaviour of entanglement asymmetry as we tune the various parameters at their extremes uncovers distinct physical phenomena. We take the chance here to briefly explain some of them. In what follows, we dial one parameter at a time, keeping all other parameters fixed at finite values. Multi-scaling limits could of course be explored as well, but we will not pursue them here.

\paragraph{Very hot and very cold.}
The regulated states \(\rho_\theta\) are pure, however their symmetrisation \eqref{eq:rho-theta-S thermal} is thermal in the Hilbert space of primary states, with a temperature
\begin{equation}
	T = \frac{1}{\beta} = \frac{1}{2\epsilon R} ~.
\end{equation}
We can thus draw a connection between asymmetry and thermal physics.
In the limit $\epsilon \to 0$, indeed, $\rho_{\theta,\sfS}$ is a highly excited state, probing all of the primary states \(\ket{k}\), even those with increasingly large energy as \(\abs{k}\) grows. It explores the full non-linearly realised \(\U(1)^{[p]}\) symmetry, and the asymmetry reflects this as
\begin{equation}
\label{eq:entas-large-T}
	\entas_{\gf{\U(1)}{p}} \,\sim\, \frac{1}{2}\log T \,+\, \text{const} ~, \qquad\qq{for} T \quad \text{large} ~.
\end{equation}
Indeed, since the seed state \(\rho_\theta\) is pure, the entanglement asymmetry exactly reduces to the von Neumann entropy of \(\rho_{\theta,\sfS}\). The behaviour in \eqref{eq:entas-large-T} is therefore entirely natural: for thermal states the von Neumann entropy coincides with the thermodynamic entropy, and at high temperature the ``partition function'' \(\Theta(q_\epsilon)\) reduces to that of a free one-dimensional gas with $\beta = 2\epsilon R$ since $q_\epsilon = \exp( - \beta \, E_{k=1})$.

In the opposite limit, when we ``cool down the system'' by taking $\epsilon \to \infty$, only the true ground state --- unique at finite volume --- contributes. Indeed: 
\begin{equation}
	\entas_{\gf{\U(1)}{p}} \to 0 ~, \qquad \qq{for} T \quad\text{small} ~.
\end{equation}
From a path-integral perspective, this behaviour is of course unsurprising, as this low temperature limit corresponds to a very long Euclidean time evolution.

We note in passing that the limit \(\n \to \infty\), conjectured to define a (semi) universal entropic quantity \cite{Ohmori:2014eia}, has a simple physical meaning in this model. To see this, recall that for R\'enyi entropies the limit \(\n\to\infty\) simply gives 
\begin{equation}
    \lim_{\n\to\infty} \ent_\n(\rho) = -\log \lambda_\t{max} ~,
\end{equation}
where \(\lambda_\t{max}\) is the largest eigenvalue of \(\rho\). Since in our setup the symmetrised state is thermal, the largest eigenvalue is just \( \lambda_\t{max} = 1/\Theta(q_\epsilon) \) from the vacuum. Hence, the entanglement asymmetry in this limit takes the form:
\begin{equation}
	\entas_{\infty} = \log \Theta\qty\big(q_\epsilon) ~.
\end{equation}
To connect this result with the broader problem, recall that partition functions of \(p\)-form gauge theories factorise into instanton and oscillator contributions \cite{Kelnhofer:2007jf, Donnelly:2016mlc} (cfr. also \cref{sec:subregion0-form,sec:subregionp-form}). The corresponding thermal partition functions were computed in \cite{Fliss:2023uiv, Hofman:2024oze}, where it was shown that the instanton sector takes precisely the form of a Jacobi theta constant --- or, more generally, a Siegel theta function (see again \cref{sec:subregion0-form,sec:subregionp-form}). Reconciling this with our present result, we conclude that the \(\n \to \infty\) R\'enyi asymmetry captures the instanton free energy of \(p\)-form gauge theory on \(\S^p_r \times \S^{d-p-1}_R\):
\begin{equation}
	\entas_\infty = \log\parti^\t{inst}\qty[\S^1_\beta\times\S^p_r\times\S^{d-p-1}_R] ~.
\end{equation}

\paragraph{Very large and very small (target).} 
Another useful control parameter is the coupling constant \(\lambda\). Upon rescaling the \(p\)-form gauge field \(\f{a}{p}\) to canonically normalise its kinetic term, one sees that \(\lambda\) controls its periods, or equivalently, the radius of the \(\U(1)\) target space. Hence, the limit \(\lambda \to \infty\) corresponds to a noncompact gauge theory with gauge group \(\R\). The entanglement asymmetry tracks this decompactification:
\begin{equation}
	\entas_{\gf{\U(1)}{p}} \,\sim\, \frac{1}{2}\log \lambda \,+\, \text{const} ~, \qquad\qq{for} \lambda\gg 1 ~.
\end{equation}
In this limit the symmetrised state is maximally mixed, and we may infer that the dimension of the Hilbert space of primary states grows as
\begin{equation}
	\dim\cH_\t{primary} \overset{ \lambda\to\infty}{\sim} \sqrt{\lambda} ~,
\end{equation}
characterising the growing degeneracy of low-energy modes in the noncompact theory.

Conversely, the limit \(\lambda \to 0\) describes a somewhat exotic theory --- a gauge field without holonomies \cite{Dymarsky:2013pqa}. Here we find 
\begin{equation}
	\entas_{\gf{\U(1)}{p}} \to 0 ~, \qquad\qq{when} \lambda\to 0 ~. 
\end{equation}
In this limit all charged states are too heavy to be excited and thus decouple. Only the symmetry-preserving vacuum, with \(k=0\), survives inside \(\rho_\theta\) and hence the asymmetry vanishes.

Finally, one may tune the size of the spatial slice. Since this directly connects to the question of spontaneous symmetry breaking, we will treat it in detail in the next section.

\subsection{The Coleman--Mermin--Wagner theorem}

Since we are interested in spontaneous symmetry breaking, we need to take a large volume limit, which as explained in \cref{ssec:quantisation} consists in taking:
\begin{equation}
    \mu R \to \infty ~, \qquad\qq{with} \mu r \quad \t{fixed} ~. 
\end{equation}
This limit yields different results, depending on $d$ and $p$. We distinguish three cases.
{\setlength{\leftmargini}{1.4em}
\begin{itemize}
    \item $\boldsymbol{d>p+2 \,.}$ The state $\ket{\theta_\epsilon}$ becomes an energy eigenstate with zero energy. Explicit computation of the $\mu R \to \infty$ limit yields$\,$%
    \footnote{In computing this limit the modular property of the Jacobi theta constant is handy:
    \begin{equation}
    \label{eq:modular}
	      \Theta \bigl( \ex{-\frac{\pi\ii}{\tau}} \bigr) = (-\ii \tau)^\frac{1}{2} \, \Theta\qty(\ex{\pi\ii\tau}) ~.
    \end{equation}}
    \begin{equation}
        \entas_{\gf{\U(1)}{p}}(\rho_\theta) \,\sim\, \frac{d-p-2}{2} \, \log(\mu R) \,+\, \t{const} ~,
    \end{equation}
    where the additive constant collects the finite subleading contributions from $r, \lambda$ and $\epsilon$. A subtlety arises if $d = 2p+2$ (and $p>0$) because the theory is conformal. In this case \(\mu\) does not exist and the relevant limit is $R/r \to \infty$. In that limit one finds:
    \begin{equation}
        \entas_{\gf{\U(1)}{p}}(\rho_\theta) \,\sim\, \frac{p}{2} \, \log\frac{R}{r} \,+\, \t{const} ~.
    \end{equation}
    We conclude that for $d>p+2$ entanglement asymmetry is nonvanishing and diverges with the radius $R$ of the spatial \((d-p-1)\)-sphere. Since $\ket{\theta_\epsilon}$ is a ground state, this signals spontaneous breaking of the \(\gf{\U(1)}{p}\) symmetry.
    
    For later comparison, we note that the coefficient of the logarithmic divergence is \(\frac{d-p-2}{2}\) times the number of broken generators, a fact that becomes manifest when one considers, for instance, a \(\U(1)^N\) gauge group.

    \item $\boldsymbol{d = p+2 \,.}$ In this critical case, $\mu R$ drops out of the expression of the energy levels and the large volume limit has no effect. Furthermore, $\ket{\theta_\epsilon}$ are not energy eigenstates. To ensure that $\ket{\theta_\epsilon}$ are ground states, we notice that the regulator in \cref{eq:regulated} is implementing a Euclidean time evolution for a time $\tau = \epsilon R$. Such an evolution will project to a ground state if $\epsilon R E_{k=1} \gg 1$, where $E_{k=1}$ is the energy gap, see \cref{eq:energies}. We thus need
    \begin{equation}
    \label{eq:epsilon-gg-2}
        \epsilon \gg \frac{2\lambda}{(\mu r)^p} ~.
    \end{equation}
    Under this assumption, the limit trivialises and one finds
    \begin{equation}
        \entas_{\gf{\U(1)}{p}}(\rho_\theta) \,\to\, 0 ~.
    \end{equation}
    We conclude that there is a unique ground state with vanishing entanglement asymmetry. This implies that, for $d = p+2$, the \(\gf{\U(1)}{p}\) symmetry is not spontaneously broken.
    
    \item $\boldsymbol{d < p+2 \,.}$ As in the previous case, $\ket{\theta_\epsilon}$ is not an energy eigenstate and a Euclidean evolution is needed to project onto a ground state. In this case we need
    \begin{equation}
    \label{eq:epsilon-gg-1}
        \epsilon \gg  \frac{2\lambda }{ (\mu r)^p \, (\mu R)^{p+2-d}} ~,
    \end{equation}
    which is automatic in the large volume limit $\mu R \to \infty$. Under this assumption, one finds again a vanishing entanglement entropy. We conclude that, for $d < p+2$, the \(\gf{\U(1)}{p}\) symmetry is not spontaneously broken.
\end{itemize}}

We finish by summarising the essential point. For a free \(p\)-form gauge theory in \(d\) spacetime dimensions, quantised on \(\Sigma = S_r^p \times S_R^{\,d-p-1}\), the electric \(p\)-form symmetry \(\gf{\U(1)}{p}\) is spontaneously broken in the
\(R \to \infty\) limit if and only if \(d > p+2\). This conclusion distils the analysis of the previous subsections and provides a direct proof of the \(p\)-form Coleman--Mermin--Wagner theorem, with entanglement asymmetry serving as an entropic order parameter.

\subsection{Non-Abelian asymmetry}
\label{subsec:nonabelian}

We will end this section by calculating the entanglement asymmetry in nonlinearly realised non-Abelian $0$-form symmetries. The methods employed here are very similar to the Abelian case, so we will be brief. 

\subsubsection{Complete symmetry breaking}
\label{sec:CompleteSSB}

Consider a quantum field theory with a continuous non-Abelian symmetry given by a compact Lie group \(G \). According to Goldstone's theorem, whenever \(G\) spontaneously breaks down to nothing, the IR effective field theory necessarily contains a massless $G$-valued scalar field $h$:
\begin{equation}
	h : X_d \rightarrow G ~.
\end{equation}
If no additional light degrees of freedom are present, or if these decouple, the low energy dynamics is described by a nonlinear $\sigma$-model (NLSM) with target $G$ --- in this case the principal chiral model --- with action:%
\footnote{In the interest of simplicity, we consider here the principal chiral model, i.e., the NLSM with bi-invariant metric on the group manifold $G$, invariant under $G \times G$, which is unique up to an overall scale for each simple factor $G_i$. The most general $G$-invariant two-derivative action, however, is written in terms of a left-invariant metric, and, up to coordinate changes, there exists an $\frac{n(n-1)}2$-dimensional space of such inequivalent metrics, where $n = \dim G_i$, for each simple factor. Using more general metrics, though, would not qualitatively change our final results.}
\begin{equation}
\label{eq:PCM action}
	S = - \frac{\lambda\, \mu^{d-2}}{2} \int \dd[d]{x} \sqrt{g} \, g^{\mu \nu}\, \tr(h^{-1}\pd_\mu h \, h^{-1} \pd_\nu h) ~.
\end{equation}
Following the discussion in the previous section, we quantise this theory on $\Sigma = \S^{d-1}_R$. 

For the time being, we restrict attention to $d >2$, ensuring that the coupling  $\lambda \mu^{d-2}$ is irrelevant and the theory is weakly coupled in the IR. Luckily for us, this model describes the massless limit of the low-energy levels of QCD, which has been meticulously studied over the years. Of particular relevance to us, \cite{Leutwyler:1987ak} argued that space-dependent modes can be treated as Gaussian, up to corrections of order $\order{\lambda^{-1}(\mu R)^{2-d}}$. We will therefore impose a weak coupling limit
\begin{equation}
\label{eq:NonAbelianAprox}
	\lambda\, (\mu R)^{d-2} \gg 1
\end{equation}
to remain in the Gaussian regime. 

Since we are looking for the lowest-lying modes, we will take the space-dependent modes to be in their ground states and focus on the dynamics of the spatially constant mode $h_0(t)$. Integrating over the spatial sphere $S^{d-1}_R$ (see footnote \ref{foot:radius}), the effective action governing this mode is given by:
\begin{equation}
\label{eq:QuantumMechanics}
	S_0 = - \frac{I}{2} \int \dd{t} \tr[\qty(h_0^{-1} \dot{h}_0)^2]  ~,
\end{equation}
where we have defined the \textit{moment of inertia}
\begin{equation}
	I = \lambda \, \mu^{d-2} \, R^{d-1} ~.
\end{equation}
The action \cref{eq:QuantumMechanics} describes the quantum mechanics of a particle on the group manifold $G$, or equivalently a rigid rotor. The corresponding Hamiltonian is
\begin{equation}
	H_0 = -\frac{\Delta_G}{2I} ~,
\end{equation}
where \(\Delta_G\) is the Laplace--Beltrami operator on the group manifold $G$:
\begin{equation}
	\Delta_G = \frac{1}{\sqrt{g}} \, \partial_a \qty( \sqrt{g}\, g^{ab} \partial_b )\,, \qq{with} g_{ab} = \frac{1}{2} \tr( \partial_a h_0^{-1} \partial_b h_0 ) ~.
\end{equation}
The Peter--Weyl theorem%
\footnote{For a textbook treatment of harmonic analysis on group manifolds see \cite{Sepanski2007}. For a physicist account, that also discusses the quantum mechanics of a particle on group manifolds, see \cite{Marinov:1979gm}.}
guarantees that the eigenfunctions of this Hamiltonian are given by the matrix elements of the Wigner \(D\)-matrices \(D^\sfR_{m,n}(g)\), namely the unitary representation matrices of $G$, where \(\sfR\) labels a finite-dimensional irreducible representation and \(m,n=1,2,\ldots,d_\sfR\), with $d_\sfR$ the dimension of $\sfR$. Moreover, these functions furnish an orthogonal basis on the space $L^2(G)$ of square-integrable functions on the group, obeying the orthogonality relations:
\begin{equation}
	\int_G \dd{g} D^\sfR_{m,n}(g)\, \qty(D^{\sfR^\prime}_{m^\prime,n^\prime}(g))^* = \frac{\delta_{m,m^\prime} \, \delta_{n,n^\prime} \, \delta^{\sfR, \sfR^\prime}}{d_\sfR} ~.
\end{equation}
Let us label the corresponding eigenstates by \(\ket{\sfR,m,n}\). Their energy levels are given by the quadratic Casimir of the representation: 
\begin{equation}
	E_{\sfR,m,n} = \frac{C_2(\sfR)}{2 I} ~,
\end{equation}
and they are \(\qty(d_\sfR)^2\)-fold degenerate. By completeness of the \(\ket{\sfR,m,n}\) basis in $L^2(G)$, we can perform an \textit{inverse Fourier transform} to build a dual basis of states:%
\footnote{The states \(\ket{\sfR,m,n}\) are the non-Abelian analogues of $\ket{k}$, which in turn correspond to the representations of $U(1)$. Likewise, the states $\ket{g}$ play the role of $\ket{\theta}$ and are therefore symmetry breaking.} 
\begin{equation}
	\ket{g} = \sum_{\sfR\in\widehat{G}} \;\; \sum_{m,m'=1}^{d_\sfR} \sqrt{d_\sfR} \; D^\sfR_{m,m'}(g) \, \ket{\sfR,m,m'} ~.
\end{equation}
Here \(\widehat{G}\) is the infinite set of finite-dimensional irreducible representations of \(G\). The states $\ket{g}$ transform by left translations under the group action:
\begin{equation}
	U(h)\ket{g} = \ket{hg} ~, \qquad\qq{for} h\in G ~,
\end{equation}
demonstrating that they break the symmetry $G$. As in the Abelian case, the states \(\ket{g}\) are only delta-function normalisable. We regulate them by the rotor Hamiltonian: 
\begin{equation}
	\ket{g_\epsilon} \coloneqq \ex{-\epsilon R \, H_0} \ket{g} ~. 
\end{equation}

We are now ready to compute the entanglement asymmetry. The non-Abelian analogue to the trace-normalised pure density matrix \cref{eq:tr-normalised-abelian-rho} is here
\begin{equation}
    \rho_g = \frac{\ketbra{g_\epsilon}}{\Theta_G\qty(q_\epsilon)} ~,
\end{equation}
where we defined 
\begin{equation}
\label{eq:NonAbelianTheta}
	q_\epsilon = \exp(-\frac{\epsilon R}{I}) ~, \quad\qq{and}\quad \Theta_G(q) = \sum\nolimits_\sfR \, (d_\sfR)^2 \;  q_\epsilon^{C_{2}(\sfR)} ~.
\end{equation}
It is straightforward to check that \(\rho_g\) is trace-normalised. Amusingly, $\Theta_G(q)$ in \cref{eq:NonAbelianTheta} is exactly the partition function of two-dimensional Yang--Mills theory on $\S^2$ \cite{Migdal:1975zg, Rusakov:1990rs, Fine:1990zz, Witten:1992xu, Blau:1991mp} under the identification $\epsilon R/I  \to g^2_\s{\t{YM}} A$, where $g_\s{\t{YM}}$ is the Yang--Mills coupling constant and $A$ the area of the $S^2$.%
\footnote{Let us make the following observation. A particle on a group is known to constitute the edge dynamics of 2d Yang--Mills \cite{Blommaert:2018oue}. Here, however, it appears related to the partition function on \(\S^2\), which cannot support edge modes. This echoes the results of \cite{Anninos:2020hfj}, where edge-mode contributions have been seen to arise in generic one-loop sphere partition functions.} Next, we use \cref{eq:rhoG-hom} to symmetrise $\rho_{g}$:
\begin{equation}
	\rho_{g,\sfS} = \int_G \dd{h}\; U(h) \, \rho_g \, \inv{U(h)} ~.
\end{equation}
It follows that
\begin{equation}
	\tr (\rho_{g,\sfS}^\n) = \frac{\Theta_G\qty(q^\n_\epsilon)}{\Theta_G\qty(q_\epsilon)^\n} ~,
\end{equation}
and hence the R\'enyi asymmetries read:
\begin{equation}
	\entas_{\n,G} = \frac{1}{1-\n}\log\frac{\Theta_G \qty(q_\epsilon^\n)}{\Theta_G\qty(q_\epsilon)^\n} ~.
\end{equation}
Finally, to interpret $\entas_{\n,G}$ as a measure of spontaneous symmetry breaking we must take the thermodynamic (or large volume) limit: 
\begin{equation}
	\mu R \rightarrow \infty ~,
\end{equation}
while ensuring that $\ketbra{g_\epsilon}$ is a ground state and the approximation in \cref{eq:NonAbelianAprox} is under control. The behaviour of the R\'enyi asymmetries crucially depends on the spacetime dimension. We distinguish three cases:
{\setlength{\leftmargini}{1.4em}
\begin{itemize}
	\item $\boldsymbol{d > 2 \,.}$ In the large volume limit, the states $\ket{g_\epsilon}$ automatically become ground states and $q_\epsilon \to 1$. Furthermore, \cref{eq:NonAbelianAprox} is trivially satisfied. Indeed this limit corresponds to $g_\s{\t{YM}}^2 A \rightarrow 0$ from the Yang--Mills perspective. It is well known that in this limit 2d Yang--Mills reduces to a topological $BF$ theory. The partition function simplifies \cite{Witten:1992xu} and, in our case, reads
	\begin{equation}
		\Theta_G\qty(q_\epsilon) \,\sim\, C_G\, \qty( \frac{ \lambda \, (\mu R)^{d-2}}{\epsilon} )^{\!\frac{\dim G}{2}} ~,
	\end{equation}
	where $C_G$ is a constant and $\dim G$ is the dimension of the group. It follows that
	\begin{equation}
		\entas_{\n,G} \,\sim\, \frac{ (d-2) \dim G }{2} \, \log (\mu R) \,+\, \t{const} ~. 
	\end{equation}
	We conclude that the entanglement asymmetry of the ground state diverges with the size of the system, signalling the spontaneous breaking of $G$ and agreeing with the Abelian computation. The coefficient of the divergence is of particular significance. We will comment on it at the end of the section.

    \item $\boldsymbol{d = 2 \,.}$ There is no scale $\mu$ in the action \cref{eq:PCM action}, however the dimensionless loop-counting coupling $\lambda^{-1}(E)$ runs with a negative 1-loop beta function \cite{Friedan:1980jm} which makes the NLSM asymptotically free in the UV and strongly coupled in the IR. We thus choose a renormalisation scale $\mu$ where the coupling $\lambda_0^{-1} = \lambda^{-1}(\mu)$ is small, and find a dynamically generated scale $\Lambda_\t{IR} = \mu \, e^{-\lambda_0 / \beta_0}$, where $-\beta_0$ is the negative 1-loop beta-function coefficient. The reader has good reasons to be pessimistic about the reach of a perturbative computation to address the properties of the ground state, but let us momentarily ignore such concerns. The states $\ket{g_\epsilon}$ are not energy eigenstates and, in order to project onto the ground state, we need
	\begin{equation}
		\lambda(R^{-1}) \,\ll\, \epsilon \, C_2 (\fund) ~,
	\end{equation}
    where $C_2(\fund)$ is the smallest quadratic Casimir among all non-trivial representations. Besides, the requirement \cref{eq:NonAbelianAprox} still holds and in this case it reads
	\begin{equation}
	\label{eq:Non-AbLimint}
		1 \ll \lambda(R^{-1}) ~.
	\end{equation}
    The first condition on $\lambda$ can always be satisfied by taking a large enough $\epsilon$; then the only contribution to $\Theta_G\qty(q^\n_\epsilon)$ comes from the singlet representation, which has $C_2( \boldsymbol{1} ) = 0$, and we find
	\begin{equation}
	 	\entas_{\n,G}  \rightarrow 0 \qquad\qquad\text{for $R$ large but $R \lesssim \Lambda_\text{IR}^{-1}$} ~.
	\end{equation}
    This result is valid at large volume, but the volume cannot be taken arbitrarily large while remaining at weak coupling. The wary reader is thus vindicated: the asymmetry vanishes up to a certain large distance $\Lambda_\text{IR}^{-1}$, but it cannot be computed perturbatively at larger distances. The principal chiral model is known to be gapped with a single vacuum at infinite volume, therefore the entanglement asymmetry should vanish, but again, this cannot be determined perturbatively. Note that the Abelian case does not suffer from this limitation: it is Gaussian and the higher modes are trivially decoupled.
    
    \item $\boldsymbol{d=1}\,.$ The low-energy effective theory is already the quantum rigid rotor, \cref{eq:QuantumMechanics}, with moment of inertia \(I = \lambda \mu^{-1}\). Energies are measured in units of \(\mu\) and there is of course no infinite volume limit, as the theory is quantised on a point. The regulated states now read:
    \begin{equation}
        \ket{g_\epsilon} = \ex{-\epsilon \, H_0/\mu} \ket{g} ~.
    \end{equation}
    Taking \(\epsilon\) large --- more specifically \( \epsilon\, C_2(\fund)/ \lambda \gg 1 \) --- has the effect of projecting out any excited states. It is immediate to see that only the singlet contributes, and we find again:
    \begin{equation}
        \entas_{\n,G} \to 0~,
    \end{equation}
    in agreement with the uniqueness of the vacuum in quantum mechanics.
\end{itemize}}

\subsubsection{Partial symmetry breaking}

Now suppose that the system has symmetry \(G\), which is spontaneously broken down to a subgroup \(H\). The resulting NLSM has in this case as target the left coset space \(G/H\) \cite{Coleman:1969sm, Callan:1969sn}. Again, we are interested in the low-energy spectrum of the NLSM on \(\S^{d-1}_R\). Following the discussion above, assuming that $\lambda \, (\mu R)^{d-2}\gg 1$, this can be well approximated by the spectrum of a quantum mechanical particle propagating on \(G/H\). The Hamiltonian in this case is:
\begin{equation}
	H_0 = - \frac{\lapl_{G/H}}{2 I} ~, \qq{with} I = \lambda \, \mu^{d-2} \, R^{d-1} ~,
\end{equation} 
where \(\lapl_{G/H}\) is the Laplacian on the homogeneous space \(G/H\), with a left-invariant metric under $G$. It is simpler to work directly in position basis, i.e., to construct the states analogue to \(\ket{g}\). Following \cite{Albert:2019ztu}, these states are described as
\begin{equation}
	\ket{xH} = \int_H \dd{h} \ket{x\,h} ~,
\end{equation}
where \(x\in G\), so \(xH\) are the left cosets of \(H\) in \(G\). It is immediate to see that the right action of \(H\) is trivial, while the left action of \(G\) is unaffected. These are the symmetry-breaking states. The regulated trace-normalised symmetry-breaking density matrix in this case is:
\begin{equation}
	\rho_x = \frac{\ex{- \epsilon R\, H_0}\ketbra{xH}\ex{- \epsilon R\, H_0}}{\Theta_{G/H}\qty(q_\epsilon)} ~,
\end{equation}
where 
\begin{equation}
	\Theta_{G/H}\qty(q_\epsilon) = \mel{xH}{\ex{- 2 \epsilon R\, H_0}}{xH} ~,
\end{equation}
with \(q_\epsilon\) as above. 

The R\'enyi asymmetries for this symmetry-breaking pattern read straightforwardly:
\begin{equation}
	\entas_{\n,G/H}\qty(\rho_x) = \frac{1}{1-\n} \, \log \, \frac{ \Theta_{G/H}\qty(q_\epsilon^\n) }{ \Theta_{G/H}\qty(q_\epsilon)^\n }~.
\end{equation}
It is not too difficult to obtain an explicit expression of \(\Theta_{G/H}(q_\epsilon)\) following the technology developed in \cite{Albert:2019ztu} and the Peter--Weyl theorem for \(G/H\) \cite{Farashahi:PW}. However, since we are interested in the large volume limit \(\mu R\to\infty\), we do not need such an expression. It suffices to notice that \(\Theta_{G/H}\) is a heat kernel and that (in \(d>2\)) the infinite volume limit corresponds to a short time expansion of the heat kernel. As such, we have
\begin{equation}
	\Theta_{G/H}(q_\epsilon) \,\sim\, C_{G/H} \qty( \frac{ \lambda \, (\mu R)^{d-2} }{ \epsilon} )^{\!\frac{\dim(G/H)}{2}}
\end{equation}
at $\mu R \gg 1$, and \(C_{G/H}\) is a constant. Hence, with \(d>2\) and in the infinite volume limit:
\begin{equation}
	\entas_{\n,\, G/H}(\rho_x) \,\sim\, \frac{ (d-2) \dim(G/H)}{2} \, \log(\mu R) \,+\, \t{const} ~.
\end{equation}
A similar behaviour of asymmetry, featuring the coefficient \(\dim(G/H)\), was reported in lattice models for matrix product states \cite{Capizzi:2023xaf}. For \(d\leq 2\) the discussion is identical to the one for $G  \to \set{1}$ and we refer the reader to it. 

We close this section by giving a unified perspective on our results. First, we have proven in the Abelian case, and provided evidence for in the non-Abelian case, that entanglement asymmetry provides a rigorous entropic version of the CMW theorem. Moreover, whenever SSB is allowed (for \(d>p+2\)), \(\entas\) exhibits a characteristic behaviour summarized as:
\begin{equation}
    \entas_{\gf{G}{p}} = \frac{\cN\, (d-p-2)}{2} \, \log(\mu R) ~, \qquad\qq{as} \mu R\gg 1 ~, 
\end{equation}
were \(\cN\) is the number of broken generators of \(\gf{G}{p}\) in the symmetry broken phase. In the 0-form case, and under the assumption that the effective theory of the spontaneously-broken phase is Lorentz invariant, \(\cN\) coincides with the number of Goldstone modes. This behaviour was conjectured by Metlitski and Grover \cite{Metlitski:2011pr} to capture the logarithmic contribution of Goldstone fields to entanglement entropy. Here we have shown that entanglement asymmetry isolates precisely this term, and moreover we have confirmed the conjecture of \cite{Metlitski:2011pr} in the 0-form case. Extending this reasoning to the higher-form case, it becomes evident that what this coefficient is actually capturing is broken generators, rather than the individual gapless degrees of freedom --- which one would naturally associate with the number of Goldstone modes \cite{Hidaka:2020ucc}. The latter would count polarisations of the \(p\)-form gauge field and would have given $\bigl( \begin{smallmatrix} d-2 \\ p \end{smallmatrix}\bigr) \, \mathcal{N}$ instead.

\section{Subregion theorem for \tps{\U(1)}{U(1)} 0-form symmetry}
\label{sec:subregion0-form}

As explained in the introduction, entropic order parameters (here entanglement asymmetry) have a key advantage over conventional diagnostics: they can probe symmetry breaking within a spatial subregion. This section makes that concrete by computing the entanglement asymmetry of a compact scalar field in $d$ spacetime dimensions in a subregion $\cA$. We will draw the ensuing conclusions in the form of a subregion CMW theorem. Our computation is closely related to the computation of entanglement entropy, which was carried out for $d=3$ in \cite{Agon:2013iva}. In \cref{sec:subregionp-form} we will extend our results to $(p+1)$-form gauge fields that describe the low-energy dynamics of spontaneously-broken $p$-form symmetries.

To set the stage, consider a compact scalar field whose action is the $p=0$ case of \cref{Eq:GenMaxwell}:
\begin{equation}
\label{eq:scalar-action}
    S[\phi] = \frac{\lambda \, \mu^{d-2}}{2}\int_{X_d} \f{f}{1} \wedge \star \, \f{f}{1} ~.
\end{equation}
Here we have denoted by $\f{f}{1} = \dd{\phi}$ the field strength of the compact scalar, which is periodically identified \(\phi\sim\phi+2\pi\) and taken canonically dimensionless. As before, $\lambda$ is a dimensionless coupling constant and $\mu$ an arbitrary energy scale. From a sigma model perspective, \( (\lambda \, \mu^{d-2})^{1/2} \) is the dimensionful target-space radius (when $d=2$, $\mu$ does not appear in the Lagrangian and $\lambda^{1/2}$ is the dimensionless radius). We will be interested in the $\U(1)$ shift symmetry
\begin{equation}
    \phi \rightarrow \phi + c ~.
\end{equation}
Let us quantise this theory on a trivial spatial slice \(\Sigma = \R^{d-1}\). We prepare a global pure state \(\rho\propto \ketbra{\theta}\) on \(\Sigma\) by specifying Dirichlet boundary conditions at infinity:%
\footnote{The same state can be prepared by a Euclidean path integral on a half of \(\R^d\) with the insertion of an operator \(\cO_\theta = \sum_{k\in\Z} \ex{\ii k \theta} V_k\), where \(V_k\) are the vertex operators of the theory \cite{Vitouladitis:2025zoy}.}
\begin{equation}
\label{eq:theta-BCs-0form}
	\phi(x) \to \theta \qq{as} \abs{x} \to \infty ~.
\end{equation}
Without loss of generality we choose \(\theta=0\) henceforth. Note that while the boundary conditions break the \(\U(1)\) symmetry, this is not enough to guarantee SSB. Indeed, in two dimensions, even when the field is fixed at a point, long-range order is destroyed by fluctuations as one moves away from that point \cite{Coleman:1973ci}.

We then restrict our state to a subregion. For simplicity, we choose a tractable, yet rich enough, setup. We take our subregion to be a ball of radius \(R\): \(\aent = \B^{d-1}_R \subset \Sigma\). On this subregion we compute the entanglement asymmetry of the resulting reduced state \(\rho_\aent\). This leads ultimately to one of the core results in this paper, \emph{the subregion CMW theorem}.

Along the way, we extend known results on the entanglement entropy of compact scalars to the settings relevant to our analysis. Before diving into the computation, we briefly collect a few elementary but recurring ingredients that will streamline what follows.

\subsection{Partition functions and winding sectors}

Suppose one is interested in the partition function of the theory defined by \cref{eq:scalar-action} on a generic spacetime manifold \(X_d\). In our applications \(X_d\) will be either \(\R^d\) or the replica manifold (see \cref{sec:higher-form-asymmetry}) and we will be interested in cases where \(X_d\) has a boundary. In those cases, one must specify boundary conditions compatible with the variational principle: either Dirichlet, \(\eval*{\f{f}{1}}_{\pd X_d}=0\), or Neumann, \(\eval*{\qty(\star\f{f}{1})}_{\pd X_d}=0\).%
\footnote{We do not include the possibility of edge modes. Allowing them enlarges the set of admissible boundary conditions, see e.g. \cite{Ball:2024hqe}.}
We proceed using Dirichlet boundary conditions, as these will turn out to be relevant in the actual computation, and we briefly comment on the other choice in the end. 

As in the case without boundary (which we explored in \cref{sec:entropic-CMW-global}), one may split \(\f{f}{1}\) as
\begin{equation}
\label{eq:f-split}
    \f{f}{1} = \f{f}{1}^{(h)} + \dd{\phi} ~.
\end{equation}
Here $\phi$ is a globally-defined noncompact scalar, while ${f_1}^{\!(h)}$ is a harmonic contribution whose role is to capture the windings.%
\footnote{Two field configurations of $\phi(x)$ that differ by a global shift of $2\pi$ are considered gauge equivalent.} With the chosen boundary conditions, ${f_1}^{\!(h)}$ is a representative of the relative cohomology \(\H^1(X_d,\pd X_d)\) (with Neumann boundary conditions it would be \(\H^1(X_d)\)). See also \cref{App:TopRepMan} for details. When \(X_d\) has no boundary, the discussion reduces to that of \cref{sec:entropic-CMW-global}.

Using the splitting \cref{eq:f-split} of the field strength and the orthogonality between harmonic and exact forms, the partition function on \(X_d\) factorises,
\begin{equation}
\label{eq:parti-split}
	\parti[X_d] = \parti^\t{inst}[X_d] \; \parti^\t{osc}[X_d] ~,
\end{equation}
where \(\parti^\t{inst}[X_d]\) captures the sum over \(\f{f}{1}^{(h)}\) (the winding sectors/instantons), while \(\parti^\t{osc}[X_d]\) is the oscillator contribution from \(\phi\). Explicitly:
\begin{align}
    \parti^\t{inst}[X_d] &= \int \DD{\f{f}{1}^{(h)}} \, \exp(-\frac{\lambda \, \mu^{d-2}}{2}\int_{X_d} \f{f}{1}^{(h)} \w \star\, \f{f}{1}^{(h)}) \qq{and} \label{eq:WindingParti} \\
    \parti^\t{osc}[X_d] &= \int \DD{\phi} \; \exp(-\frac{\lambda \, \mu^{d-2}}{2}\int_{X_d} \dd{\phi}\w\star \dd{\phi}) ~.
\end{align}
The oscillator piece is a standard one-loop determinant that is invariant under shift symmetry and will therefore play no role in the computations below. The instanton contribution deserves closer attention.

By flux quantisation, the path integral over winding sectors reduces to a discrete sum. Let \(\set{\tau_i}\) be a basis of Dirichlet harmonic 1-forms on \(X_d\) (i.e., such that $\tau_i \big| {}_{\partial X_d} = 0$ and with integer periods) with \(i=1,2,\ldots,b_1\) and where \( b_1 =\dim\H^1(X_d,\pd X_d)\). For details, see \cref{App:TopRepMan}.
Then ${f_1}^{\!(h)}$ is expressed as 
\begin{equation}
	\f{f}{1}^{(h)} = 2\pi w^i \tau_i ~, \qquad w^i\in\Z ~,
\end{equation}
with implicit summation over \(i\). The integers \(w^i\) label the winding numbers on each (relative) cycle and are summed over in the path integral. The instanton action, appearing in the exponent of \cref{eq:WindingParti}, becomes 
\begin{equation}
	S^\t{inst} = \frac{\lambda \, \mu^{d-2}}{2}\int_{X_d} \f{f}{1}^{(h)}\w \star \, \f{f}{1}^{(h)} = 2\pi^2 \lambda \, \mu^{d-2}\; w^i\, \bbM_{ij}\, w^j ~,
\end{equation}
where the matrix \(\bbM_{ij}\) denotes the Gram matrix of this basis:
\begin{equation}
	\bbM_{ij} = \int_{X_d}\tau_i \w\star \, \tau_j ~.
\end{equation}
We will refer to \(\bbM_{ij}\) as the \emph{instanton matrix} of \(X_d\). This is manifestly a symmetric and strictly positive-definite real matrix. In \cref{app:electrostatics} we perform a careful computation of \(\bbM_{ij}\) for the relevant replica geometry by translating it to an equivalent problem of calculating a capacitance matrix in electrostatics, building upon \cite{Agon:2013iva}.

With \(\bbM_{ij}\) specified, the instanton partition function becomes:
\begin{align}
\label{eq:inst-0form}
	\parti^\t{inst}[X_d] = \sum\nolimits_{ \substack{ w^i \, \in \, \Z \\[.1em] i=1,\ldots,b_1}} \, \exp(-2\pi^2 \lambda \, \mu^{d-2}\; w^i\, \bbM_{ij}\, w^j)= \Theta\qty( 2\pi\lambda \, \mu^{d-2} \, \bbM_{ij}) ~.
\end{align}
In the last step we introduced the Siegel theta function. For a strictly positive-definite symmetric \(n\times n\) matrix \(\bbA\), it is defined as:
\begin{equation}
\label{eq:Theta-def}
	\Theta(\bbA) = \sum\nolimits_{\vec{w} \,\in\, \Z^n} \, \ex{-\pi \, \vec{w}^\mathsf{T} \bbA\, \vec{w}} ~.
\end{equation}
Finally, we note that in trivial topology \(\f{f}{1} = \dd{\phi}\) globally and hence \(\parti^\t{inst}[X_d]=1\).

\subsection{R\'enyi entropies of a compact scalar in \tps{d}{d} dimensions}
\label{ssec:Rényi-0form}

\begin{figure}[tbp]
    \centering
    \noindent\makebox[\textwidth]{
    \def\svgwidth{\paperwidth}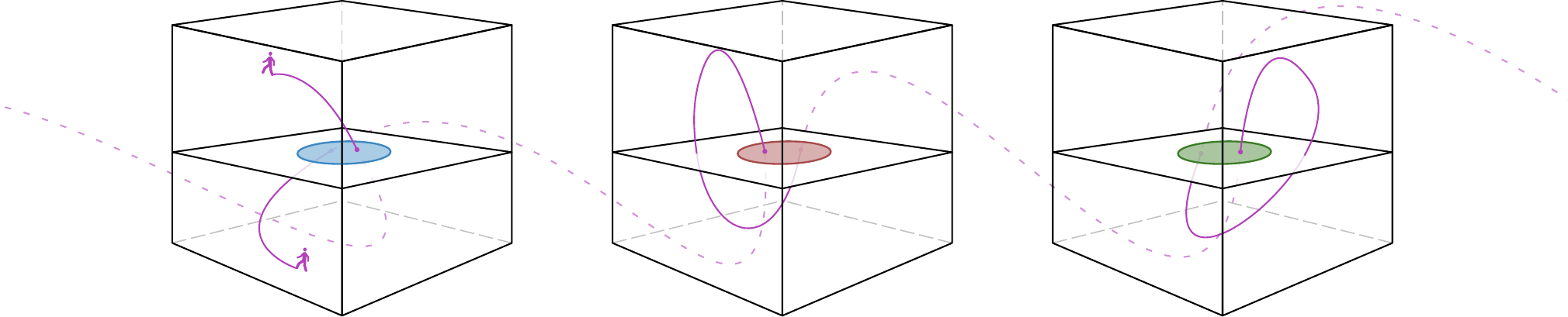}
    \caption{A cartoon of the replica manifold \(X_\n\). The coloured discs denote the location of the branch cuts $\B_R^{d-1}$, which are identified cyclically.}
    \label{fig:portal}
\end{figure}

We build up towards entanglement asymmetry by first computing the R\'enyi entropies of a compact scalar field on \(\mathbb{R}^d\). Our analysis follows and extends \cite{Agon:2013iva}, which treated the case \(d = 3\) in detail and sketched partial results for general \(d\). Consider the subregion \(\cA = \B^{d-1}_R\). We aim to evaluate the Rényi entropies:
\begin{equation}
   \ent_\n \coloneqq \frac{1}{1-\n} \, \log \frac{\tr(\rho_\aent^\n) }{ (\tr\rho_\aent)^\n} ~.
\end{equation}
Here the traces of \(\rho_\aent\) and its powers are computed via a Euclidean path integral, as explained in \cref{sec:pathintegralmethods}. Owing to the factorisation in \cref{eq:parti-split}, the R\'enyi entropies naturally decompose into winding and oscillator contributions:
\begin{equation}
    \ent_\n = \ent_\n^{\t{inst}} + \ent_\n^{\t{osc}} ~.
\end{equation}
As explained earlier, for our purposes it will suffice to compute the winding contribution alone. With that in mind, we evaluate
\begin{equation}
\label{eq:SN-wind-def}
    \ent_\n^{\t{inst}} = \frac{1}{1-\n} \, \log \frac{\parti^{\t{inst}} [X_\n] }{ \qty(\parti^{\t{inst}}[X_1])^\n} ~.
\end{equation}
The geometry in question is the replica manifold \(X_\n\). It is constructed by taking \(\n\) copies of \(\mathbb{R}^d\) and cyclically gluing them along the \(\n\) balls \(\B^{d-1}_R\), as illustrated in \cref{fig:portal}. The result is an \(\n\)-fold branched cover of \(\mathbb{R}^d\), whose branch locus (generalization of the branch points in $d=2$) is the entangling surface \(\pd\aent = \S^{d-2}_R\). Note here that \(X_\n\) possesses \(\n\) asymptotic boundaries --- which we shall denote as \(B_i\) (\(i=1,\ldots,\n\)) --- one on each sheet. We will refer to the full asymptotic boundary of \(X_\n\) as 
\begin{equation}
    B_\t{asym} = B_1 \sqcup \cdots \sqcup B_\n ~.
\end{equation}
Further details on the topology of \(X_\n\) are provided in \cref{App:TopRepMan}.

As discussed in \cref{sec:pathintegralmethods}, we must regulate the entangling surface by excising a tubular neighbourhood of radius $\epsilon$ around it. This procedure introduces an additional \textit{entanglement boundary}, on top of the asymptotic boundaries already present on the replica manifold: 
\begin{equation}
    B_\t{ent} = S^1_{\n\epsilon} \times S^{d-2}_R ~.
\end{equation}
On the entanglement boundary we impose boundary conditions chosen to ensure a clean factorisation of the Hilbert space. At this stage, for the computation of R\'enyi entropies, these are given by Dirichlet or Neumann boundary conditions. Both are equally valid. However, once we turn to the calculation of R\'enyi asymmetries, there is an additional requirement: they must allow the involved topological operators to end topologically on the entanglement boundary. Such conditions are referred to as symmetric boundary conditions in the literature.

In general, $\ent_\n$ has divergent contributions as the cutoff \(\epsilon\) is removed \cite{Ohmori:2014eia}. However, as we show in \cref{app:smooth-Bent}, the winding contribution remains well behaved in this limit. Therefore, from here on, we take $\epsilon \to 0$ and remove the cutoff. In addition, we need to impose boundary conditions on the asymptotic boundaries. Specifically, in accordance with the discussion at the beginning of this section, we impose
\begin{equation}
    \phi \big|_{B_i} = 0
\end{equation}
at each asymptotic boundary \(B_i\). This choice is consistent with the replica construction. In particular, since $\parti[X_\n] = \tr(\rho_\aent^\n)$ and $\rho_\aent = \tr_{\coaent} \ketbra{0}$, the field \(\phi\) must take the same value across all asymptotic boundaries.

For consistency, imposing Dirichlet conditions for \(\phi\) also enforces Dirichlet conditions on the harmonic piece ${f_1}^{\!(h)}$. This allows us to compute the winding contribution using the formalism developed earlier. Topologically, \(X_\n\) is a \(d\)-dimensional sphere with \(\n\) balls excised (corresponding to the $\n$ asymptotic boundary components). This geometry harbours \(\n-1\) nontrivial relative 1-cycles that connect those balls. Central to the computation is the \((\n-1) \times (\n-1)\) instanton matrix \(\bbM_{\n,d}\) associated with this geometry (rescaled so as to set $R=1$). This matrix turns out to have a surprisingly interesting structure, as we explain in \cref{app:ssec:M-structure}. It is a Toeplitz matrix and its components are given by:
\begin{equation}
\label{eq:MNd-components}
    \qty[\bbM_{\n,d}]_{j\ell} = \frac{1}{\n} \, \sum_{k=1}^{\n-1} \, \ex{2\pi \ii \frac{(j-\ell)k}{\n}} \, C_d\qty(\frac{k}{\n}) ~.
\end{equation}
The function \(C_d\qty(k/\n)\) is computed by solving the Laplace equation in \(d\) dimensions with monodromies that depend on \(k/\n\) along the entangling surface. We devote \cref{app:ssec:solving-laplace,app:assemblingpieces,subsec:numericaleval} to this computation. When the dust settles, we find:
\begin{equation}
\label{eq:expr-cd}
	C_d \left(\frac{k}{\n}\right) = \frac{ 4 \, \pi^{\frac{d-1}2} \, \Gamma \bigl( \frac{d-1}{2} \bigr) }{ \Gamma \bigl( \frac{d-2}{2} \bigr) \, \Gamma\bigl( \frac{d}{2} \bigr)} \; u^{\sfT} \qty[\mathbf{1} + \cot[2](\frac{\pi\, k}{\n})\, T^{(d)}]^{-1} u ~.
\end{equation}
Here $T^{(d)}$ is an infinite-dimensional matrix defined in \cref{subsec:numericaleval}, $\mathbf{1}$ is the identity matrix, whereas $u$ is a unit vector that projects to the component in the upper-left corner. In \cite{Agon:2013iva} an expression for $C_d\bigl( \frac{k}{\n} \bigr)$ in \(d = 3\) was proposed:
\begin{equation}
\label{eq:c0-ahjk}
    C_{d=3}\left( \frac{k}{\n} \right) = 4\pi \left( 1-\frac{2k}{\n} \right) \tan(\frac{\pi k}{\n}) ~.
\end{equation}
Here we propose an expression for the case $d=4$:
\begin{equation}
\label{eq:c0-d=4}
    C_{d=4}\left( \frac{k}{\n} \right) = 8\pi^2 \, \frac{k}{\n} \, \left( 1 - \frac{k}{\n} \right) ~.
\end{equation}
We motivate these proposals in \cref{app:ssec:analytic}. We do not have closed formulas for other values of $d$, but numerical evaluation is possible, see \cref{subsec:numericaleval}.

Let us proceed, since we will see that most of our results do not depend on the form of $C_d\bigl( \frac{k}{\n}\bigr)$. From the denominator of \cref{eq:expr-cd} we see that something special happens in \(d=2\): the instanton matrix reduces identically to the  zero matrix. This signals that all instanton configurations contribute equally and are unsuppressed, each carrying a vanishing action. We will argue in \cref{ssec:theorem-subregion0-form} that this feature provides a way to prove the Coleman--Mermin--Wagner theorem. For the time being, we restrict our attention to \(d>2\). The two-dimensional case requires some care, and we defer its discussion to \cref{ssec:theorem-subregion0-form}.

Proceeding with the computation, we now have all the ingredients to determine the instanton partition function. It reads:
\begin{equation}
\label{eq:partition-0form-exact}
    \parti^\t{inst}[X_\n] = \Theta\qty(2\pi\lambda \, (\mu R)^{d-2} \, \bbM_{\n,d}) ~.
\end{equation}
Note here the explicit dependence on the radius \(R\) of the subregion $\cA$. In the appendix the branch locus was placed on a unit sphere; rescaling it to a sphere of radius \(R\) accounts for the factor of \((\mu R)^{d-2}\) appearing above.%
\footnote{In this and the following section, contrary to \cref{sec:entropic-CMW-global}, we adopt standard conventions for the radius $R$ of the subregion along the noncompact directions, so that a $(d-2)$-dimensional sphere $S^{d-2}_R$ has $\text{area} = \Omega_{d-2} R^{d-2}$ with $\Omega_{d-2} = 2 \, \pi^{(d-1)/2} / \Gamma\bigl( \frac{d-1}2 \bigr)$.}
Moreover, when \(\n = 1\) the manifold is topologically trivial and the instanton contribution to the partition function in the denominator of \cref{eq:SN-wind-def} reduces to 1. Hence, the instanton contribution to the R\'enyi entropy becomes:
\begin{equation}
\label{eq:Rényi-0form-exact}
    \ent_\n^\t{inst} = \frac{1}{1-\n} \, \log\Theta\qty( 2\pi\lambda \, (\mu R)^{d-2} \, \bbM_{\n,d}) ~.
\end{equation}
It is sometimes useful to rewrite \cref{eq:Rényi-0form-exact} in a different way. A natural generalisation of \cref{eq:modular} yields \cite{Mumford1982TataLO}:
\begin{equation}
    \Theta \bigl(\bbA \bigr) = \frac{1}{\sqrt{\det\bbA}}\; \Theta\qty(\inv{\bbA}) ~.
\end{equation}
Using this identity we get the equivalent expression:
\begin{equation}
\label{eq:Rényi-0form-exact-poisson}
    \ent_\n^\t{inst} = \frac{1}{2 \, (\n-1)} \log\det( 2\pi \lambda \, \qty(\mu R)^{d-2} \,\bbM_{\n,d}) + \frac{1}{1-\n} \log \Theta\qty( \frac{1}{2\pi \lambda \, (\mu R)^{d-2}} \, \bbM_{\n,d}^{-1}) ~.
\end{equation}

\subsubsection{Simplifying limits and entanglement entropy}
\label{sssec:entanglement-entropy}

While \cref{eq:Rényi-0form-exact,eq:Rényi-0form-exact-poisson} give exact expressions for the R\'enyi entropies, they can be rather unwieldy when it comes to extracting the entanglement entropy, because analytically continuing a Siegel theta function is generally intractable. Nevertheless, each of \cref{eq:Rényi-0form-exact,eq:Rényi-0form-exact-poisson} is useful in different limiting regimes of the subregion radius \(R\).

In the regime \(\mu R \gg 1\), we use the asymptotic behaviour
\begin{equation}
    \lim_{s\to\infty} \Theta(s \bbA) = 1
\end{equation}
to deduce, from \cref{eq:Rényi-0form-exact}, that the instanton contribution to the R\'enyi entropy vanishes:
\begin{equation}
    \ent_\n^\t{inst} \to 0 ~, \qquad\qq{as} \mu R\to \infty ~.
\end{equation}
In contrast, in the opposite regime \(\mu R \ll 1\), we appeal to \cref{eq:Rényi-0form-exact-poisson} which gives:
\begin{equation}
    \ent_\n^\t{inst} \sim \frac{1}{2} \log(2\pi \lambda \, \qty(\mu R)^{d-2}) + \frac{1}{2 \, (\n-1)}\log\det\bbM_{\n,d}~.
\end{equation}
Assuming that the matrix \(\bbM_{\n,d}\) is positive definite, we obtain:
\begin{equation}
    \ent_\n^\t{inst} \sim \frac{d-2}{2} \, \log (\mu R) \,+\, \t{const} ~,
\end{equation}
where the additive constant depends on the details of the function \(C_d\) and therefore on $\n$ (for $d=3,4$ we compute $\det\bbM_{\n,d}$, from which the constant follows, in \cref{app:ssec:determinant}). These limits give us a handle on the entanglement entropy, since in both regimes the continuation \(\n \to 1\) is straightforward. Altogether, we find:
\begin{equation}
    \ent^\t{inst} = \lim_{\n\to 1} \, \ent_\n^\t{inst} \sim 
    \begin{cases}
    0 ~, & \mu R \gg 1 ~, \\
    \frac{d-2}2 \log (\mu R) \,+\, \t{const} ~, & \mu R \ll 1 ~.
    \end{cases}
\end{equation}
Note that \(\ent^{\t{inst}}\) is negative for all \(\mu R\). This is easy to see at \(\mu R \ll 1\), while for general \(\mu R\) it follows from monotonicity of the Siegel theta function. This feature is not a cause for concern: the full entanglement entropy of the theory is \(\ent = \ent^{\t{inst}} + \ent^{\t{osc}}\), where \(\ent^{\t{osc}}\) denotes the oscillator contribution. As discussed above, the oscillator contribution must be computed in the presence of a UV regulator and typically diverges as a power of the cutoff. Consequently, \(\ent^{\t{inst}}\) is always subleading. In this light, \(\ent^{\t{inst}}\) is best seen as a negative universal subleading correction, much like the topological entanglement entropy
\cite{Kitaev:2005dm, Levin:2006zz} in topologically ordered phases.

\subsection{R\'enyi asymmetries of a compact scalar in \tps{d}{d} dimensions}

We now turn to the central object of interest for us: the R\'enyi asymmetries, defined in \cref{eq:Rényis} as
\begin{equation}
    \entas_{\n,\U(1)} = \ent_{\n,\sfS} - \ent_\n ~.
\end{equation}
Here \(\ent_{\n,\sfS}\) denotes the contribution from the symmetrised state \(\rho_{\aent,\sfS}\):
\begin{equation}
    \ent_{\n,\sfS} = \frac{1}{1-\n} \, \log\parti_{\n,\sfS} = \frac{1}{1-\n} \, \log\tr(\rho_{\aent,\sfS}^\n)~.
\end{equation}
This relation also defines the symmetrised partition function \(\parti_{\n,\sfS}\). In terms of the charged moments \(\mathcal{E}_\n(\vec{\alpha})\) defined in \cref{eq:ChargedMoments}, the symmetrised partition function takes the form
\begin{equation}
\label{eq:integrated-charged-moment}
    \parti_{\n,\sfS} = \, \prod_{i=1}^\n \, \int_0^{2\pi} \frac{\dd{\alpha_i}}{2\pi} \; \cE_\n(\vec{\alpha}) ~. 
\end{equation}
To evaluate the charged moments, we proceed as follows. We begin by inserting symmetry operators, as shown schematically in \cref{eq:ChargedMoment}, into the path integral computed in the previous subsection. It is important to note that, in order to define the charged moments as in \cref{eq:ChargedMoment}, it was essential to regulate the entangling surface and impose symmetric boundary conditions there.

Symmetric boundary conditions require that the topological operators
\begin{equation}
    U_\alpha(\gamma) = \exp(\ii \alpha \int_\gamma \star \, \f{f}{1})
\end{equation} 
can end topologically on the entangling boundary. It is then evident that this requirement is satisfied precisely by imposing Neumann boundary conditions on \(B_\t{ent}\). Once this is done, the topological operators may be smoothly deformed so that they encircle the punctures corresponding to the asymptotic boundaries. Following the discussion in \cref{ssec:Rényi-0form}, the entangling boundary can then be safely shrunk away. Pictorially, the resulting path integral takes the form
\begin{equation}
    \cE_\n(\vec{\alpha}) \,= \;\; \vcenter{\hbox{\def\svgwidth{0.35\textwidth}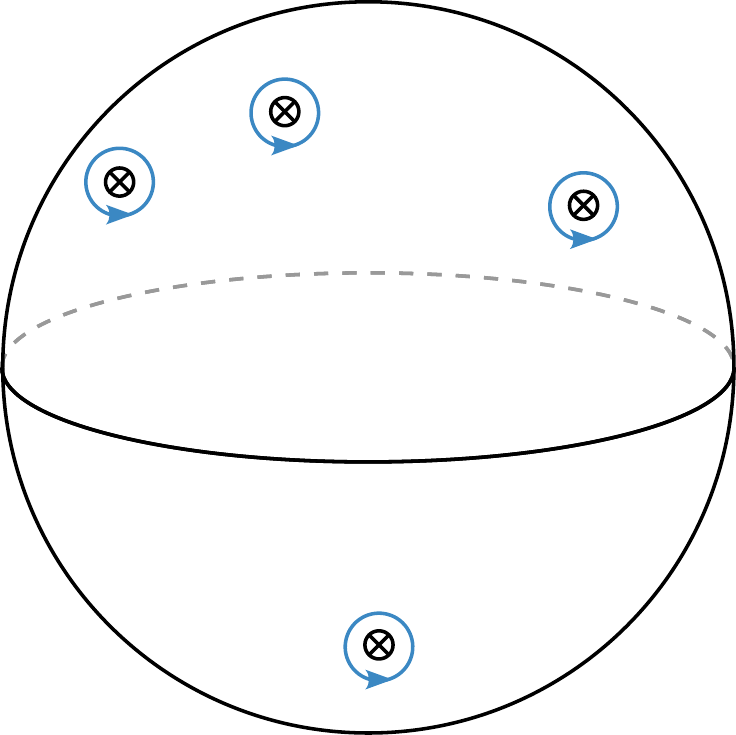}} ~.
\end{equation}

The symmetry operators act on the scalar field by inducing constant shifts. Consequently, their effect on the boundary conditions is to modify the asymptotic behaviour of \(\phi\) to
\begin{equation}
    \phi \big|_{B_i} = \alpha_i ~.
\end{equation}
It follows that the symmetry operators shift the fluxes through the relative one-cycles \(\gamma_i\) of the replica manifold to the new values
\begin{equation}
    \frac{1}{2\pi}\int_{\gamma_i} \f{f}{1} = w^i + s^i ~,
\end{equation}
where the fractional shifts \(s^i \in \ropen{0}{1}\) are defined as
\begin{equation}
    s_i = \frac{1}{2\pi}\qty(\alpha_i - \alpha_\n) ~.
\end{equation}
These modifications affect only the topological (instanton) sector of the theory, leaving the oscillator part of the partition function untouched. Hence, the charged moments factorise again:
\begin{equation}
    \cE_\n(\vec{s}) = \cE_\n^\t{inst}\qty(\vec{s}) \; \parti^\t{osc}\qty[X_\n] ~,
\end{equation}
where the instanton contribution is given by
\begin{equation}
    \cE_\n^\t{inst}(\vec{s}) = \sum\nolimits_{\vec{w} \,\in\, \Z^{\n-1}} \exp( - 2\pi^2 \lambda \, (\mu R)^{d-2} \; \qty(\vec{w}+\vec{s})^{\!\mathsf{T}} \, \bbM_{\n,d} \, \qty(\vec{w}+\vec{s})) ~.
\end{equation}
Here we have assembled the flux shifts \(s_i\) into a vector \(\vec{s} \in I^{\n-1}\) with \(I = \ropen{0}{1}\). 

The symmetrised partition function is obtained by integrating over the arguments \(\alpha_i\) of the topological operators. Equivalently, this can be expressed as an integral over \(\vec{s}\), yielding
\begin{align}
    \parti_{\n,\sfS}^\t{inst} &= \int_{I^{\n-1}} \dd{\vec{s}} \; \sum\nolimits_{\vec{w} \,\in\, \Z^{\n-1}} \, \exp(-2\pi^2 \lambda \, (\mu R)^{d-2} \; \qty(\vec{w}+\vec{s})^{\!\mathsf{T}} \, \bbM_{\n,d} \, \qty(\vec{w}+\vec{s})) \nn 
    &= \int_{\R^{\n-1}} \dd{\vec{v}} \; \exp(-2\pi^2 \lambda \, (\mu R)^{d-2}\; \vec{v}^{\mathsf{T}} \, \bbM_{\n,d} \, \vec{v}) \nn  
    &= \qty[\det(2\pi \lambda \, (\mu R)^{d-2} \, \bbM_{\n,d})]^{-1/2} ~.
\label{eq:Z_symm}
\end{align}
In the second line we combined the continuous shifts \(\vec{s} \in I^{\n-1}\) with the discrete jumps \(\vec{w} \in \mathbb{Z}^{\n-1}\) into a single vector \(\vec{v} \in \mathbb{R}^{\n-1}\) and performed the resulting Gaussian integral.

The full symmetrised partition function is therefore
\begin{equation}
    \parti_{\n,\sfS} = \parti_{\n,\sfS}^\t{inst} \; \parti^\t{osc}\qty[X_\n] ~.
\end{equation}
The oscillator contributions cancel in the R\'enyi asymmetry, leaving
\begin{equation}
    \entas_\n = \frac{1}{1-\n} \, \log\frac{\parti_{\n,\sfS}^\t{inst} }{ \parti^\t{inst}[X_\n]} ~.
\end{equation}
Finally, substituting from \cref{eq:partition-0form-exact} and \cref{eq:Z_symm}, we obtain a compact closed form for the R\'enyi asymmetries:
\begin{equation}
\label{eq:asymmetry-0form-exact}
    \entas_\n = \frac{1}{2 \, (\n-1)} \, \log\det(2\pi \lambda \, (\mu R)^{d-2} \, \bbM_{\n,d}) + \frac{1}{\n-1} \, \log \Theta \qty( 2\pi \lambda \, (\mu R)^{d-2}\,\bbM_{\n,d}) ~.
\end{equation}
For completeness, we also note an equivalent representation obtained upon Poisson resummation of the theta function, namely exploiting \cref{eq:Rényi-0form-exact-poisson}:
\begin{equation}
\label{eq:asymmetry-0form-exact-poisson}
    \entas_\n = \frac{1}{\n-1} \, \log \Theta \qty( \frac{1}{2\pi \lambda \, (\mu R)^{d-2}} \, \bbM_{\n,d}^{-1}) ~.
\end{equation}

As with the R\'enyi entropies, also the R\'enyi asymmetries involve the Siegel theta function, from which it is cumbersome to extract explicit information. To understand how the asymmetries behave as functions of the subregion size, it is therefore instructive to examine the limiting cases where the subregion is either very large or very small.

In the case \(\mu R \gg 1\) of a large subregion, we may appeal to  the asymptotic behaviour of the Siegel theta function, or equivalently our previous result that \(\ent_\n^\t{inst}\) vanishes in that regime. In particular, from \cref{eq:asymmetry-0form-exact} we find:
\begin{equation}
\label{eq:EntasBigmur}
    \entas_\n \sim \frac{d-2}{2} \, \log (\mu R) \,+\, \t{const} ~, \qquad\qq{as} \mu R \gg 1 ~,
\end{equation}
up to exponentially small corrections. The constant can be written in terms of $\lambda$ and the determinant of $\bbM_{\n,d}$. Conversely, in the limit \(\mu R \ll 1\) of a small subregion, we can use \cref{eq:asymmetry-0form-exact-poisson} to obtain
\begin{equation}
    \entas_\n \,\sim\, 0 ~, \qquad\qq{as} \mu R \ll 1 ~,  
\end{equation}
up to exponentially small corrections. Beyond these two simplifying regimes, using the fact that \(\bbM_{\n,d}\) is positive definite, one can show that the R\'enyi asymmetries increase monotonically with the radius \(R\).

\begin{figure}[p]
    \centering  
    \includegraphics[width=0.9\textwidth]{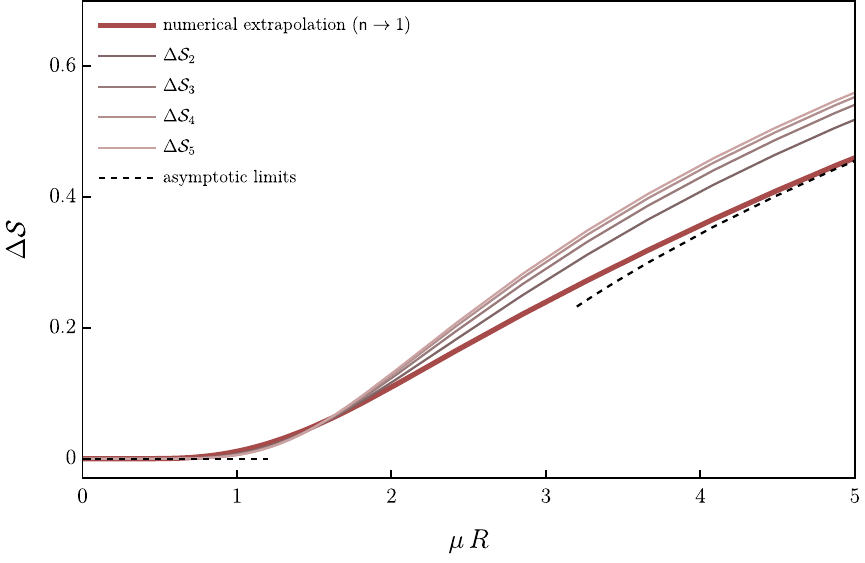}   
    \caption{\label{fig:EntasFinal}%
    Entanglement asymmetry for a compact scalar field in $d=3$ dimensions, with $\lambda = \frac{1}{(4 \pi )^{3/2}}$. We show the R\'enyi asymmetries $\entas_\n$ for $\n = 2,3,4,5$, the numerical extrapolation to $\n \rightarrow 1$, and the asymptotic behaviours that the $\n \rightarrow 1$ limit must reproduce. In this case, for the numerical extrapolation we also used the Rényi asymmetries for $\n = 6,7$ and an improved ansatz $f(\n) = a + b \, \n + c \, \n^{2} + d \, \n^{-1} + e \, \n^{-2}$. }
\end{figure}

\begin{figure} 
    \centering  
    \includegraphics[width=0.9\textwidth]{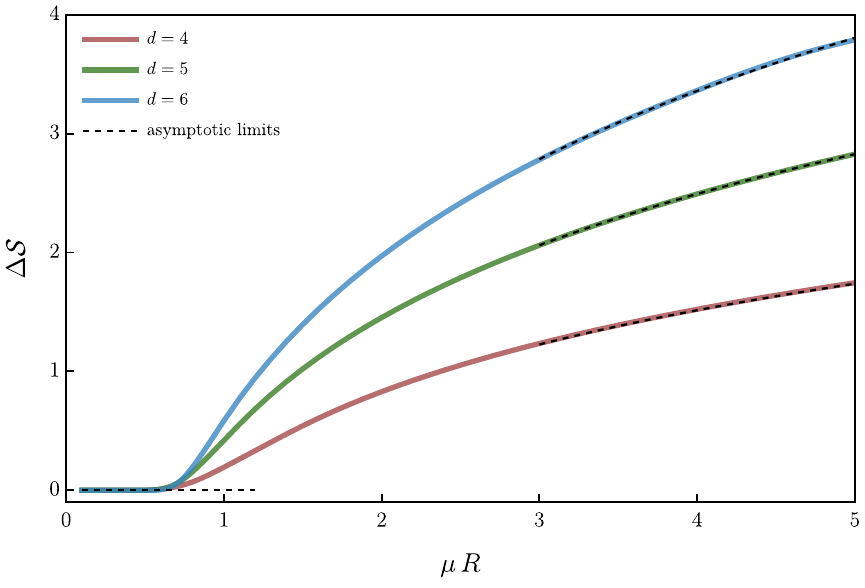}   
    \caption{\label{fig:EntasGend}%
    Entanglement asymmetry for a compact scalar field in $d=4,5,6$, with $\lambda = \frac{1}{(4 \pi )^{3/2}}$. We show the numerical extrapolations to $\n \to 1$ and the asymptotic limits. When drawing the asymptotic limits at large $\mu R$, the additive constants in \cref{eq:EntasBigmur} were fitted for $d=5,6$.}
\end{figure}

Let us now consider the interpolating behaviour of $\entas_\n$ and $\entas$ away from the aforementioned limits, using the methods described in \cref{subsec:numericaleval}. First, we compute the R\'enyi asymmetries $\entas_\n$ for several values of $\n$. In $d=3$ and $d=4$ dimensions this can be done analytically using the explicit form \cref{eq:MNd-components} of the instanton matrix $\bbM_{\n,d}$, obtained from the function $C_d(\beta)$. In other dimensions, the computation can be done numerically by truncating the infinite-dimensional matrix $T^{(d)}$ in \cref{eq:expr-cd}. Once $\entas_\n$ is computed for several values of $\n$, one can numerically extrapolate to $\n = 1$ as pioneered in \cite{Agon:2013iva} and further explored in \cite{DeNobili:2015dla}. We computed $\entas_\n$ for $\n=2,3,4,5$ in various dimensions and numerically extrapolated to $\n=1$ using the fit ansatz $f(\n) = a + b \, \n + c \, \n^{-1} + d \, \n^{-2}$. Note that the ansatz is slightly different from the one used in \cite{Agon:2013iva, DeNobili:2015dla}. In $d=3$ dimensions, the explicit expression \cref{eq:c0-ahjk} for the function $C_{d=3}(\beta)$ allows us to analytically determine the constant in the asymptotic expansion \cref{eq:EntasBigmur} at large $\mu R$. We find (see \cref{app:ssec:determinant}):
\begin{equation}
    \det \bbM_{\n, d=3} = \frac1\pi \, \biggl( \frac{8\pi}{\n} \biggr)^{\!\! \n-1} \, \Gamma\biggl( \frac{\n}{2} \biggr)^{\!\! 2} ~.
\end{equation}
By taking the limit $\n\to1$ we obtain the constant in the asymptotic expansion of the entanglement asymmetry as well:
\begin{equation}
\label{eq:asymp-ent-asymm-d3}
    \entas \,\sim\, \frac{1}{2} \log (\mu R) + \frac12 \log \bigl( 4\pi^2 \lambda \bigr) - \frac{\gamma_\mathrm{E}}2 \qquad\qquad \text{for $d=3$} ~.
\end{equation}
Here $\gamma_\mathrm{E}$ is the Euler--Mascheroni constant. Similarly, in $d=4$ we use \cref{eq:c0-d=4} to determine
\begin{equation}
    \det \bbM_{\n, d=4} = \frac{ (8\pi^2)^{\n-1} \, \Gamma(\n)^2 }{ \n^{2\n-1} } ~,
\end{equation}
and from here the asymptotic behaviour of the entanglement asymmetry:
\begin{equation}
\label{eq:asymp-ent-asymm-d4}
    \entas \,\sim\, \log (\mu R) + \frac12 \log \bigl( 16 \pi^3 \lambda \bigr) - \gamma_\mathrm{E} - \frac12 \qquad\qquad \text{for $d=4$} ~.
\end{equation}
In \cref{fig:EntasFinal} we present our results for $d=3$, namely $\entas_\n$ for $\n = 2,3,4,5$ and the numerically extrapolated entanglement asymmetry $\entas$. We also indicate the two asymptotic limits with dashed lines. In \cref{fig:EntasGend} we present the extrapolated entanglement asymmetries for $d=4,5,6$. This time, for $d=5,6$ we do not have an analytic computation of the constant in \cref{eq:EntasBigmur}, therefore the constants have been fitted when drawing the asymptotic limits at large $\mu R$. Notice that the large $\mu R$ limits are reached faster in $d>3$ than in $d=3$: the reason is clear from \cref{eq:asymmetry-0form-exact}.

Let us interpret these results. Since we have taken the subregion to be a ball, varying its size is, in a precise sense, an RG flow in which the relevant energy scale is $R^{-1}$. We argued in \cref{eq:entasgrowth} that \(\entas\) increases along such RG flow from small to large $R$. Our results agree with this expectation: in the IR ($\mu R \gg 1$) the symmetry is broken by the ground state, while in the UV ($\mu R \ll 1$) the symmetry is restored. Furthermore, in the IR \(\entas\) captures the logarithmic growth of asymmetry for a global spatial slice as calculated in \cref{sec:entropic-CMW-global}. An important comment is in order: our computation has been done using a free Goldstone theory. If such a theory is taken to be the EFT of a spontaneously broken symmetry, its validity should end around the scale $\mu$, where arbitrary higher-derivative corrections become relevant. In this sense, \cref{fig:EntasFinal,fig:EntasGend} are only valid until $\mu R \sim 1$, precisely where the symmetry is restored. However, even if we must remain agnostic about the shape of \(\entas\) for $\mu R \lesssim 1$, it is clear that, as long as the EFT is valid, the symmetry is being restored. Indeed, even when the EFT is no longer trustworthy,  the monotonicity of \(\entas\) guarantees that the amount of symmetry breaking cannot increase as $\mu R$ decreases.

Let us close with a few comments. We have evaluated the R\'enyi asymmetries for a spherical subregion. However, the conceptual picture remains unchanged for subregions \(\aent\) of arbitrary geometry. The only difference lies in solving the corresponding Laplace equation with monodromy conditions upon encircling \(\pd\aent\). This naturally changes the form of the instanton matrix \(\bbM_{\n,d}\), and therefore the fine details of the asymmetry. Nevertheless, if we take \(R\) to serve as a proxy for the characteristic size of \(\aent\), the asymmetry retains precisely the same functional dependence as in \cref{eq:asymmetry-0form-exact}. In this sense, our conclusions concerning both the limiting behaviour of \(\entas\) for small and large subregions, and the monotonicity of R\'enyi asymmetries with respect to the subregion size, are universal.

\subsection{The theorem}
\label{ssec:theorem-subregion0-form}

We now return to the subtle case of \(d=2\) and examine the fate of asymmetries in two dimensions. As established above, in this case all components of the instanton matrix vanish identically. It follows that the instanton partition function \(\parti^\t{inst}\qty[X_\n]\) is left invariant by the symmetriser and entanglement asymmetry vanishes. Since \(\parti^\t{inst}\qty[X_\n]\) is formally divergent, though, let us be more careful by reintroducing the regulating tube \(\B^2_{\n \epsilon}\times\S^{d-2}_R\) around the entangling surface, always taken with Neumann boundary conditions for the reasons discussed earlier. Its role here is to regularise the instanton contribution:
\begin{equation}
	\parti^\t{inst}\qty[X_\n] = \Theta\qty( 2\pi\lambda \, (\mu R)^{d-2} \, \epsilon^2 \, \bbA_\n) ~,
\end{equation}
where \(\bbA_\n\) is an \((\n-1)\times(\n-1)\) matrix with entries of order one, encoding the corrections arising from the entanglement boundary, such as \cref{eq:epsilon-correction}. With \(\epsilon\ll 1\) we then obtain, to leading order in \(\epsilon\):
\begin{equation}
	\parti^\t{inst}\qty[X_\n] = \qty[\det( 2\pi\lambda \, (\mu R)^{d-2} \, \epsilon^2\, \bbA_\n)]^{-1/2} ~.
\end{equation}
By the same token, and from eqn.~\cref{eq:Z_symm}, the symmetrised partition function \(\parti_{\n,\sfS}^\t{inst}\) is given by the same expression. Hence, their ratio is unity and we find
\begin{equation}
	\entas_{\n,\U(1)} = 0
\end{equation}
in \(d=2\), for any value of the radius \(R\). Finally, in \(d<2\) the situation trivialises. Now we do not have the possibility of a meaningful subregion, as the full spatial slice is at most a collection of points. For any subregion (a collection of fewer points) we can appeal to \cref{sec:entropic-CMW-global} and conclude that \(\entas_{\n,\U(1)} = 0\) in this case too. 

This result allows us to formulate a sharp, entropic version of the CMW theorem. Phrased loosely, it states that no matter how one tunes the system in an attempt to realise spontaneous symmetry breaking (SSB), it is simply not allowed in \(d\leq 2\).
Specifically, we consider a theory describing a would-be Goldstone boson. If the symmetry were spontaneously broken, this would be the effective theory of the light degrees of freedom. We then prepare a global pure state on \(\Sigma = \R^{d-1}\) by imposing boundary conditions that explicitly break the symmetry, \(\phi(x)\to\theta\) as \(\abs{x}\to\infty\).
Next, we restrict attention to a subregion \(\aent = \bbB_R^{d-1}\) and compute the R\'enyi asymmetries of the reduced state \(\rho_\aent\). As explained above, these asymmetries provide a faithful measure of spontaneous symmetry breaking: if SSB were present, the R\'enyi asymmetries would detect it. And we find that
\begin{equation}
\label{eq:DeltaS-CMW-subregion}
	\entas_{\n,\U(1)}(R) =
	\begin{cases}
		0 ~,                                                                                  & d\leq 2 ~, \\[6pt]
		\dfrac{1}{\n-1} \, \log \Theta\qty(\dfrac{\bbM_{\n,d}^{-1}}{2\pi \lambda \, (\mu R)^{d-2}}) ~, & d>2 ~,
	\end{cases}
\end{equation}
from which the following statements immediately follow:
\begin{itemize}
	\item Even when restricting attention to a finite portion of spacetime, spontaneous symmetry breaking does not occur in \(d \leq 2\).
	\item When SSB does occur (for \(d > 2\)), the expression \cref{eq:DeltaS-CMW-subregion} provides a quantitative entropic measure of its strength within a subregion.
	\item The degree of symmetry breaking is monotonically increasing with the size of the subregion. Equation \cref{eq:DeltaS-CMW-subregion} interpolates smoothly and monotonically between the two limiting values,
	      \begin{equation}\begin{aligned}
		      \entas_{\n,\U(1)}(0) & = 0~, \qq{and} \\ 
            \entas_{\n,\U(1)}(R) & \sim \, \frac{d-2}{2} \, \log(\mu R) ~, \qquad\qq{as} R \gg \inv{\mu} ~.
	      \end{aligned}\end{equation}
\end{itemize}

\section{Subregion theorem for \tps{\U(1)}{U(1)} \tps{p}{p}-form symmetry}
\label{sec:subregionp-form}

In this section we extend our subregion theorem to the case of \(p\)-form symmetry breaking. We will be relatively brief, as the core argument closely parallels the 0-form case, with appropriate modifications.

To start, we consider a low-energy effective theory in which a \(p\)-form \(\U(1)\) symmetry is realised nonlinearly. The simplest such --- and sufficient for our purposes --- is \(p\)-form Maxwell theory, as in \cref{sec:entropic-CMW-global} with action \cref{Eq:GenMaxwell} repeated here for convenience:
\begin{equation}
	S = \frac{\lambda\, \mu^\Delta}{2} \int_{X_d} \f{f}{p+1} \wedge \star \, \f{f}{p+1} ~,
\end{equation}
where again \(\Delta=d-2p-2\), and \(\f{f}{p+1}\) is the curvature of a \(p\)-form connection \(\f{a}{p}\).

As in the 0-form case, the partition function on a generic manifold --- possibly with boundary, so long as one imposes Dirichlet or Neumann boundary conditions --- factorises \cite{Kelnhofer:2007jf, Donnelly:2016mlc}:
\begin{equation}
	\parti[X_d] = \parti^\t{inst}[X_d] \; \parti^\t{osc}[X_d] ~.
\end{equation}
The oscillator contribution reduces to an alternating product of one-loop determinants of the gauge field and its associated ghost-for-ghost tower. As before, this sector will not concern us further as its contribution cancels in the computation of the entanglement asymmetry.

The instanton contribution mirrors the structure of \cref{eq:inst-0form}, but now the sum runs over integral \((p+1)\)-form fluxes, or equivalently over the lattice spanned by \(\H^{p+1}(X_d, \pd X_d;\Z)\).\footnote{This corresponds to Dirichlet boundary conditions on \(\pd X_d\), which are the physically relevant ones in our setup. For Neumann boundary conditions, the sum would instead run over absolute cohomology.} In this case, and for the reasons discussed in \cref{sec:entropic-CMW-global}, the spatial slice of interest is
\begin{equation}
	\Sigma_{d-1} = \S^p_r \times \R^{d-p-1} ~,
\end{equation}
where we have taken the compact manifold to be a $d$-sphere \(\S^p_r\) of radius $r$ for simplicity. We do stress, however, that our computations do not depend crucially on that choice. We consider pure states \(\rho\propto \ketbra{\theta}\) on \(\Sigma\) prepared by a Euclidean path integral on \(\R\times \Sigma\), with boundary conditions that naturally generalise \cref{eq:theta-BCs-0form}:
\begin{equation}
	\int_{\S^p_r\times\set{x}} \f{a}{p} \;\to\, \theta ~, \qq{as} \abs{x}\to\infty~,
\end{equation}
where \(x\) here denotes a point in \(\R^{d-p-1}\). Without loss of generality we will consider boundary conditions with \(\theta=0\). 

Subsequently we restrict to a subregion, chosen here to be
\begin{equation}
	\aent = S^p_r \times \B^{d-p-1}_R ~,
\end{equation}
where \(\B^{d-p-1}_R\) is a \((d-p-1)\)-dimensional ball of radius \(R\). On the resulting reduced state \(\rho_\aent\) we compute the entanglement asymmetry using the techniques of \cref{sec:subregion0-form} extended to the present \(p\)-form setting.

\subsection{R\'enyi entropies of \tps{p}{p}-form Maxwell theory in \tps{d}{d} dimensions}

To calculate the Rényi entropies, the primary ingredient is \(\tr(\rho_\aent^\n)\). As before, this trace is realised by a Euclidean path integral over the replica manifold:
\begin{equation}
	\tr \qty(\rho_\aent^\n) = \parti[X_{\n}]~,
\end{equation}
where now the replica geometry has topology
\begin{equation}
	X_{\n} \simeq \S^p \times Y_{\n}~, \qq{with} Y_{\n} \coloneqq \S^{d-p} \smallsetminus \set{B_1,\cdots,B_{\n}} ~,
\end{equation}
where each \(B_i\) labels an asymptotic boundary of the \(\n\)-fold cover of the noncompact directions. It follows immediately that the relevant relative cohomology is
\begin{equation}
	\H^{p+1}\qty(X_{\n},\pd X_{\n};\Z) = \Z^{\n-1} ~.
\end{equation}
Strictly speaking, this holds for \(d\neq 2p+1\). When \(d=2p+1\) there is an additional contribution from the volume form of \(\S^p\). To avoid unnecessary complications, we will assume \(d\neq 2p+1\), though everything goes through also in this special case with a bit of care. This cohomology group is generated by \(\tau_i \w \omega_p\) (\(i=1,\ldots,\n-1\)), where \(\tau_i\) are the generators of \(\H^1(Y_\sfn,\pd Y_\sfn)\) (see \cref{App:TopRepMan}), while \(\omega_p\) is the normalised volume form on the \(p\)-sphere:
\begin{equation}
	\int_{\S^p_r} \omega_p = 1 ~.
\end{equation}

With this basis, the instanton contribution to the replica partition function follows:
\begin{equation}
\label{eq:inst-pform}
	\parti^\t{inst}\qty[X_\n] = \sum\nolimits_{\vec{w}\in\Z^{\n-1}} \; \exp( - 2\pi^2 \lambda \, \mu^\Delta \; \vec{w}^\sfT \, \bbM^{(p)} \, \vec{w}) ~,
\end{equation}
where
\begin{equation}
	\bigl[ \bbM^{(p)} \bigr]_{ij} = \int_{X_\n} \qty(\tau_i \w \omega_p) \w \star \, \qty(\tau_j \w \omega_p) = \frac{1}{r^p} \int_{Y_\n} \tau_i \w \star \, \tau_j = \frac{R^{d-p-2}}{r^p} \, \qty[\bbM_{\n,d-p}]_{ij} ~.
\end{equation}
Here we took the volume of $S^p_r$ to be $r^p$, in agreement with our conventions in \cref{sec:entropic-CMW-global}, and denoted \(\bbM_{\n,d-p}\) the instanton matrix as given in \cref{eq:MNd-components}, now evaluated in \((d-p)\) dimensions. Note again that the factor of \(R^{d-p-2}\) reflects the placement of the entangling surface at radius \(R\).

Collecting all pieces and re-expressing \cref{eq:inst-pform} in terms of the Siegel theta function \cref{eq:Theta-def}, we obtain
\begin{equation}
	\parti^\t{inst}\qty[X_\n] = \Theta\qty(\frac{ 2\pi \lambda \, \qty(\mu R)^{d-p-2}}{ \qty(\mu r)^p} \; \bbM_{\n,d-p}) ~.
\end{equation}
For completeness note that in the unreplicated case there are no nontrivial relative
\((p+1)\)-cycles, so the instanton partition function reduces to unity.

Altogether, the global contribution to the R\'enyi entropies of \(p\)-form Maxwell theory gives:
\begin{equation}
	\ent_\n^\t{inst} = \frac{1}{1-\n} \, \log\Theta\qty(\frac{2\pi \lambda \, \qty(\mu R)^{d-p-2}}{ \qty(\mu r)^p} \; \bbM_{\n,d-p}) ~,
\end{equation}
or equivalently in a Poisson-resummed form:
\begin{equation}
	\ent_\n^\t{inst} = \frac{1}{2 \, (\n-1)} \, \log\det(\frac{2\pi \lambda \, \qty(\mu R)^{d-p-2}}{ \qty(\mu r)^p} \; \bbM_{\n,d-p}) + \frac{1}{1-\n}\log \Theta\qty(\frac{ \qty(\mu r)^p}{2\pi \lambda \, (\mu R)^{d-p-2}} \; \bbM_{\n,d-p}^{-1}) .
\end{equation}

\subsection{R\'enyi asymmetries of \tps{p}{p}-form Maxwell theory in \tps{d}{d} dimensions}

This brings us to the computation of the Rényi asymmetries. The only missing piece is the integrated charged moments \(\parti_{\n,\sfS}\), defined as in \cref{eq:integrated-charged-moment}. The sole modification from the zero-form case is that the charged moments \(\cE_{\n}(\vec{\alpha})\) carry a \(p\)-sphere over each point. Pictorially,
\begin{equation}
	\cE_\n(\vec{\alpha}) = \vcenter{\hbox{\def\svgwidth{0.4\textwidth}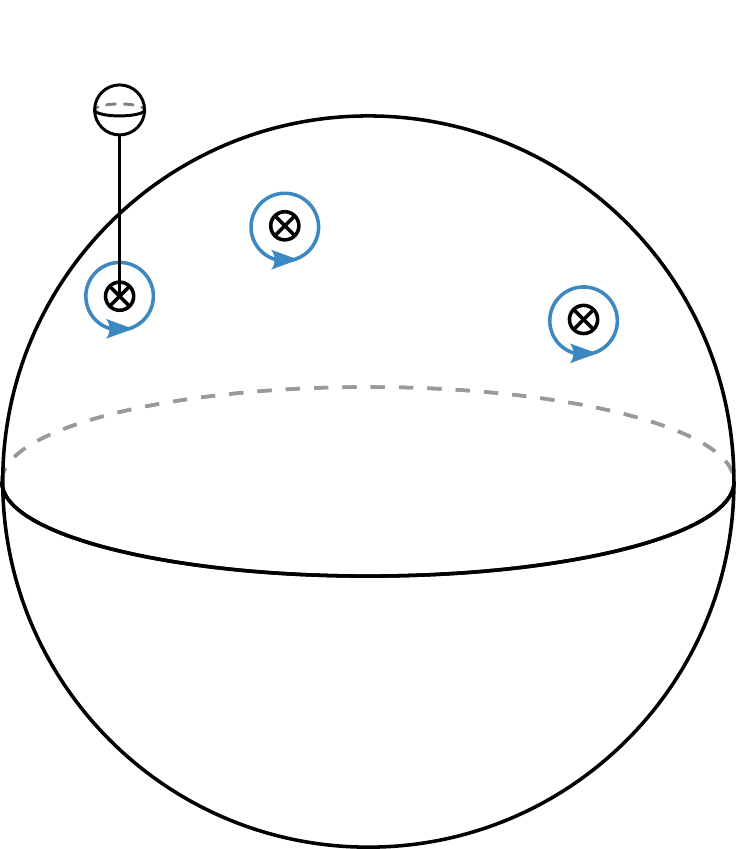}} ~.
\end{equation}

The effect of introducing topological operators is identical to the zero-form case. The boundary conditions are shifted to
\begin{equation}
	\int_{\S^p\times\set{B_i}} \f{a}{p} = \alpha_i ~,
\end{equation}
while the fluxes through the relative \((p+1)\)-cycles are modified according to
\begin{equation}
	\int_{\gamma_{p+1}^i} \f{f}{p+1} = w^i + s^i,\qq{with} s_i = \frac{1}{2\pi}(\alpha_{i}-\alpha_\n) ~.
\end{equation}
It follows immediately that the integrated charged moments take the form
\begin{equation}
	\parti_{\n,\sfS} = \qty[\det(\frac{2\pi \lambda (\mu R)^{d-p-2}}{ (\mu r)^{p}} \bbM_{\n,d-p})]^{-\frac{1}{2}} ~.
\end{equation}

Finally, substituting these quantities into the definition \cref{eq:Rényis} yields the entanglement asymmetry of \(p\)-form Maxwell theory,
\begin{equation}
\label{eq:asymmetry-pform-exact}
	\entas_\n = \frac{1}{2(\n-1)}\log\det(\frac{2\pi \lambda (\mu R)^{d-p-2}}{ (\mu r)^{p}} \bbM_{\n,d-p}) + \frac{1}{\n-1}\log \Theta\qty(\frac{2\pi \lambda (\mu R)^{d-p-2}}{ (\mu r)^{p}} \bbM_{\n,d-p})~,
\end{equation}
or equivalently:
\begin{equation}
	\entas_\n = \frac{1}{\n-1}\log \Theta\qty(\frac{ \qty(\mu r)^p}{2\pi \lambda (\mu R)^{d-p-2}}\bbM_{\n,d-p}^{-1})~.
\end{equation}

\subsection{The theorem}

In \(d = p + 2\) dimensions, the same important remark arises as in the zero-form case at \(d = 2\): the matrix \(\bbM_{\n,d}^{(p)} = \bbM_{\n,d-p}\) vanishes identically. The resolution is identical in spirit. One must again fatten the entangling region to \(\B^2_{\n\epsilon} \times \pd \aent\) in order to regulate the calculation. Upon removing the cutoff, it follows that
\begin{equation}
	\entas_{\n,\gf{\U(1)}{p}} = 0
\end{equation}
in \(d = p + 2\) dimensions. Similarly, the asymmetries also vanish identically in \(d<p+2\).

This result brings us naturally to the final expression for the Rényi asymmetries of higher-form symmetries. We find
\begin{equation}\label{eq:DeltaS-CMW-subregion-pform}
	\entas_{\n,\gf{\U(1)}{p}} =  
		\begin{cases}
		0 ~,                                                                                   & d\leq p+2~, \\[6pt]
		\dfrac{1}{\n-1}\log \Theta\qty(\dfrac{ \qty(\mu r)^p}{2\pi \lambda (\mu R)^{d-p-2}}\bbM_{\n,d-p}^{-1})~, & d>p+2 ~.
	\end{cases}
\end{equation}
In short, this computation confirms that even when restricted to a finite subregion, the higher-form analogue of the CMW theorem \cite{Gaiotto:2014kfa,Lake:2018dqm} continues to hold: a continuous \(p\)-form symmetry cannot undergo spontaneous breaking in spacetime dimensions \(d \leq p + 2\). All further consequences of \cref{eq:DeltaS-CMW-subregion} in the zero-form case --- such as the monotonic growth of symmetry breaking with the subregion size --- carry over verbatim to the higher-form setting encapsulated by \cref{eq:DeltaS-CMW-subregion-pform}. We therefore refrain from repeating them here and refer the reader to the discussion in \cref{sec:subregion0-form}.

In the limit \(\mu R \gg 1\), where the entangling region effectively covers the entire spatial slice, the entanglement spectrum flattens out --- that is, the Rényi entropies become independent of the replica index \(\n\). In this regime, extracting the entanglement asymmetry is straightforward. For \(d > p + 2\), we find
\begin{equation}
	\entas_{\gf{\U(1)}{p}} \sim \frac{d-p-2}{2}\log(\mu R) + \t{const} \qquad\qq{as} \mu R\gg 1 ~,
\end{equation}
while, as expected, it vanishes identically for \(d < p + 2\). For completeness, in the case that \(d = 2p + 2\), where the theory becomes conformal, the above limit ceases to apply. The appropriate scaling limit is instead \(R / r \gg 1\), in which case the asymmetry becomes:
\begin{equation}
		\entas_{\gf{\U(1)}{p}} \sim \frac{d-p-2}{2}\log(\frac{R}{r}) + \t{const} \qquad\qq{as} \frac{R}{r}\gg 1~.
\end{equation}    
The additive constant (slightly different between the generic and conformal cases) denotes a subleading correction, dependent on the geometry of the entangling surface \(\pd\aent\) and on the coupling \(\lambda\). In complete analogy with the \(p=0\) case, it can be computed exactly in \(d=p+3\) and \(d=p+4\), by virtue of the analytic expressions \cref{eq:c0-ahjk,eq:c0-d=4}. In all instances, this asymptotic regime happily reproduces the results of \cref{sec:entropic-CMW-global}, where the entanglement asymmetry was computed for the entire spatial slice.

\section{Discussion}\label{sec:discussion}

In this work we have defined entanglement asymmetry for higher-form symmetries, and explored its behaviour in phases with spontaneously-broken continuous invertible \(p\)-form symmetries. Along the way we encountered a number of entropic statements about symmetry breaking and the renormalisation group. There are several natural directions to extend our analysis.

A first, technical generalisation concerns the definition of higher-form entanglement asymmetry itself, introduced in \cref{sec:higher-form-asymmetry}. Throughout this work we assumed, for simplicity, that the spatial slice has no torsion in its homology. For many physical settings this is a harmless assumption, but torsional cycles are known to carry subtle information. In supersymmetric gauge theories, for example, they are sensitive to the global structure of the gauge group \cite{Razamat:2013opa}. It would therefore be interesting to understand whether entanglement asymmetry can diagnose more delicate features of symmetry-breaking patterns by probing such torsional cycles. There is a natural extension of our definition that incorporates torsion. In the main text, we reduced all sums over \(\H_{d-p-1}(\Sigma;\U(1))\) to integrals over \(\U(1)^{b_p(\Sigma)}\). However, in the presence of torsion, the universal coefficient theorem gives \(\H_{d-p-1}(\Sigma;\U(1)) \cong \U(1)^{b_p(\Sigma)} \oplus \operatorname{Tor}\qty(\H_{d-p-2}(\Sigma;\Z))\) (non-canonically). Thus, in our definition of entanglement asymmetry \cref{eq:rho-sym-def}, we would have a sum over torsional defects, on top of the integrals we considered here. We expect this to be the correct generalisation of our proposal, though we have not attempted a detailed analysis here. We leave an understanding of the role of torsion cycles in entanglement asymmetry as an open question.

A second natural extension concerns the behaviour of entanglement asymmetry for continuous non-Abelian symmetries when evaluated on subregions. In this work we restricted our non-Abelian analysis to the full spatial slice, where the construction is comparatively straightforward. For subregions one expects the entanglement asymmetry to approach the value for the global state (computed in \cref{subsec:nonabelian}) as the region becomes large, while in the opposite limit of small subregions it should vanish, reflecting the symmetry restoration in the UV. Understanding the details of this crossover would be interesting.

A more ambitious programme is to understand entanglement asymmetry for more general symmetry structures. One natural direction is that of higher-group symmetries \cite{Cordova:2018cvg, Benini:2018reh}. This case is especially interesting, as such systems exhibit a hierarchy of symmetry-breaking scales \cite{Cordova:2018cvg}. It would be interesting to understand how entanglement asymmetry detects this, particularly in light of its interpretation as an RG monotone. A closely related arena is that of noninvertible symmetries, where analogous hierarchies of breaking patterns are known \cite{Cordova:2022ieu, Choi:2022fgx, Garcia-Valdecasas:2023mis}. In several examples the associated Goldstone theory is already understood \cite{Cordova:2018cvg, Antinucci:2024bcm}, opening the door to an analysis similar to the one developed here. Along these lines, continuous noninvertible (higher-form) symmetries --- appearing, for
instance, in the orbifold branch of a compact scalar or in \(\O(2)\) gauge theory --- offer another fertile testing ground.
 
We hope that our work has convinced the reader that entanglement asymmetry --- and, more broadly, entropic order parameters --- are of great interest for quantum field theory. Surprisingly, we have only started uncovering their possibilities. We leave the further exploration of their applications as an invitation to the interested reader.

\section*{Acknowledgments}

We thank Valentin Benedetti, Kostas Chalas, Giovanni Galati, and Andy O'Bannon for useful discussions.
We also thank the organisers and participants of the workshop ``Entanglement asymmetry across energy scales'' for stimulating discussions.
S.V. would like to acknowledge the Isaac Newton Institute for Mathematical Sciences (Cambridge) for support and hospitality during the programme ``Quantum field theory with boundaries, impurities, and defects'', where work on this paper was undertaken, supported by EPSRC grant EP/Z000580/1.
F.B. is supported by the ERC-COG grant NP-QFT \textnumero~864583 “Non-perturbative dynamics of quantum fields: from new deconfined phases of matter to quantum black holes”, by the MUR-FARE grant EmGrav \textnumero~R20E8NR3HX “The Emergence of Quantum Gravity from Strong Coupling Dynamics”, by the MUR-PRIN2022 grant \textnumero~2022NY2MXY, and by the INFN “Iniziativa Specifica ST\&FI”. 
E.G.-V. is supported by MIUR PRIN Grant 2020KR4KN2 “String Theory as a bridge between Gauge Theories and Quantum Gravity” and by INFN Iniziativa Specifica ST\&FI.
S.V. is supported by a Marina Solvay fellowship and by the Fonds de la Recherche Scientifique (FNRS) under grant \textnumero~4.4503.15.

\appendix
\crefalias{section}{appendix}
\crefalias{subsection}{appendix}

\section{The topology of the replica manifold}
\label{App:TopRepMan}

In this appendix we collect some basic topological facts about the replica manifold \(X_\n\) introduced in \cref{sec:subregion0-form}. Recall that \(X_\n\) is defined as the \(\n\)-fold branched cover of \(\R^d\) with branch cut along a unit \((d-1)\)-ball. Topologically this space is equivalent to a \(d\)-sphere with \(\n\) punctures (corresponding to the $\n$ points at infinity), or equivalently with \(\n\) \(d\)-balls excised.

For the calculation of R\'enyi entropies, we require its homology and relative homology groups. For our purposes real coefficients will suffice. The absolute homology follows directly from the observation that \(X_\n\) is homotopically equivalent to a wedge sum of \((\n-1)\) \((d-1)\)-spheres:
\begin{equation}
    X_\n \cong \S^d \smallsetminus \set{\B^d_1 \sqcup \cdots \sqcup \B^d_\n} \overset{\t{\tiny hom}}{\simeq} \bigvee^{\n-1} \S^{d-1} ~.
\end{equation}
From this equivalence the homology groups are immediate:
\begin{equation}
\label{eq:hom-Mn}
    \H_k\qty(X_\n) = 
    \begin{cases}
        \R & k=0 ~, \\
        0 & 0<k<d-1 ~, \\
        \R^{\n-1} & k=d-1 ~, \\ 
        0 & k=d ~.
    \end{cases}
\end{equation}

We now turn to the relative homology of the replica manifold with respect to its boundary \(\pd X_\n = \S^{d-1}_1 \sqcup \cdots \sqcup \S^{d-1}_\n\). A standard route is to use the relative homology long exact sequence:
\begin{equation}
    \cdots \to \H_k(\pd X_\n) \to \H_k(X_\n) \to \H_k(X_\n,\pd X_\n) \to \H_{k-1}(\pd X_\n) \to \cdots ~,
\end{equation}
but in our case it is much simpler to relate relative homology to reduced homology:
\begin{equation}
    \H_k(X_\n,\pd X_\n) = \widetilde{\H}_k(X_\n / \pd X_\n) ~, 
\end{equation}
where \(\widetilde{H}_k\) denotes the reduced homology, and the quotient \(X_\n/\pd X_\n\) is the space obtained from \(X_\n\) by collapsing its boundary to a single point. See \cite{hatcher} for details. Noting that
\begin{equation}
    X_\n / \pd X_\n \overset{\t{\tiny hom}}{\simeq} \S^d\vee\bigvee^{\n-1} \S^{1} ~,
\end{equation}
we immediately obtain:
\begin{equation}
    \H_k(X_\n,\pd X_\n) = 
    \begin{cases} 
        0 & k=0 ~, \\
        \R^{\n-1} & k=1 ~, \\
        0 & 1<k<d ~, \\
        \R & k=d ~. \\
    \end{cases}
\end{equation}

For later use, we also record the cohomology groups, obtained by Poincaré–Lefschetz duality:
\begin{equation}
    \H_k(X_\n) \cong \H^{d-k}(X_\n,\pd X_\n) \quad\qq{and}\quad \H_k(X_\n,\pd X_\n) \cong \H^{d-k}(X_\n) ~.    
\end{equation} 
Thus we have:
\begin{equation}
\H^{k}(X_\n) = 
    \begin{cases} 
        \R & k=0~, \\
        0 & 0<k<d-1~, \\
        \R^{\n-1} & k=d-1~, \\
        0 & k=d~.
    \end{cases} \qquad\quad
    \H^k\qty(X_\n,\pd X_\n) = 
    \begin{cases}
        0 & k=0~, \\
        \R^{\n-1} & k=1~, \\
        0 & 1<k<d~, \\ 
        \R & k=d~. \\
    \end{cases}
\end{equation}

Finally, recall that on manifolds with boundary the de Rham picture requires some care. It is no longer true that \(\ker\lapl = \ker\dd \cap \ker\cdd\) \cite{Cappell:2005wcg}. Following \cite{Ball:2024xhf}, we refer to forms in \(\ker\dd \cap \ker\cdd\) as strongly harmonic, and we distinguish two natural boundary conditions: 
\begin{align}
\label{eq:dirichletforms}
    \operatorname{Harm}_\t{D}^k(M) &\coloneqq \set{\omega\in \Omega^k(M) \suchthat \dd\omega = 0 ~,\;\; \cdd\omega = 0 ~,\ \t{and} \;\; \omega\big|_{\pd M} = 0} \\
    \operatorname{Harm}_\t{N}^k(M) &\coloneqq \set{\omega\in \Omega^k(M) \suchthat \dd\omega = 0 ~,\;\; \cdd\omega = 0 ~,\ \t{and} \;\; \star \, \omega\big|_{\pd M} = 0} ~.
\end{align} 
These are the Dirichlet and Neumann (strongly) harmonic forms, respectively. One has the identifications \cite{Cappell:2005wcg}:
\begin{equation}
    \operatorname{Harm}_\t{D}^k(M) \cong \H^k(M,\pd M) \quad\qq{and}\quad \operatorname{Harm}_\t{N}^k(M) \cong \H^k(M) ~.
\end{equation}
Applied to the replica manifold, this means: 
\begin{itemize}
\item one Neumann harmonic 0-form: the constant function, 
\item \((\n-1)\) Dirichlet harmonic 1-forms: \(\tau_i\), \(i=1,\cdots,\n-1\), normalised against a basis of relative one-cycles \(\set{\alpha_i}\) by 
\begin{equation}
    \int_{\alpha_i} \tau_j = \delta_{ij}~.    
\end{equation}
\item \((\n-1)\) Neumann harmonic \((d-1)\)-forms. A basis of them is given by \(\star \, \tau_i\). 
\item one Dirichlet harmonic top-form: the volume form \(\vol_{X_\n} = \star \, 1\). 
\end{itemize}


\section{The instanton matrix from electrostatics}
\label{app:electrostatics}

In this appendix we derive useful expressions for the matrix \(\bbM_{\n,d}\) that appears in the main text (in particular in \cref{sec:subregion0-form,sec:subregionp-form}) in the calculation of the instanton partition function. 

Let \(\set{\tau_i}_{i=1}^{\n-1}\) be a basis of Dirichlet harmonic forms on \(X_\n\), as defined in \cref{App:TopRepMan}. Define \(\bbM_{\n,d}\) to be the Gram matrix of the above basis:
\begin{equation}
\label{eq:M_n-def-app}
	\qty[\bbM_{\n,d}]_{ij} \coloneqq \int_{X_\n} \tau_i \w \star \, \tau_j ~.
\end{equation}
By the general discussion of \cref{sec:subregion0-form,sec:subregionp-form}, this symmetric matrix controls the instanton contribution to the partition function and hence the entanglement asymmetry. The goal of this appendix is to determine \(\bbM_{\n,d}\).

\subsection{An equivalent electrostatics problem}
\label{app:ssec:equivalent-electrostatics}

It is helpful to map the above calculation to a capacitance problem in electrostatics. Denote each of the asymptotic boundaries of \(X_\n\) as \(B_j\) (\(j=1,\ldots,\n\)) and consider \((\n-1)\) electrostatic scalar potentials \(v_i\) (\(i=1,\ldots,\n-1\)) solving
\begin{equation}
\label{eq:laplace-problem}
	\lapl v_i = 0 \qq{with boundary conditions} 
	v_i \big|_{B_j} = \begin{cases} 
		\delta_{ij} ~, & j=1,\ldots,\n-1 ~, \\ 0, & j=n ~. 
	\end{cases}
\end{equation}
Here \(\lapl\) denotes the scalar Laplacian on \(X_\n\). The \(\n\)-th boundary acts as ``ground'', reflecting the fact that \(X_\n\) carries only \((\n-1\)) non-trivial one-cycles.

We may then define the one-forms \(\tau_i\) as 
\begin{equation}
	\tau_i \coloneqq \dd{v_i}~, \qquad i=1,\ldots, \n-1~. 
\end{equation}
These are automatically closed. They are also coclosed, as \(v_i\) satisfy the Laplace equation. Furthermore, they vanish on \(\pd X_\n\) since \(v_i\) are constant on each boundary component of \(X_\n\). Hence they form a basis of \(\H^1(X_\n,\pd X_\n)\). 
\begin{figure}[tbp]
	\centering
	\def\svgwidth{0.4\textwidth}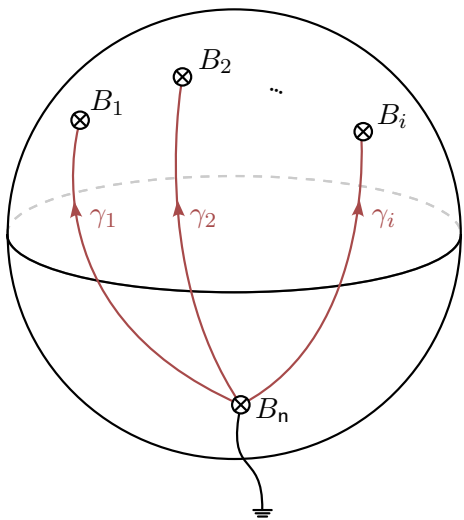
	\caption{\label{fig:homology-basis}%
    \(X_\n\) drawn as a \(d\)-sphere with \(\n\) boundaries, as explained in \cref{App:TopRepMan}. The boundaries, denoted by crossed circles, get mapped to the conductors. Of these, \(B_\n\) is grounded. The relevant basis \(\set{\gamma_i}\) of relative homology cycles is drawn in red.}
\end{figure}
Moreover, let \(\set{\gamma_i}\) be a basis of relative 1-cycles in \(H_1(X_\n,\pd X_\n)\), consisting of arcs stretching from \(B_\n\) to \(B_i\) as depicted in \cref{fig:homology-basis}. Then
\begin{equation}
	\int_{\gamma_j} \tau_i = \int_{\gamma_j} \dd{v_i} = v_i \big|_{B_j} - v_i \big|_{B_\n} = \delta_{ij} ~,
\end{equation}
and hence \(\tau_i\) are precisely the harmonic forms entering in \cref{eq:M_n-def-app}.

With this identification, the desired matrix reads 
\begin{equation}
	\qty[\bbM_{\n,d}]_{ij} = \int_{X_\n} \dd{v_i} \w \star \dd{v_j} = \int_{\pd X_\n} v_i \w\star\dd{v_j}
	= \sum_{\ell=1}^{\n-1} v_i \big|_{B_\ell} \int_{B_\ell} \star\dd{v_j} = \int_{B_i} \star\dd{v_j} ~,
\end{equation}
where we integrated by parts and used the constancy of \(v_i\) along the boundary components of \(X_\n\). To summarise, the Gram matrix computes the capacitance matrix for the multi-conductor geometry \(X_\n\). See also \cite{gross2004electromagnetic, li2025} for related approaches.

\subsection{Unfolding and the structure of \tps{\bbM_{\n,d}}{M(n,d)}}
\label{app:ssec:M-structure}

To completely determine the matrix \(\bbM_{\n,d}\) we must solve the Laplace problem posed in \cref{eq:laplace-problem}. This task becomes considerably more tractable if we ``unfold'' the problem to an equivalent one on \(\R^d\). In the following we will let the branch cut be a spherical ball $\B^{d-1}_1$ of unit radius, whereas we will reinstate a generic radius $R$ at the end.

We begin by contracting the harmonic basis vectors \(v_i\) with an arbitrary constant real vector \(\beta^i\) (with \(i = 1, \dotsc, \n - 1\)), in order to avoid carrying free indices around. This yields a harmonic scalar field:
\begin{equation}
    v_\beta\coloneqq \sum\nolimits_{i=1}^{\n-1} \beta^i v_i : X_\n\to \R ~,
\end{equation}
subject to the following boundary conditions on the asymptotic components of the replica manifold:
\begin{equation}
    v_\beta \big|_{B_i} = \beta^i~ \qquad\qq{and}\qquad v_\beta \big|_{B_\n} = 0 ~.
\end{equation}
Now, let \(x\) be a point in \(\mathbb{R}^d\) (excluding the branch cut), and denote by \(x_j\) its image on the \(j\)-th sheet of \(X_\n\). We then define a set of complex scalar potentials on \(\mathbb{R}^d\):
\begin{equation}
    \widetilde{v}_k: \R^d \to\C ~,
\end{equation} 
indexed by \(k=1,\ldots, \n\) and given explicitly by:
\begin{equation}
\label{eq:vk-from-vbeta}
    \widetilde{v}_k(x) \coloneqq \frac{1}{\cN_k} \, \sum_{j=1}^\n \; \ex{2\pi i \frac{j k}{\n}} \, v_\beta(x_j), \quad\qq{with}\quad \cN_k \coloneqq \sum_{j=1}^{\n-1} \; \ex{2\pi i \frac{j k}{\n}} \, \beta^j ~.
\end{equation}
Each \(\widetilde{v}_k\) is thus a harmonic function on \(\mathbb{R}^d\), and is held at unit potential at the asymptotic boundary:
\begin{equation}
\label{eq:asymptotics-vtildek}
    \widetilde{v}_k \big|_{\S^{d-1}_\infty} = 1 ~.
\end{equation}

Let us now observe a few key properties of the potentials \(\widetilde{v}_k\). First, by construction, each \(\widetilde{v}_k\) satisfies the Laplace equation on \(\mathbb{R}^d\). In the replicated geometry, traversing from one sheet to the next amounts to:
\begin{equation}
    v_\beta(x_j) \mapsto v_\beta(x_{j+1}) ~.
\end{equation}
In the unfolded picture, this induces a monodromy in the complex potential:
\begin{equation}
\label{eq:monodromy-vtildek}
    \widetilde{v}_k(x) \,\mapsto\, \ex{-2\pi i \, k/\n} \; \widetilde{v}_k(x)
\end{equation}
upon encircling the entangling surface \(\S^{d-2}_1\) of unit radius. The properties \cref{eq:asymptotics-vtildek} and \cref{eq:monodromy-vtildek} completely fix the harmonic functions $\widetilde{v}_k$, therefore they do not depend on the vector $\beta^i$ used in their construction. It also immediately follows that \(\widetilde{v}_\n\) is invariant under the transformation \cref{eq:monodromy-vtildek} and hence is the constant function 1. We then define the positive numbers
\begin{equation}
    C_d\qty(\frac{k}{\n}) = \int_{\R^d} \qty(\dd{\widetilde{v}_k})^* \w \star \dd{\widetilde{v}_k} \qquad\t{for}\quad k = 1, \ldots, \n-1 ~,
\end{equation}
which represent the energy stored in the fields \(\widetilde{v}_k\). By the same reasoning used before, this integrals can be equivalently evaluated as a flux through the boundary at infinity:
\begin{equation}
\label{eq:I-flux}
	C_{d}\qty(\frac{k}{\n}) = \int_{\S^{d-1}_\infty} \star \, \dd\widetilde{v}_k ~.  
\end{equation} 

We now have all the ingredients needed to determine the matrix \(\bbM_{\n,d}\). On the one hand, by definition,
\begin{equation}
    \int_{X_\n} \dd{v_\beta}\w\star\dd{v_\beta} = \sum\nolimits_{j,\ell = 1}^{\n-1} \qty[ \bbM_{\n,d} ]_{j\ell} \; \beta^j \, \beta^\ell ~.
\end{equation}
On the other hand, by unfolding the integral across the sheets:
\begin{align}
    \int_{X_\n} \dd{v_\beta} \w \star \dd{v_\beta} &= \sum_{j=1}^\n \, \int_{\t{\(j\)-th sheet}} \dd{v_\beta(x_j)}\w\star \dd{v_\beta(x_j)}
    = \frac{1}{\n} \, \sum_{k=1}^{\n-1} \, \abs{\cN_k}^2 \, C_d \qty( \frac{k}{\n}) \nn 
    &= \sum_{j,\ell=1}^{\n-1} \, \qty[ \, \frac{1}{\n} \, \sum_{k=1}^{\n-1} \; \ex{2\pi i \, \frac{(j-\ell)k}{\n}} \, C_d\qty(\frac{k}{\n}) \, ] \; \beta^j \, \beta^\ell ~.
\end{align}
In the second equality we inverted \cref{eq:vk-from-vbeta}, used that $\widetilde v_{-k}(x) = \widetilde v_k(x)^*$, and that \(\widetilde{v}_\n\) is trivial and does not contribute. Since \(\beta^j\) is arbitrary, it follows that the elements of the instanton matrix are given by:
\begin{equation}
\label{eq:M-structure}
    \qty[\bbM_{\n,d}]_{j\ell}=\frac{1}{\n} \, \sum_{k=1}^{\n-1} \; \ex{2\pi i \, \frac{(j-\ell)k}{\n}} \, C_d\qty(\frac{k}{\n}) ~.
\end{equation}

Let us also offer an alternative viewpoint. The matrix \(\bbM_{\n,d}\) as obtained in \cref{eq:M-structure} has the structure of a symmetric Toeplitz matrix. More precisely, it can be viewed as the leading principal submatrix of an \(\n\times\n\) symmetric circulant matrix. This is precisely the form expected for the capacitance matrix in cyclic arrangements of conductors \cite{fang2005}. Grounding one of the conductors, thereby reducing the number of independent potentials, we recover exactly our \(\bbM_{\n,d}\).  

Let us pause to take stock of what happened. We have shown that solving the Laplace problem on \(X_\n\) as in \cref{eq:laplace-problem} is equivalent to solving the following problem on \(\R^d\):
\begin{equation}
\label{eq:laplace-problem-2}
	\lapl \, \widetilde{v}_k = 0 \quad\qq{with}\quad \widetilde{v}_k \,\mapsto\, \ex{-2\pi\ii \: \frac{k}{\n}} \, \widetilde{v}_k \quad\qq{around the unit \(\S^{d-2}\).}
\end{equation}
In short, we have traded a partial differential equation on a complicated geometry with simple boundary conditions, for one on flat space with a nontrivial monodromy.

\subsection{Hyperfunky coordinates}

In the following section we will go on to solve the Laplace problem we set up above. That problem will simplify greatly, upon introducing the correct coordinate system. We take a moment here to do so.

\subsubsection{Hypercylindrical coordinates}

First let us introduce \emph{hypercylindrical coordinates} on \(\R^d\). They consist of a height \(\tau\in(-\infty,\infty)\), a radius \(\rho\geq 0\), and \(d-2\) angles, that we collectively denote as \(\vec{\theta}\), covering a \((d-2)\)-dimensional sphere. In these coordinates the metric of flat space takes the form
\begin{equation}
    \dd{s^2} = \dd{\tau^2} + \dd{\rho^2} + \rho^2 \dd{\Omega_{d-2}^2} ~,
\end{equation} 
where \(\dd{\Omega_{d-2}^2}\) is the line element of the unit sphere \(\S^{d-2}\). The Laplacian takes the form:
\begin{equation}
    \lapl = \pd_\tau^2 + \frac{1}{\rho^{d-2}} \, \pd_\rho\qty(\rho^{d-2} \, \pd_\rho) + \frac{1}{\rho^2} \, \lapl_{\S^{d-2}} ~,
\end{equation}
with \(\lapl_{\S^{d-2}}\) the Laplacian on a unit sphere \(\S^{d-2}\). 

\subsubsection{Oblate hyperspheroidal coordinates}

It is simple to pass to an elliptic coordinate system specified by an \emph{oblate} radial coordinate \(\zeta\geq 0\) and a \emph{hyperbolic} polar coordinate \(\eta\in[-1,1]\) covering the \((\tau,\rho)\) plane. The angular coordinates are left untouched. The relation to the cylinder variables is:
\begin{equation}
    \tau = \zeta\, \eta ~, \qquad\qquad \rho = \sqrt{\qty(1+\zeta^2)\qty(1-\eta^2)} ~.
\end{equation}
In this coordinate system the branch cut $\B^{d-1}_1$ is covered by $\{\zeta = 0,\, 0 \leq \eta \leq 1\}$ when approached from above (from positive values of $\tau$), and by $\{\zeta = 0,\, -1 \leq \eta \leq 0 \}$ when approached from below (from negative values of $\tau$), while the entangling surface is located at $\{\zeta = 0, \eta = 0\}$.
The metric looks as follows:
\begin{equation}
    \dd{s^2} = \frac{\zeta^2+\eta^2}{1+\zeta^2} \, \dd{\zeta^2} \,+\, \frac{\zeta^2+\eta^2}{1-\eta^2} \, \dd{\eta^2} \,+\, \qty(1+\zeta^2)\qty(1-\eta^2) \dd{\Omega_{d-2}^2} ~.
\end{equation}
The form of the Laplacian can be easily worked out to be
\begin{align}
    \lapl &= \frac{1}{\zeta^2+\eta^2} \Biggl[ \bigl( 1+\zeta^2 \bigr)^{\frac{3-d}{2}} \pd_\zeta \biggl( \bigl( 1+\zeta^2 \bigr)^{\frac{d-1}{2}} \pd_\zeta \biggr) + \bigl( 1-\eta^2 \bigr)^{\frac{3-d}{2}} \pd_\eta \biggl( \bigl(1-\eta^2 \bigr)^{\frac{d-1}{2}} \pd_\eta \biggr) \Biggr] \nn  
    &\quad + \frac{1}{\qty(1+\zeta^2)\qty(1-\eta^2)} \, \lapl_{\S^{d-2}} ~.
\end{align}

\subsection{Solving the Laplace equation}
\label{app:ssec:solving-laplace}

In the coordinates introduced above, the boundary value problem \cref{eq:laplace-problem-2} with $\lapl \, \widetilde{v}_k = 0$ acquires the following prescribed boundary conditions at \(\zeta\to\infty\) and \(\zeta=0\):
\begin{equation}
	\lim_{\zeta\to\infty} \widetilde{v}_k(\zeta,\eta, \vec{\theta}) = 1~, \qq{and}
	\left\lbrace \begin{aligned}
		      \widetilde{v}_k(0,\eta,\vec{\theta})          & = \ex{2\pi\ii\,\frac{k}{\n}} \, \widetilde{v}_k(0,-\eta,\vec{\theta})          \\
		      \pd_\zeta\widetilde{v}_k(0,\eta,\vec{\theta}) & = - \ex{2\pi\ii\,\frac{k}{\n}} \, \pd_\zeta\widetilde{v}_k(0,-\eta,\vec{\theta})
	      \end{aligned} \right. \quad \t{for }
	\eta>0 ~.
\end{equation}
Since both the equation and the boundary conditions are axisymmetric, we can assume the same for the solution. We will henceforth drop any dependence on \(\vec{\theta}\). 

Under separation of variables,
\begin{equation}
	\widetilde{v}_k(\zeta,\eta) = Z(\zeta) \, H(\eta) ~,
\end{equation}
the Laplace equation becomes
\begin{alignat}{3}
	& \qty[ \bigl(1+\zeta^2 \bigr) \, \pd_\zeta^2 + (d-1) \, \zeta \, \pd_\zeta - \ell \, (\ell+d-2) ] \, Z(\zeta) &&= 0 ~, \label{eq:Gegenbauer-z} \\
	& \qty[ \bigl( 1-\eta^2 \bigr) \, \pd_\eta^2 - (d-1) \, \eta \, \pd_\eta + \ell \, (\ell+d-2)] \, H(\eta) &&= 0 ~. \label{eq:Gegenbauer-h}
\end{alignat}
These are of the form of (hyperbolic) Gegenbauer equations. The solutions to \cref{eq:Gegenbauer-h} that are regular at \(\eta=\pm 1\) are given by the Gegenbauer polynomials \(\cC_\ell^{(d)}(\eta)\),%
\footnote{In the literature, see e.g. \cite{abramowitz1964handbook}, Gegenbauer polynomials are usually denoted as \(C_\ell^{(\alpha)}(\eta)\), where \(\alpha=\frac{d-2}{2}\). We label them by \(d\) in order not to clutter the notation. In other words: \(\cC_\ell^{(d)} |_\text{here} = C_\ell^{((d-2)/2)} |_\text{there}\). A similar comment holds for the Gegenbauer functions of the second kind; see for instance \cite{common1996bi}.}
while the solutions to \cref{eq:Gegenbauer-z} are \(\cC_\ell^{(d)}(\ii\zeta)\) and \(\cQ_\ell^{(d)}(\ii\zeta)\), where \(\cQ_\ell^{(d)}\) are the Gegenbauer functions of the second kind. Both \(\cC^{(d)}_\ell\) and \(\cQ^{(d)}_\ell\) can be expressed in terms of hypergeometric functions. Explicitly we have:
\begin{align}
    \cC^{(d)}_\ell(x) &=  \frac{(\ell+d-3)! }{ \ell! \, (d-3)!} \; {}_2F_1\qty( -\ell ,\, \ell + d -2 ;\, \frac{d-1}{2} ;\, \frac{1-x}{2}) ~, \\ 
    \cQ^{(d)}_\ell(x) &= \frac{\sqrt{\pi} \; \Gamma\qty(\frac{2\ell + d-1}{2}) }{ \Gamma\qty(\frac{2\ell + d}{2}) \, (2 x)^{\ell + d -2}} \; {}_2F_1\qty( \frac{\ell + d-2}{2},\, \frac{\ell + d-1}{2};\, \frac{2\ell + d}{2};\, \frac{1}{x^2}) ~.
\end{align}  
We chose the normalisations so that \(\cC_\ell^{(3)}(x)\) and \(\cQ_\ell^{(3)}(x)\) reduce to Legendre polynomials and Legendre functions of the second kind, respectively. Morover, $\cC_0{}^{\!(d)}(\eta) = 1$. Except for \(\ell=0\), the Gegenbauer polynomials blow up at \(\zeta\to\infty\), while the Gegenbauer functions of the second kind are regular for $d>2$. Hence the generic solution to \cref{eq:Gegenbauer-z,eq:Gegenbauer-h} in $d>2$ is given by 
\begin{equation}
\label{eq:modes}
    \widetilde{v}_k(\zeta,\eta) = 1 + \ii^d \, \sum_{\ell=0}^{\infty} \, c_\ell(k) \; \cQ_\ell^{(d)}(\ii\zeta) \; \cC_\ell^{(d)}(\eta) ~,
\end{equation}   
where \(c_\ell(k)\) are coefficients to be fixed by the boundary conditions. The factor $\ii^d$ has been introduced so that the $c_\ell(k)$ are real, because $\widetilde{v}_k$ is invariant under reflection $\eta \to -\eta$ accompanied by complex conjugation. In $d=2$ the discussion still applies with a single caveat: we are missing a solution in the $\ell=0$ case since $\cC^{(2)}_0(x)$ and $\cQ^{(2)}_0(x)$ are both constant, i.e. they are the same solution to \cref{eq:Gegenbauer-z}. Instead, the missing solution takes the form $Z(\zeta) \propto \operatorname{arcsinh}(\zeta)$, which however is not regular at $\zeta \to \infty$. The generic solution for $d=2$ is therefore \cref{eq:modes} with the $\ell=0$ term removed.

\subsection{Assembling the pieces}
\label{app:assemblingpieces}

Having determined the solutions \(\widetilde{v}_k(\zeta,\eta)\), we now need to plug them back in \cref{eq:I-flux} to calculate \(C_d \bigl( \frac{k}{\n} \bigr)\). The orthogonality among Gegenbauer polynomials implies
\begin{equation}
    \int_{-1}^1 \! d\eta \;\; C_\ell^{(d)}(\eta) \; \bigl( 1-\eta^2 \bigr)^{\frac{d-3}2} = 0 \qquad\text{for } \ell > 0 ~,
\end{equation}
from which it follows that only the \(\ell=0\) term contributes. Since for $d=2$ the $l=0$ term is absent, this automatically implies the vanishing of the instanton matrix in $d=2$. For $d>2$ note that, for \(\zeta\to\infty\), the Gegenbauer functions of the second kind with \(\ell=0\) behave as 
\begin{equation}
    \cQ^{(d)}_0(\ii\zeta) = \frac1{\zeta^{d-2}} \, \qty[ \frac{ \sqrt{\pi} \; \Gamma \bigl( \frac{d-1}{2} \bigr) }{ (2i)^{d-2} \, \Gamma \bigl( \frac{d}{2} \bigr) } \,+\, \order{\zeta^{-2}} ] ~.
\end{equation}
Hence, the flux at infinity reads 
\begin{equation}
\label{eq:C-and-c}
    C_d\qty(\frac{k}{\n}) = \int_{S^{d-1}_\infty} \! \star \, \dd\widetilde{v}_k = \int_{\S^{d-1}} \! \dd{\Omega_{d-1}} \eval{\qty(\zeta^{d-1} \, \pd_\zeta \widetilde{v}_k)}_{\zeta\gg 1} = \frac{ c_0(k) \, \pi^{\frac{d+1}2} \, \Gamma\bigl( \frac{d-1}2 \bigr) }{ 2^{d-4} \, \Gamma\bigl( \frac{d-2}2 \bigr) \, \Gamma\bigl( \frac d2 \bigr) }
\end{equation}
where \(d\Omega_{d-1}\) denotes the volume form on the unit sphere \(\S^{d-1}\). Thus  $C_d\bigl( \frac{k}{\n} \bigr)$ is completely determined by the boundary conditions. In \cite{Agon:2013iva} evidence was presented for the form of $C_d\bigl(\frac{k}{\n} \bigr)$ in $d=3$:
\begin{equation}
	C_3\left(\frac{k}{\n}\right) = 4\pi \qty(1-\frac{2k}{\n})\tan(\frac{\pi k}{\n}) ~,
\end{equation} 
using that $C_{d=3}(\beta) = 2 J_\text{there}(\beta)$.
This allows an explicit evaluation of the Gram matrix, which reads:
\begin{equation}
	\qty[\bbM_{\n,3}]_{j\ell} = \frac{4\pi}{\n}\sum_{k=1}^{\n-1} \, \ex{2\pi \ii \frac{(j-\ell)k}{\n}} \qty(1-\frac{2k}{\n})\tan(\frac{\pi k}{\n}) ~.
\end{equation}
This matrix is real and positive definite. Below we will present a conjecture for $C_4(\beta)$.

To reinstate the dimensionful radius $R$ of the subregion $\cA = \B^{d-1}_R$, we simply rescale the metric by a factor of $R^2$. The Dirichlet forms $\tau_i$ and the potentials $v_i$ are dimensionless, while $\star \, \tau_i$ have dimension of $(\t{length})^{d-2}$ and get rescaled by $R^{d-2}$. It follows that the instanton matrix becomes $R^{d-2} \, [\bbM_{\n,d}]_{ij}$.

\subsection{Numerical evaluation}
\label{subsec:numericaleval}

We have not been able to find a closed expression for $C_d\qty(k/\n)$ in the general case. We can, however, compute it numerically following appendix B.3 of \cite{Agon:2013iva} to which we refer for details. One may show that, formally,
\begin{equation}
\label{eq:GenDimc0}
	C_d\left(\frac{k}{\n}\right) = \frac{4 \, \pi^{\frac{d-1}2} \, \Gamma\bigl( \frac{d-1}{2} \bigr) }{ \Gamma\bigl( \frac{d-2}{2} \bigr) \, \Gamma\bigl( \frac{d}{2} \bigr)} \; u^{\sfT} \qty[\mathbf{1} + \cot[2](\frac{\pi\,k}{\n})\, T^{(d)}]^{-1} u ~.
\end{equation}
Here $T^{(d)}$ is an infinite-dimensional matrix, $\mathbf{1}$ is the identity matrix, whereas $u = \left(\delta_{\ell\, 0}\right)_{\ell \text{ even}}$ is an infinite-dimensional vector with a single $1$ in the first position, which extracts the element in the upper-left corner of the inverse matrix in brackets. 
The matrix $\bigl[ T^{(d)} \bigr]_{\ell m}$ (for $\ell,m$ even) is given by
\begin{equation}
\label{eq:Td-def}
    T^{(d)} \coloneqq S^{(d)} \, Q^{(d)}_- \, \qty(S^{(d)})^\sfT \, \qty(Q^{(d)}_+)^{-1} ~,
\end{equation}
whose definition in turn involves the diagonal matrices $Q^{(d)}_+$ (for $\ell$ even) and $Q^{(d)}_-$ (for $\ell$ odd), and the matrix $S^{(d)}$ (for $\ell$ even, $m$ odd). Their matrix elements are given by:
\begin{align}
	\qty[Q^{(d)}_+]_{\ell \ell } &\coloneqq \ii \, \frac{\cQ^{(d)}_\ell(0)}{\cQ^{(d)\prime}_\ell(0)} \quad (\ell \text{ even}) ~,
    \qquad\qquad
	\qty[Q^{(d)}_-]_{\ell \ell} \coloneqq \ii \, \frac{\cQ^{(d)}_\ell(0)}{\cQ^{(d)\prime}_\ell(0)} \quad (\ell \text{ odd}) ~, \\
	\qty[S^{(d)}]_{\ell m} &\coloneqq \left(K_\ell^{(d)}\,K_{m}^{(d)}\right)^{-1/2}
	\int_{-1}^1 \! \dd{\eta}\; \mathrm{sgn}(\eta) \; w_d(\eta) \; \cC^{(d)}_\ell(\eta) \; \cC^{(d)}_{m}(\eta) ~
	\quad (\ell \text{ even} ,\ m \text{ odd}) ~. \nonumber
\end{align}
In the above we defined:
\begin{equation}
	w_d(\eta)\coloneqq \qty(1-\eta^2)^{\frac{d-3}{2}} ~, \qquad\qq{and} K_\ell^{(d)} \coloneqq \int_{-1}^{1} \dd{\eta} \, w_d(\eta) \, \qty(\cC^{(d)}_\ell(\eta))^2 ~.
\end{equation}
The explicit expressions for the matrix elements of \(Q_\pm^{(d)}\) and \(S^{(d)}\) are given by:
\begin{equation}
	\qty[Q^{(d)}_\pm]_{\ell \ell} = \frac{\Gamma \bigl( \frac{\ell + 1}{2} \bigr) \, \Gamma\bigl( \frac{\ell + d - 2}{2} \bigr) }{ 2 \, \Gamma\bigl( \frac{\ell + 2}{2} \bigr) \, \Gamma\bigr( \frac{ \ell + d -1 }{2} \bigr) } ~, \qquad
    K_\ell^{(d)} = \frac{ \pi \, 2^{4-d} \, (\ell + d -3)! }{ (2\ell + d - 2) \, \ell! \, \Gamma\bigl( \frac{d-2}2 \bigr)^2 } ~,
\end{equation}
and
\begin{equation}
    \qty[S^{(d)}]_{\ell m} = \frac{ 2^{d-2} \, i \, \cA_\ell \, \cB_m }{ \pi \, (\ell + m + d - 2) (\ell - m) } \qquad (\ell \text{ even} ,\ m \text{ odd}) ~, 
\end{equation}
where we defined
\begin{equation}
     \cA_\ell \coloneqq \ii^\ell \, \frac{\Gamma\bigl( \frac{\ell+d-2}2 \bigr) }{ \Gamma\bigl( \frac{\ell + 2}2 \bigr) } \, \sqrt{ \frac{ (2\ell+d-2) \, \ell!}{ (\ell+d-3)! }} ~, \quad
    \cB_m \coloneqq \ii^m \, \frac{ \Gamma\bigl( \frac{m+d-1}2 \bigr) }{ \Gamma\bigl( \frac{m+1}2 \bigr) } \, \sqrt{ \frac{ (2m+d-2) \, m! }{ (m+d-3)! }} ~.
\end{equation}
The entries of \(T^{(d)}\) follow from \cref{eq:Td-def} after performing one infinite sum:
\begin{equation}
T^{(d)}_{\ell m} = \begin{cases}
\displaystyle \frac{ - 2^{d-2} \, \cA_\ell \, \cB_m }{ \pi^2 \, (\ell + m + d - 2) } \; \frac{ \psi\bigl( \frac{1-\ell}2 \big) + \psi\bigl( \frac{\ell + d-1}2 \bigr) - \psi\bigl( \frac{1-m}2 \big) - \psi\bigl( \frac{m + d-1}2 \bigr) }{ \ell - m } \quad & \ell \neq m \\[1.2em]
\displaystyle \frac1{ \pi^2 } \; \Bigl[ \psi_1 \bigl( \tfrac{1-\ell}2 \bigr) - \psi_1 \bigl( \tfrac{\ell + d - 1}2 \bigr) \Bigr] & \ell = m
\end{cases}
\end{equation}
for $\ell, m$ even, where $\psi(x) = \partial_x \log\Gamma(x)$ is the digamma function, while $\psi_1(x) = \partial_x^2  \log\Gamma(x)$ is the trigamma function.%
\footnote{Here, for $d=3$, we correct a typo in eqn.~(B.36) of \cite{Agon:2013iva}.}
Note that $T_{\ell\ell}$ is the $m \to \ell$ limit of $T_{\ell m}$. One could use the reflection formula $\psi(1-z) - \psi(z) = \pi \cot(\pi z)$ to rewrite $\psi\bigl( \frac{1-\ell}2 \bigr) = \psi\bigl( \frac{\ell + 1}2 \bigr)$ for $\ell \in 2\mathbb{Z}$.

The formal solution \cref{eq:GenDimc0} shows that $C_d(x)$ is a real function and that it satisfies $C_d(x) = C_d(1-x)$. In turn, this guarantees that $\bbM_{\n,d}$ in \cref{eq:M-structure} is a real symmetric matrix. Provided that $C_d(k/\n)>0$ for all $k=1, \ldots, \n-1$ (which is not obvious from \cref{eq:GenDimc0}), one also easily proves that $\bbM_{\n,d}$ is strictly positive definite.

\subsection{Conjectured analytic expressions}
\label{app:ssec:analytic}

The inverse matrix in \cref{eq:GenDimc0} can be written as a series expansion:
\begin{equation}
    F_d(\lambda) \coloneqq  \qty[ \qty( \mathbf{1} + \lambda^2 \, T^{(d)} )^{-1}]_{00} = 1 + \sum\nolimits_{p=1}^\infty (-\lambda^2)^p \, \qty[ T^{(d)\, p} ]_{00} \;.
\end{equation}
Thus, one can attempt to compute the upper-left element of the powers of the matrix $T$.

For $d=3$, with computer assistance, we computed the first three powers analytically: $[T]_{00} = 1/3$, $\bigl[ T^2 \bigr]_{00} = 1/5$, $\bigl[ T^3 \bigr]_{00} = 1/7$. It was indeed proposed in \cite{Agon:2013iva} that
\begin{equation}
    \bigl[ T^p \bigr]_{00} = \frac{1}{2p+1} \qquad\qquad\text{for } d=3 ~.
\end{equation}
We numerically computed higher powers, up to $p=10$, by truncating $T^{(3)}$ to a matrix of size $N=2000$, and found an accurate agreement with the proposed pattern. Further evidence was presented in \cite{Agon:2013iva}. The series can be resummed:
\begin{equation}
    F_3(\lambda) = \frac{\arctan(\lambda)}{\lambda} = \frac\pi2 \, (1-2\beta) \tan(\pi \beta) ~,
\end{equation}
where we used $\lambda = \cot(\pi \beta)$. As a further check, notice that $\lim_{\lambda\to\infty} F(\lambda) = 0$, as it should be since $e^{2\pi i \beta}$ is the monodromy around the entangling surface, and when the monodromy is trivial the harmonic function is trivial as well.

For $d=4$, with computer assistance, we also computed the first three powers analytically. That prompted us to conjecture the following pattern:
\begin{equation}
    \bigl[ T^p \bigr]_{00} = \frac{4 \, H^\text{odd}_p}{\pi^2 \, p} \;,\qquad H^\text{odd}_p = \sum_{j=1}^p \frac1{2j-1} = \frac12 \, \Bigl( \psi\bigl( p + \tfrac12 \bigr) - \psi\bigl( \tfrac12 \bigr) \Bigr) \qquad \text{for } d=4 ~.
\end{equation}
We checked it numerically up to $p=10$ with matrices of size $N=2000$, obtaining agreement with accuracy around $10^{-5}$. The series can be resummed:
\begin{equation}
    F_4(\lambda) = 1 - \frac{4}{\pi^2} \arctan^2(\lambda) = 4 \beta(1-\beta) ~.
\end{equation}
Also in this case, $\lim_{\lambda\to\infty} F(\lambda) = 0$. It would be nice to prove this conjecture.

\subsection{Matrix determinant in \tps{d=3,4}{d=3,4}}
\label{app:ssec:determinant}

For a square matrix of size $(\n-1)$ of the form
\begin{equation}
    \bigl[ \bbM_\n \bigr]_{j\ell} = \frac1{\n} \, \sum_{k=1}^{\n-1} \, e^{2\pi i (j-\ell)k/\n} \; \zeta_k
\end{equation}
as in \cref{eq:MNd-components}, the determinant is $\det \bbM_\n = \frac1{\n} \prod_{k=1}^{\n-1} \zeta_k$. The function $C_{d=3}(\beta)$ in \cref{eq:c0-ahjk} for $\beta = \frac12$ is defined as a limit, therefore
\begin{equation}
    \zeta_k = \begin{cases} 4\pi \bigl( 1 - \frac{2k}{\n}\bigr) \tan\bigl( \frac{\pi k}{\n} \bigr) &\text{for } k \neq \n/2 \;,\\ 8 & \text{for } k = \n/2 \;. \end{cases}
\end{equation}
Notice that $\zeta_{\n-k} = \zeta_k$. When computing the product of the $\zeta_k$'s, the following identities are useful:
\begin{equation}
    \text{For $\n$ odd:} \quad \prod_{k=1}^{(\n-1)/2} \tan\biggl( \frac{\pi k}{\n} \biggr) = \sqrt \n ~. \qquad
    \text{For $\n$ even:} \quad \prod_{k=1}^{\n/2 -1} \tan\biggl( \frac{\pi k}{\n} \biggr) = 1 ~.
\end{equation}
We thus find that, for $d=3$, the determinant of $\bbM_{\n,3}$ is:
\begin{equation}
    \det \bbM_{\n,3} = \begin{cases} \dfrac{ (4\pi)^{\n-1} \, \bigl( (\n-2)!! \bigr)^2 }{ \n^{\n-1} } \qquad &\text{for $\n$ odd} ~, \\[1em]
    \dfrac{ 8 \, (4\pi)^{\n-2} \, \bigl( (\n-2)!! \bigr)^2 }{ \n^{\n-1}} \qquad &\text{for $\n$ even} ~. \end{cases}
\end{equation}
By writing the double factorial in terms of a gamma function, one obtains an analytic continuation in $\n$ that captures both expressions:
\begin{equation}
    \det \bbM_{\n,3} = \frac1\pi \, \qty( \frac{8\pi}{\n} )^{\n-1} \, \Gamma\qty( \frac{\n}2 )^2 ~.
\end{equation}
The Taylor expansion of this expression for $\n \to 1$ is
\begin{equation}
    \det \bbM_{\n,3} = 1 + \bigl( \log(2\pi) - \gamma_\mathrm{E} \bigr)(\n-1) + \order{\qty(\n-1)^2} ~,
\end{equation}
which we used to obtain the asymptotic entanglement asymmetry \cref{eq:asymp-ent-asymm-d3}. Those expressions can also be used to determine the R\'enyi and entanglement entropies in \cref{sssec:entanglement-entropy}.

The case $d=4$ is simpler. We have $C_4(\beta) = 8\pi^2 \beta(1-\beta)$ as in \cref{eq:c0-d=4}, therefore
\begin{equation}
    \det \bbM_{\n,4} = \frac{(8\pi^2)^{\n-1} \, \Gamma(\n)^2 }{ \n^{2\n-1} } ~.
\end{equation}
The expansion for $\n \to 1$ is
\begin{equation}
    \det \bbM_{\n,4} = 1 + \Bigl( \log(8\pi^2) - 2 \gamma_\mathrm{E} - 1 \Bigr) (\n-1) + \order{\qty(\n-1)^2} ~,
\end{equation}
which we used to obtain the asymptotic entanglement asymmetry \cref{eq:asymp-ent-asymm-d4}.

\section{Shrinking the entanglement boundary}
\label{app:smooth-Bent}

In this appendix we argue that the winding/instanton contribution to the replica partition function remains smooth upon shrinking the entanglement boundary. As a reminder, in order to compute this contribution, one needs to evaluate the instanton matrix \(\bbM_\n\), i.e. the induced metric on \(\H^1(X_\n,\pd X_\n)\). Here we show that this matrix depends smoothly on the length \(\epsilon\) of the entanglement boundary \(B_\t{ent}\) when \(\epsilon\ll 1\).

We begin by noting that \(B_\t{ent}\) introduces an additional relative one-cycle in the homology \(\H_1(X_\n,\pd X_\n)\), on which we impose ``clear-cut'' boundary conditions \cite{Ohmori:2014eia}. Those are given by Dirichlet or Neumann BCs. Let us examine each case separately.

First consider the case of Dirichlet BCs
\begin{equation}
    \f{f}{1} \Big|_{B_\t{ent}} = 0 ~,
\end{equation}
and denote by \(\bbM_\n\) the instanton matrix of \(X_\n\) prior to regulating. After introducing the cutoff, the relevant matrix is the Gram matrix of \(\H^1(X_\n,\pd X_\n)\). It takes the augmented form:
\begin{equation}
    \widetilde{\bbM}_\n = \mqty(\bbM_\n & \vec{m} \\[4pt] \vec{m}^\sfT & \widetilde{m}) ~.
\end{equation}
The new entries, the vector \(\vec{m}\) and the number \(\widetilde{m}\), arise due to the cutoff and depend on \(\epsilon\).%
\footnote{Strictly speaking, \(\bbM_\n\) also receives corrections from the introduction of the cutoff. However these are subleading and vanish in the limit \(\epsilon\to 0\), so we neglect them for this argument.}
Our aim is to show that both \(\vec{m}\) and \(\widetilde{m}\) vanish in the \(\epsilon \to 0\) limit. 

In \cref{app:ssec:equivalent-electrostatics} we reinterpret \(\bbM_\n\) as the capacitance matrix of a system of \(\n\) conducting \((d - 1)\)-spheres. Once the regulating surface is introduced, the problem becomes that of \(\n\) spherical conductors alongside a very thin, toroidal conductor: the excised tube \(\B^2_{\n\epsilon} \times S^{d-2}_R\). Thus, \(\widetilde{m}\) corresponds to the self-capacitance of this toroidal capacitor. Given that \(\epsilon \ll R\), the geometry is effectively cylindrical. In \(d = 3\), this is nothing but a coaxial cable, whose capacitance behaves as:
\begin{equation}
    \widetilde{m}_{d=3} \,\propto\, \frac{1}{\log\epsilon} \qq{as} \epsilon\to 0 ~.
\end{equation} 
For \(d > 3\), a similar calculation yields a power-law decay:
\begin{equation}
    \widetilde{m}_{d>3} \,\propto\, \epsilon^{d-3} \qq{as} \epsilon\to 0 ~.
\end{equation} 
The \(d=2\) case is more subtle. The regulator becomes disconnected, forming two round plates of radius \(\n\epsilon\) separated by a distance \(R \gg \n\epsilon\), both held at unit potential. The outer boundary (one of the asymptotic boundaries) is grounded. The relevant quantity is the total charge on these plates, which again decays logarithmically \cite{miyoshi2023estimating}:
\begin{equation}
    \widetilde{m}_{d=2} \,\propto\, \frac{1}{\log\epsilon} \qq{as} \epsilon\to 0 ~.
\end{equation}
Hence, in all cases, the self-capacitance \(\widetilde{m}\) vanishes as the cutoff is removed. It remains to show that \(\vec{m}\) decays as well. This follows directly from the Cauchy--Schwarz inequality (since \(\widetilde{\bbM}_\n\) is a Gram matrix):
\begin{equation}
    \abs{m_i} \leq \widetilde{m}\, [\bbM_\n]_{ii}\to 0 \qq{as} \epsilon\to 0 ~.
\end{equation} 
In other words, we see that with Dirichlet BCs, the instanton matrix reduces effectively to 
\begin{equation}
    \widetilde{\bbM}_{n,d} = \mqty(\bbM_{\n,d} & \vec{0} \\[4pt] \trans{\vec{0}} & 0)
\end{equation}
in the limit \(\epsilon\to 0\). 

The other option is to impose Neumann BCs on the entanglement boundary:
\begin{equation}
    \qty(\star\f{f}{1}) \Big|_{B_\t{ent}} = 0 ~.
\end{equation}
This choice of boundary conditions does not require as elaborate an argument to ensure regularity as the brick wall is shrunk to zero. The instanton configurations are again captured by the Gram matrix of \(\H^1\qty(X_\n, B_\t{asym})\), which is precisely the matrix \(\bbM_\n\) introduced earlier, and calculated in \cref{app:electrostatics}. The only subtlety lies in confirming that \(\bbM_\n\) does not depend on \(\epsilon\) in a dangerous way.

Continuing with the electrostatics analogy, the presence of the regulating tube corresponds to vanishing electric permittivity inside \(\B^2_{\n \epsilon}\times \S^{d-2}_R\). Consider one of the diagonal elements of \(\bbM_\n\), say the \(ii\) component, which computes the self-capacitance of \(B_i\):
\begin{equation}
    M_i = \int_{X_\n} \tau_i \w\star \, \tau_i ~.
\end{equation}
The difference from the unregulated case is then given by
\begin{equation}
\label{eq:epsilon-correction}
    \delta M_i = \int_{\B^2_{\n \epsilon}\times \S^{d-2}_R} \tau_i\w\star \, \tau_i \;\sim\, \epsilon^2 R^{d-2} + \order{\epsilon^4} ~,
\end{equation} 
where we used that the regulating tube is very small, so \(\tau_i\) are approximately constant in that region. The corrections to the off-diagonal terms exhibit the same behaviour with \(\epsilon\).

\printbibliography

\end{document}

%% file: preamble.tex
\newcommand{\ent}{\cS}
\newcommand{\entas}{\Delta\ent}
\newcommand{\relent}[2]{\cS\qty(#1\,\middle\Vert\, #2)}

\newcommand{\aent}{\cA}
\newcommand{\co}[1]{#1'}
\newcommand{\coaent}{\co\aent}

\usepackage{bbm}
\usepackage[normalem]{ulem}

\newcommand{\n}{\sfn}

\newcommand{\fund}{\vcenter{\hbox{\ensuremath{\Box}}}}

\newcommand{\tps}[2]{\texorpdfstring{\ensuremath{#1}}{#2}}

\RequirePackage[top=12mm,bottom=12mm,left=30mm,right=30mm,head=12mm,includeheadfoot]{geometry}
\RequirePackage{tocloft}

\RequirePackage[nottoc,notlot,notlof]{tocbibind}

\RequirePackage{titlesec}
\titlespacing*{\section}{0pt}{1.8\baselineskip}{\baselineskip}

\usepackage{lmodern}
\usepackage{amsfonts,amssymb,amsthm}
\usepackage{upgreek,mathrsfs}
\usepackage{bold-extra}
\usepackage{microtype}

\makeatletter
\g@addto@macro\bfseries{\boldmath}
\makeatother

\usepackage[dvipsnames]{xcolor}
\definecolor{maroon}{RGB}{167,74,74}
\definecolor{magreen}{RGB}{59,125,37}
\definecolor{mablue}{RGB}{59,136,195}
\definecolor{mayellow}{RGB}{242,147,24}
\colorlet{mabrown}{maroon!50!magreen}
\colorlet{maorange}{maroon!50!mayellow}
\colorlet{macyan}{magreen!50!mablue}

\colorlet{red}{maroon}
\colorlet{green}{magreen}
\colorlet{blue}{mablue}
\colorlet{yellow}{mayellow}
\colorlet{brown}{mabrown}
\colorlet{orange}{maorange}
\colorlet{cyan}{macyan}

\RequirePackage{hyperref}
\hypersetup{colorlinks,
	linkcolor=mablue,
	citecolor=magreen,
	urlcolor=maorange,
	anchorcolor=mablue,
	filecolor=mablue,
	menucolor=mablue}

\makeatletter
\renewcommand\section{\@startsection{section}{1}{\z@}%
  {-3.5ex \@plus -1.3ex \@minus -.7ex}%
  {2.3ex \@plus.4ex \@minus .4ex}%
  {\large\bfseries}}
\renewcommand\subsection{\@startsection{subsection}{2}{\z@}%
	{-2.3ex\@plus -1ex \@minus -.5ex}%
	{1.2ex \@plus .3ex \@minus .3ex}%
	{\normalsize\bfseries}}
\renewcommand\subsubsection{\@startsection{subsubsection}{3}{\z@}%
	{-2.3ex\@plus -1ex \@minus -.5ex}%
	{1ex \@plus .2ex \@minus .2ex}%
	{\normalsize\bfseries}}
\renewcommand\paragraph{\@startsection{paragraph}{4}{\z@}%
	{1.75ex \@plus1ex \@minus.2ex}%
	{-1em}%
	{\normalsize\bfseries}}
\renewcommand\subparagraph{\@startsection{subparagraph}{5}{\parindent}%
	{1.75ex \@plus1ex \@minus .2ex}%
	{-1em}%
	{\normalsize\bfseries}}
\makeatother

\usepackage[style=numeric-comp, eprint=true, natbib, backend=biber, maxbibnames=10, giveninits=true,isbn=false, url=false, doi=false, sorting=none, useprefix=true]{biblatex}
\usepackage{hyperref}

\DeclareFieldFormat*{title}{\emph{#1}}

\DeclareFieldFormat*{journaltitle}{#1}

\DeclareFieldFormat{pages}{#1}

\DeclareFieldFormat{labelalpha}{\textsc{#1}}

\renewbibmacro{in:}{}

\AtEveryBibitem{\clearfield{urldate}}
\AtEveryBibitem{\clearfield{month}}
\AtEveryBibitem{\clearfield{day}}
\AtEveryBibitem{\clearfield{issn}}
\AtEveryBibitem{\clearfield{version}}

\DeclareFieldFormat{doilink}{%
	\iffieldundef{doi}{%
		\iffieldundef{url}{%
			\iffieldundef{isbn}{%
				\iffieldundef{issn}{%
					#1%
				}{%
					\href{http://books.google.com/books?vid=ISSN\thefield{issn}}{#1}%
				}%
			}{%
				\href{http://books.google.com/books?vid=ISBN\thefield{isbn}}{#1}%
			}%
		}{%
			\href{\thefield{url}}{#1}%
		}%
	}{%
		\href{http://dx.doi.org/\thefield{doi}}{#1}%
	}%
}
\DeclareBibliographyDriver{article}{%
	\usebibmacro{bibindex}%
	\usebibmacro{begentry}%
	\usebibmacro{author/translator+others}%
	\setunit{\labelnamepunct}\newblock
	\usebibmacro{title}\addcomma\space%
	\newunit
	\printlist{language}%
	\newunit\newblock
	\usebibmacro{byauthor}%
	\newunit\newblock
	\usebibmacro{bytranslator+others}%
	\newunit\newblock
	\printfield{version}%
	\newunit\newblock
	\printtext[doilink]{%
		\usebibmacro{journal+issuetitle}%
		\newunit
		\usebibmacro{byeditor+others}%
		\newunit
		\usebibmacro{note+pages}%
	}%
	\newunit\newblock
	\iftoggle{bbx:isbn}
	{\printfield{issn}}
	{}%
	\newunit\newblock
	\usebibmacro{doi+eprint+url}%
	\newunit\newblock
	\usebibmacro{addendum+pubstate}%
	\setunit{\bibpagerefpunct}\newblock
	\usebibmacro{pageref}%
	\usebibmacro{finentry}}

\usepackage{comment}
\usepackage[inline]{enumitem}
\usepackage{tikz-cd}
\usepackage{appendix}
\usepackage{empheq}

\newcommand{\nn}{\nonumber \\}

\renewcommand{\geq}{\geqslant}
\renewcommand{\leq}{\leqslant}

\usepackage{physics}
\newcommand{\parti}{\mathcal{Z}}

\let\f\relax
\newcommand{\f}[2]{{#1}_{#2}} 
\newcommand{\gf}[2]{{#1}^{\qty[#2]}} 

\usepackage{stackengine} 

\newcommand{\suchthat}{\;\middle|\;}
\newcommand{\pd}{\partial} 

\newcommand{\w}{\mathbin{\scalebox{0.8}{$\wedge$}}}

\newcommand{\inv}[1]{{{#1}^{-1}}} 
\newcommand{\trans}[1]{{{#1}^{\sfT}}} 

\renewcommand{\vec}{\boldsymbol}
\newcommand{\ex}[1]{\mathrm{e}^{#1}}
\newcommand{\ii}{i} 

\newcommand{\set}[1]{\qty{#1}}

\let\C\undefined
\let\U\undefined
\newcommand{\R}{\mathbb{R}} 
\newcommand{\C}{\mathbb{C}} 
\newcommand{\Z}{\mathbb{Z}} 
\renewcommand{\S}{S} 
\newcommand{\B}{\mathbb{B}} 

\newcommand{\U}{\mathrm{U}}

\renewcommand{\O}{\mathrm{O}}

\newcommand*{\id}{\text{\usefont{U}{bbold}{m}{n}1}}

\renewcommand{\H}{H} 

\renewcommand{\t}[1]{{\text{#1}}}
\newcommand{\s}[1]{{\scriptscriptstyle #1}}

\usepackage{amsthm}
\theoremstyle{definition}

\def \cA {\mathcal{A}}
\def \cB {\mathcal{B}}
\def \cC {\mathcal{C}}

\def \cE {\mathcal{E}}

\def \cH {\mathcal{H}}

\def \cN {\mathcal{N}}
\def \cO {\mathcal{O}}

\def \cQ {\mathcal{Q}}
\def \cR {\mathcal{R}}
\def \cS {\mathcal{S}}

\def \bbA {\mathbb{A}}
\def \bbB {\mathbb{B}}

\def \bbM {\mathbb{M}}

\def \bbZ {\mathbb{Z}}
\def \bb1 {{\mathbb{1}}}

\usepackage[outline]{contour}

\def \sfR {\mathsf{R}}
\def \sfS {\mathsf{S}}
\def \sfT {\mathsf{T}}

\def \sfb {\mathsf{b}}

\def \sfn {\mathsf{n}}

\newcommand{\cdd}{\dd^{\dagger}}

\usepackage{csquotes}
\usepackage{import}

\newcommand{\lapl}{\Delta}

\def\fdiffd{\mathrm{D}}
\DeclareDocumentCommand\fdifferential{ o g d() }{ 
	\IfNoValueTF{#2}{
		\IfNoValueTF{#3}
		{\fdiffd\IfNoValueTF{#1}{}{^{#1}}}
		{\mathinner{\fdiffd\IfNoValueTF{#1}{}{^{#1}}\argopen(#3\argclose)}}
	}
	{\mathinner{\fdiffd\IfNoValueTF{#1}{}{^{#1}}#2} \IfNoValueTF{#3}{}{(#3)}}
}
\DeclareDocumentCommand\DD{}{\fdifferential}

\DeclareDocumentCommand\variation{ o g d() }{ 
	\IfNoValueTF{#2}{
		\IfNoValueTF{#3}
		{\updelta \IfNoValueTF{#1}{}{^{#1}}}
		{\mathinner{\updelta \IfNoValueTF{#1}{}{^{#1}}\argopen(#3\argclose)}}
	}
	{\mathinner{\updelta \IfNoValueTF{#1}{}{^{#1}}#2} \IfNoValueTF{#3}{}{(#3)}}
}

\def\sdiffd{\mathbf{d}}
\DeclareDocumentCommand\sdifferential{ o g d() }{ 
	\IfNoValueTF{#2}{
		\IfNoValueTF{#3}
		{\sdiffd\IfNoValueTF{#1}{}{^{#1}}}
		{\mathinner{\sdiffd\IfNoValueTF{#1}{}{^{#1}}\argopen(#3\argclose)}}
	}
	{\mathinner{\sdiffd\IfNoValueTF{#1}{}{^{#1}}#2} \IfNoValueTF{#3}{}{(#3)}}
}

\DeclareMathOperator{\volume}{vol}
\DeclareDocumentCommand\vol{}{\opbraces{\volume}}

\DeclareMathOperator{\kernel}{ker}
\DeclareDocumentCommand\ker{}{\opbraces{\kernel}}

\RequirePackage{interval}
\intervalconfig{soft open fences}
\newcommand{\ropen}[2]{\interval[open right]{#1}{#2}}

\newcommand{\closed}[2]{\interval{#1}{#2}}

\usepackage[noabbrev]{cleveref}

\crefname{subsection}{subsection}{subsections}
\crefname{equation}{}{}
\crefname{appsec}{appendix}{appendices}

\numberwithin{equation}{section}

%% file: figures/rho.pdf_tex
\begingroup%
  \makeatletter%
  \providecommand\color[2][]{%
    \errmessage{(Inkscape) Color is used for the text in Inkscape, but the package 'color.sty' is not loaded}%
    \renewcommand\color[2][]{}%
  }%
  \providecommand\transparent[1]{%
    \errmessage{(Inkscape) Transparency is used (non-zero) for the text in Inkscape, but the package 'transparent.sty' is not loaded}%
    \renewcommand\transparent[1]{}%
  }%
  \providecommand\rotatebox[2]{#2}%
  \newcommand*\fsize{\dimexpr\f@size pt\relax}%
  \newcommand*\lineheight[1]{\fontsize{\fsize}{#1\fsize}\selectfont}%
  \ifx\svgwidth\undefined%
    \setlength{\unitlength}{450.79079936bp}%
    \ifx\svgscale\undefined%
      \relax%
    \else%
      \setlength{\unitlength}{\unitlength * \real{\svgscale}}%
    \fi%
  \else%
    \setlength{\unitlength}{\svgwidth}%
  \fi%
  \global\let\svgwidth\undefined%
  \global\let\svgscale\undefined%
  \makeatother%
  \begin{picture}(1,1.40697802)%
    \lineheight{1}%
    \setlength\tabcolsep{0pt}%
    \put(0,0){\includegraphics[width=\unitlength,page=1]{rho.pdf}}%
    \put(0.44385249,1.23780855){\color[rgb]{0,0,0}\makebox(0,0)[lt]{\lineheight{1.25}\smash{\begin{tabular}[t]{l}$\scriptstyle\Sigma$\end{tabular}}}}%
    \put(0.37512081,0.06971084){\color[rgb]{0,0,0}\makebox(0,0)[lt]{\lineheight{1.25}\smash{\begin{tabular}[t]{l}$\overline{\scriptstyle\Sigma}$\end{tabular}}}}%
    \put(0.53176783,0.55600591){\color[rgb]{0,0,0}\makebox(0,0)[lt]{\lineheight{0}\smash{\begin{tabular}[t]{l}\footnotesize $\dagger$\end{tabular}}}}%
  \end{picture}%
\endgroup%

%% file: figures/ReducedDensity.pdf_tex
\begingroup%
  \makeatletter%
  \providecommand\color[2][]{%
    \errmessage{(Inkscape) Color is used for the text in Inkscape, but the package 'color.sty' is not loaded}%
    \renewcommand\color[2][]{}%
  }%
  \providecommand\transparent[1]{%
    \errmessage{(Inkscape) Transparency is used (non-zero) for the text in Inkscape, but the package 'transparent.sty' is not loaded}%
    \renewcommand\transparent[1]{}%
  }%
  \providecommand\rotatebox[2]{#2}%
  \newcommand*\fsize{\dimexpr\f@size pt\relax}%
  \newcommand*\lineheight[1]{\fontsize{\fsize}{#1\fsize}\selectfont}%
  \ifx\svgwidth\undefined%
    \setlength{\unitlength}{219.76856142bp}%
    \ifx\svgscale\undefined%
      \relax%
    \else%
      \setlength{\unitlength}{\unitlength * \real{\svgscale}}%
    \fi%
  \else%
    \setlength{\unitlength}{\svgwidth}%
  \fi%
  \global\let\svgwidth\undefined%
  \global\let\svgscale\undefined%
  \makeatother%
  \begin{picture}(1,1.66146648)%
    \lineheight{1}%
    \setlength\tabcolsep{0pt}%
    \put(0,0){\includegraphics[width=\unitlength,page=1]{ReducedDensity.pdf}}%
    \put(0.98113195,0.35052705){\color[rgb]{0.66666667,0,0}\makebox(0,0)[lt]{\lineheight{1.25}\smash{\begin{tabular}[t]{l}\footnotesize $\mathcal{A}$\end{tabular}}}}%
    \put(0.99425355,1.09102803){\color[rgb]{0.4,0.50196078,0}\makebox(0,0)[lt]{\lineheight{1.25}\smash{\begin{tabular}[t]{l}\footnotesize $\overline{\mathcal{A}}$\end{tabular}}}}%
    \put(0.3561891,1.33358403){\color[rgb]{0,0,0}\makebox(0,0)[lt]{\lineheight{0}\smash{\begin{tabular}[t]{l}\footnotesize $\dagger$\end{tabular}}}}%
  \end{picture}%
\endgroup%

%% file: figures/Traces.pdf_tex
\begingroup%
  \makeatletter%
  \providecommand\color[2][]{%
    \errmessage{(Inkscape) Color is used for the text in Inkscape, but the package 'color.sty' is not loaded}%
    \renewcommand\color[2][]{}%
  }%
  \providecommand\transparent[1]{%
    \errmessage{(Inkscape) Transparency is used (non-zero) for the text in Inkscape, but the package 'transparent.sty' is not loaded}%
    \renewcommand\transparent[1]{}%
  }%
  \providecommand\rotatebox[2]{#2}%
  \newcommand*\fsize{\dimexpr\f@size pt\relax}%
  \newcommand*\lineheight[1]{\fontsize{\fsize}{#1\fsize}\selectfont}%
  \ifx\svgwidth\undefined%
    \setlength{\unitlength}{369.87852562bp}%
    \ifx\svgscale\undefined%
      \relax%
    \else%
      \setlength{\unitlength}{\unitlength * \real{\svgscale}}%
    \fi%
  \else%
    \setlength{\unitlength}{\svgwidth}%
  \fi%
  \global\let\svgwidth\undefined%
  \global\let\svgscale\undefined%
  \makeatother%
  \begin{picture}(1,0.98718383)%
    \lineheight{1}%
    \setlength\tabcolsep{0pt}%
    \put(0,0){\includegraphics[width=\unitlength,page=1]{Traces.pdf}}%
    \put(0.34689386,0.8983109){\color[rgb]{0,0,0}\makebox(0,0)[lt]{\lineheight{0}\smash{\begin{tabular}[t]{l}$\dagger$\end{tabular}}}}%
    \put(0.87803486,0.70227022){\color[rgb]{0,0,0}\makebox(0,0)[lt]{\lineheight{0}\smash{\begin{tabular}[t]{l}$\dagger$\end{tabular}}}}%
    \put(0.73281907,0.17619865){\color[rgb]{0,0,0}\makebox(0,0)[lt]{\lineheight{0}\smash{\begin{tabular}[t]{l}$\dagger$\end{tabular}}}}%
    \put(0.16705861,0.33001286){\color[rgb]{0,0,0}\makebox(0,0)[lt]{\lineheight{0}\smash{\begin{tabular}[t]{l}$\dagger$\end{tabular}}}}%
  \end{picture}%
\endgroup%

%% file: figures/ChargedMoment.pdf_tex
\begingroup%
  \makeatletter%
  \providecommand\color[2][]{%
    \errmessage{(Inkscape) Color is used for the text in Inkscape, but the package 'color.sty' is not loaded}%
    \renewcommand\color[2][]{}%
  }%
  \providecommand\transparent[1]{%
    \errmessage{(Inkscape) Transparency is used (non-zero) for the text in Inkscape, but the package 'transparent.sty' is not loaded}%
    \renewcommand\transparent[1]{}%
  }%
  \providecommand\rotatebox[2]{#2}%
  \newcommand*\fsize{\dimexpr\f@size pt\relax}%
  \newcommand*\lineheight[1]{\fontsize{\fsize}{#1\fsize}\selectfont}%
  \ifx\svgwidth\undefined%
    \setlength{\unitlength}{369.87852562bp}%
    \ifx\svgscale\undefined%
      \relax%
    \else%
      \setlength{\unitlength}{\unitlength * \real{\svgscale}}%
    \fi%
  \else%
    \setlength{\unitlength}{\svgwidth}%
  \fi%
  \global\let\svgwidth\undefined%
  \global\let\svgscale\undefined%
  \makeatother%
  \begin{picture}(1,0.98718383)%
    \lineheight{1}%
    \setlength\tabcolsep{0pt}%
    \put(0,0){\includegraphics[width=\unitlength,page=1]{ChargedMoment.pdf}}%
    \put(0.31039548,0.88614456){\color[rgb]{0,0,0}\makebox(0,0)[lt]{\lineheight{0}\smash{\begin{tabular}[t]{l}$\dagger$\end{tabular}}}}%
    \put(0.87803486,0.7428247){\color[rgb]{0,0,0}\makebox(0,0)[lt]{\lineheight{0}\smash{\begin{tabular}[t]{l}$\dagger$\end{tabular}}}}%
    \put(0.75715106,0.17619865){\color[rgb]{0,0,0}\makebox(0,0)[lt]{\lineheight{0}\smash{\begin{tabular}[t]{l}$\dagger$\end{tabular}}}}%
    \put(0.15023096,0.2947756){\color[rgb]{0,0,0}\makebox(0,0)[lt]{\lineheight{0}\smash{\begin{tabular}[t]{l}$\dagger$\end{tabular}}}}%
    \put(0,0){\includegraphics[width=\unitlength,page=2]{ChargedMoment.pdf}}%
    \put(0.59277831,0.75356299){\color[rgb]{0.66666667,0,0}\makebox(0,0)[lt]{\lineheight{1.25}\smash{\begin{tabular}[t]{l}$U_{a_1}$\end{tabular}}}}%
    \put(0.71686168,0.59909346){\color[rgb]{0.66666667,0,0}\makebox(0,0)[lt]{\lineheight{1.25}\smash{\begin{tabular}[t]{l}$U^\dagger_{a_1}$\end{tabular}}}}%
    \put(0.20850738,0.60264443){\color[rgb]{0.66666667,0,0}\makebox(0,0)[lt]{\lineheight{1.25}\smash{\begin{tabular}[t]{l}$U_{a_2}$\end{tabular}}}}%
    \put(0.31022334,0.7306338){\color[rgb]{0.66666667,0,0}\makebox(0,0)[lt]{\lineheight{1.25}\smash{\begin{tabular}[t]{l}$U^\dagger_{a_2}$\end{tabular}}}}%
  \end{picture}%
\endgroup%

%% file: figures/replica3d.pdf_tex
\begingroup%
  \makeatletter%
  \providecommand\color[2][]{%
    \errmessage{(Inkscape) Color is used for the text in Inkscape, but the package 'color.sty' is not loaded}%
    \renewcommand\color[2][]{}%
  }%
  \providecommand\transparent[1]{%
    \errmessage{(Inkscape) Transparency is used (non-zero) for the text in Inkscape, but the package 'transparent.sty' is not loaded}%
    \renewcommand\transparent[1]{}%
  }%
  \providecommand\rotatebox[2]{#2}%
  \newcommand*\fsize{\dimexpr\f@size pt\relax}%
  \newcommand*\lineheight[1]{\fontsize{\fsize}{#1\fsize}\selectfont}%
  \ifx\svgwidth\undefined%
    \setlength{\unitlength}{969.14312744bp}%
    \ifx\svgscale\undefined%
      \relax%
    \else%
      \setlength{\unitlength}{\unitlength * \real{\svgscale}}%
    \fi%
  \else%
    \setlength{\unitlength}{\svgwidth}%
  \fi%
  \global\let\svgwidth\undefined%
  \global\let\svgscale\undefined%
  \makeatother%
  \begin{picture}(1,0.20224048)%
    \lineheight{1}%
    \setlength\tabcolsep{0pt}%
    \put(0,0){\includegraphics[width=\unitlength,page=1]{replica3d.pdf}}%
  \end{picture}%
\endgroup%

%% file: figures/charged.pdf_tex
\begingroup%
  \makeatletter%
  \providecommand\color[2][]{%
    \errmessage{(Inkscape) Color is used for the text in Inkscape, but the package 'color.sty' is not loaded}%
    \renewcommand\color[2][]{}%
  }%
  \providecommand\transparent[1]{%
    \errmessage{(Inkscape) Transparency is used (non-zero) for the text in Inkscape, but the package 'transparent.sty' is not loaded}%
    \renewcommand\transparent[1]{}%
  }%
  \providecommand\rotatebox[2]{#2}%
  \newcommand*\fsize{\dimexpr\f@size pt\relax}%
  \newcommand*\lineheight[1]{\fontsize{\fsize}{#1\fsize}\selectfont}%
  \ifx\svgwidth\undefined%
    \setlength{\unitlength}{353.24382019bp}%
    \ifx\svgscale\undefined%
      \relax%
    \else%
      \setlength{\unitlength}{\unitlength * \real{\svgscale}}%
    \fi%
  \else%
    \setlength{\unitlength}{\svgwidth}%
  \fi%
  \global\let\svgwidth\undefined%
  \global\let\svgscale\undefined%
  \makeatother%
  \begin{picture}(1,0.99842932)%
    \lineheight{1}%
    \setlength\tabcolsep{0pt}%
    \put(0,0){\includegraphics[width=\unitlength,page=1]{charged.pdf}}%
    \put(0.18408152,0.77259924){\color[rgb]{0,0,0}\makebox(0,0)[lt]{\lineheight{1.25}\smash{\begin{tabular}[t]{l}$B_1$\end{tabular}}}}%
    \put(0.42126439,0.86428992){\color[rgb]{0,0,0}\makebox(0,0)[lt]{\lineheight{1.25}\smash{\begin{tabular}[t]{l}$B_2$\end{tabular}}}}%
    \put(0.81292893,0.74284034){\color[rgb]{0,0,0}\makebox(0,0)[lt]{\lineheight{1.25}\smash{\begin{tabular}[t]{l}$B_i$\end{tabular}}}}%
    \put(0.556895,0.09288935){\color[rgb]{0,0,0}\makebox(0,0)[lt]{\lineheight{1.25}\smash{\begin{tabular}[t]{l}$B_\n$\end{tabular}}}}%
    \put(0.57391269,0.81495442){\color[rgb]{0,0,0}\rotatebox{1.6246088}{\makebox(0,0)[lt]{\lineheight{1.25}\smash{\begin{tabular}[t]{l}.\end{tabular}}}}}%
    \put(0.58424198,0.81258862){\color[rgb]{0,0,0}\rotatebox{1.6246088}{\makebox(0,0)[lt]{\lineheight{1.25}\smash{\begin{tabular}[t]{l}.\end{tabular}}}}}%
    \put(0.59411164,0.81004001){\color[rgb]{0,0,0}\rotatebox{1.6246088}{\makebox(0,0)[lt]{\lineheight{1.25}\smash{\begin{tabular}[t]{l}.\end{tabular}}}}}%
    \put(0.11549435,0.66862539){\color[rgb]{0.23137255,0.53333333,0.76470588}\makebox(0,0)[lt]{\lineheight{1.25}\smash{\begin{tabular}[t]{l}$\alpha_1$\end{tabular}}}}%
    \put(0.33996985,0.76261947){\color[rgb]{0.23137255,0.53333333,0.76470588}\makebox(0,0)[lt]{\lineheight{1.25}\smash{\begin{tabular}[t]{l}$\alpha_2$\end{tabular}}}}%
    \put(0.74592827,0.63531752){\color[rgb]{0.23137255,0.53333333,0.76470588}\makebox(0,0)[lt]{\lineheight{1.25}\smash{\begin{tabular}[t]{l}$\alpha_i$\end{tabular}}}}%
    \put(0.46794099,0.02588695){\color[rgb]{0.23137255,0.53333333,0.76470588}\makebox(0,0)[lt]{\lineheight{1.25}\smash{\begin{tabular}[t]{l}$\alpha_\n$\end{tabular}}}}%
  \end{picture}%
\endgroup%

%% file: figures/charged-pform.pdf_tex
\begingroup%
  \makeatletter%
  \providecommand\color[2][]{%
    \errmessage{(Inkscape) Color is used for the text in Inkscape, but the package 'color.sty' is not loaded}%
    \renewcommand\color[2][]{}%
  }%
  \providecommand\transparent[1]{%
    \errmessage{(Inkscape) Transparency is used (non-zero) for the text in Inkscape, but the package 'transparent.sty' is not loaded}%
    \renewcommand\transparent[1]{}%
  }%
  \providecommand\rotatebox[2]{#2}%
  \newcommand*\fsize{\dimexpr\f@size pt\relax}%
  \newcommand*\lineheight[1]{\fontsize{\fsize}{#1\fsize}\selectfont}%
  \ifx\svgwidth\undefined%
    \setlength{\unitlength}{353.24382019bp}%
    \ifx\svgscale\undefined%
      \relax%
    \else%
      \setlength{\unitlength}{\unitlength * \real{\svgscale}}%
    \fi%
  \else%
    \setlength{\unitlength}{\svgwidth}%
  \fi%
  \global\let\svgwidth\undefined%
  \global\let\svgscale\undefined%
  \makeatother%
  \begin{picture}(1,1.15355911)%
    \lineheight{1}%
    \setlength\tabcolsep{0pt}%
    \put(0,0){\includegraphics[width=\unitlength,page=1]{charged-pform.pdf}}%
    \put(0.18891977,1.04330036){\color[rgb]{0,0,0}\makebox(0,0)[lt]{\lineheight{1.25}\smash{\begin{tabular}[t]{l}$\scriptstyle\S^p_r$\end{tabular}}}}%
    \put(0.18408152,0.77259924){\color[rgb]{0,0,0}\makebox(0,0)[lt]{\lineheight{1.25}\smash{\begin{tabular}[t]{l}$B_1$\end{tabular}}}}%
    \put(0.42126439,0.86428992){\color[rgb]{0,0,0}\makebox(0,0)[lt]{\lineheight{1.25}\smash{\begin{tabular}[t]{l}$B_2$\end{tabular}}}}%
    \put(0.81292893,0.74284035){\color[rgb]{0,0,0}\makebox(0,0)[lt]{\lineheight{1.25}\smash{\begin{tabular}[t]{l}$B_i$\end{tabular}}}}%
    \put(0.57391269,0.81495442){\color[rgb]{0,0,0}\rotatebox{1.6246088}{\makebox(0,0)[lt]{\lineheight{1.25}\smash{\begin{tabular}[t]{l}.\end{tabular}}}}}%
    \put(0.58424198,0.81258862){\color[rgb]{0,0,0}\rotatebox{1.6246088}{\makebox(0,0)[lt]{\lineheight{1.25}\smash{\begin{tabular}[t]{l}.\end{tabular}}}}}%
    \put(0.59411164,0.81004001){\color[rgb]{0,0,0}\rotatebox{1.6246088}{\makebox(0,0)[lt]{\lineheight{1.25}\smash{\begin{tabular}[t]{l}.\end{tabular}}}}}%
    \put(0.11549435,0.6686254){\color[rgb]{0.23137255,0.53333333,0.76470588}\makebox(0,0)[lt]{\lineheight{1.25}\smash{\begin{tabular}[t]{l}$\alpha_1$\end{tabular}}}}%
    \put(0.33996985,0.76261947){\color[rgb]{0.23137255,0.53333333,0.76470588}\makebox(0,0)[lt]{\lineheight{1.25}\smash{\begin{tabular}[t]{l}$\alpha_2$\end{tabular}}}}%
    \put(0.74592827,0.63531752){\color[rgb]{0.23137255,0.53333333,0.76470588}\makebox(0,0)[lt]{\lineheight{1.25}\smash{\begin{tabular}[t]{l}$\alpha_i$\end{tabular}}}}%
    \put(0,0){\includegraphics[width=\unitlength,page=2]{charged-pform.pdf}}%
    \put(0.41339078,1.13899772){\color[rgb]{0,0,0}\makebox(0,0)[lt]{\lineheight{1.25}\smash{\begin{tabular}[t]{l}$\scriptstyle\S^p_r$\end{tabular}}}}%
    \put(0,0){\includegraphics[width=\unitlength,page=3]{charged-pform.pdf}}%
    \put(0.8193492,1.01169578){\color[rgb]{0,0,0}\makebox(0,0)[lt]{\lineheight{1.25}\smash{\begin{tabular}[t]{l}$\scriptstyle\S^p_r$\end{tabular}}}}%
  \end{picture}%
\endgroup%

%% file: figures/homology-basis.pdf_tex
\begingroup%
  \makeatletter%
  \providecommand\color[2][]{%
    \errmessage{(Inkscape) Color is used for the text in Inkscape, but the package 'color.sty' is not loaded}%
    \renewcommand\color[2][]{}%
  }%
  \providecommand\transparent[1]{%
    \errmessage{(Inkscape) Transparency is used (non-zero) for the text in Inkscape, but the package 'transparent.sty' is not loaded}%
    \renewcommand\transparent[1]{}%
  }%
  \providecommand\rotatebox[2]{#2}%
  \newcommand*\fsize{\dimexpr\f@size pt\relax}%
  \newcommand*\lineheight[1]{\fontsize{\fsize}{#1\fsize}\selectfont}%
  \ifx\svgwidth\undefined%
    \setlength{\unitlength}{353.24528503bp}%
    \ifx\svgscale\undefined%
      \relax%
    \else%
      \setlength{\unitlength}{\unitlength * \real{\svgscale}}%
    \fi%
  \else%
    \setlength{\unitlength}{\svgwidth}%
  \fi%
  \global\let\svgwidth\undefined%
  \global\let\svgscale\undefined%
  \makeatother%
  \begin{picture}(1,1.12559194)%
    \lineheight{1}%
    \setlength\tabcolsep{0pt}%
    \put(0,0){\includegraphics[width=\unitlength,page=1]{homology-basis.pdf}}%
    \put(0.18408284,0.89975847){\color[rgb]{0,0,0}\makebox(0,0)[lt]{\lineheight{1.25}\smash{\begin{tabular}[t]{l}$B_1$\end{tabular}}}}%
    \put(0.42126473,0.99144877){\color[rgb]{0,0,0}\makebox(0,0)[lt]{\lineheight{1.25}\smash{\begin{tabular}[t]{l}$B_2$\end{tabular}}}}%
    \put(0.81292764,0.8699997){\color[rgb]{0,0,0}\makebox(0,0)[lt]{\lineheight{1.25}\smash{\begin{tabular}[t]{l}$B_i$\end{tabular}}}}%
    \put(0.5467257,0.22121708){\color[rgb]{0,0,0}\makebox(0,0)[lt]{\lineheight{1.25}\smash{\begin{tabular}[t]{l}$B_\n$\end{tabular}}}}%
    \put(0.5739124,0.94211347){\color[rgb]{0,0,0}\rotatebox{1.6246088}{\makebox(0,0)[lt]{\lineheight{1.25}\smash{\begin{tabular}[t]{l}.\end{tabular}}}}}%
    \put(0.58424164,0.93974769){\color[rgb]{0,0,0}\rotatebox{1.6246088}{\makebox(0,0)[lt]{\lineheight{1.25}\smash{\begin{tabular}[t]{l}.\end{tabular}}}}}%
    \put(0.59411126,0.93719908){\color[rgb]{0,0,0}\rotatebox{1.6246088}{\makebox(0,0)[lt]{\lineheight{1.25}\smash{\begin{tabular}[t]{l}.\end{tabular}}}}}%
    \put(0.18408284,0.65241949){\color[rgb]{0.65490196,0.29019608,0.29019608}\makebox(0,0)[lt]{\lineheight{1.25}\smash{\begin{tabular}[t]{l}$\gamma_1$\end{tabular}}}}%
    \put(0.40279701,0.65241949){\color[rgb]{0.65490196,0.29019608,0.29019608}\makebox(0,0)[lt]{\lineheight{1.25}\smash{\begin{tabular}[t]{l}$\gamma_2$\end{tabular}}}}%
    \put(0.8026028,0.65241949){\color[rgb]{0.65490196,0.29019608,0.29019608}\makebox(0,0)[lt]{\lineheight{1.25}\smash{\begin{tabular}[t]{l}$\gamma_i$\end{tabular}}}}%
  \end{picture}%
\endgroup%